\documentclass[article,nojss]{jss}
\usepackage[utf8]{inputenc}
\usepackage[T1]{fontenc}
\usepackage{times}
\usepackage[english]{babel}
\usepackage{amsmath}
\usepackage{amssymb}
\usepackage{amsthm}
\usepackage{tikz}
\usepackage{caption}
\usepackage{subcaption}
\author{Daniel Kosiorowski\\Cracow University of Economics \And
Zygmunt Zawadzki \\Cracow University of Economics}

\title{\pkg{DepthProc}: An \pkg{R} Package for Robust Exploration of Multidimensional Economic Phenomena}

\Plainauthor{Daniel Kosiorowski, Zygmunt Zawadzki} 
\Plaintitle{DepthProc An R Package for Robust Exploration of Multidimensional Economic
Phenomena} 
\Shorttitle{DepthProc An R Package for...} 

\Address{
Daniel Kosiorowski\\
Department of Statistics\\
Faculty of Management\\
Cracow University of Economics\\
31-510 Cracow, Poland\\
E-mail: \email{daniel.kosiorowski@uek.krakow.pl}\\
URL: \url{https://e-uczelnia.uek.krakow.pl/course/view.php?id=137}
}
\usepackage[utf8]{inputenc}
\usepackage[T1]{fontenc}
\Abstract{Data depth concept offers a variety of powerful and user-friendly tools for the robust exploration and inference of complex socio economic phenomena. Because of their nonparametric nature, offered techniques may be successfully used in cases of our lack of knowledge on parametric models that generate data. This paper presents an \proglang{R} package \pkg{DepthProc}, which is available
 under GPL-2 licences on CRAN and R-forge servers for Windows, Linux, and OS X platform. The package consists of successful implementations of several depth-based techniques involving multivariate quantile-quantile plots, multivariate scatter estimators, local Wilcoxon tests for multivariate as well as functional data, and robust regressions. In order to show the package capabilities, real datasets concerning \emph{The Fourth Millennium Goal of The United Nations } realization evaluation, a relationship between  the minimal wage and the unemployment rate in France, the Internet users activity, and the air pollution in cities Katowice and Cracow in Poland within a day and night are used.}
\Plainkeywords{statistical depth function, robust data analysis, multivariate methods, R} 

\begin{document}
\section{Introduction}
 Modern economics crucially depend on advances in applications of recent developments in statistics. Let us take, for instance, a theory and practice of a portfolio optimization, a practice of credit scoring, an evaluation of results of governmental aid programs, a creation of a taxation system, an assessment of attractiveness of candidates on a labor market, the monitoring of the concentration of dangerous particles in the atmosphere on a particular day and night, and a reconciliation of the electricity supply and  demand during a hot summer.

Unfortunately, in economics, we often cannot use powerful tools of the classical multivariate statistics based on the mean vector, the covariance matrix, and the normality assumptions. In great part, economic phenomena cannot be modeled by means of elliptically contoured light-tailed one-modal distributions. Usually, our knowledge of economic laws is not sufficient for an efficient parametric modeling of the economic systems. Moreover, "today economics" significantly differs from "tomorrow economics", because of the technological development and/or due to the appearance of the new social phenomena. Additionally, the datasets under our consideration consist of outliers and/or inliers of various kind and/or we have to cope with the missing data phenomenon. Data arrive to an observer in packages of different sizes in unequally spaced time periods (\cite{JMS}). Because of the existence of outliers within an income or expenditures data, analyses conducted using classical measures of social inequalities based on the Lorenz curve may lead to wrong political decisions (\cite{kosiorowski2014income}).

Robust statistics aims at identifying a tendency represented by an influential majority of data and at detecting the observations departing from that tendency (\cite{Marona:2006}, \cite{Wilcox}, \cite{Chebana}). Nonparametric and robust statistical procedures are especially useful in economics, where an activity of the majority of the influential agents determines the behavior of a market, the closeness to a financial crash, etc. From a conceptual point of view, robust statistics is closely related to well-known economic ideas like \textit{Pareto's effectiveness} or \textit{Nash's equilibrium} (\cite{Mizera:2002}, \cite{Kotz:2003}).

The main aim of this paper is to present an \pkg{R} package (\cite{R}) \pkg{DepthProc} consisting of successful implementations of a selection of nonparametric and robust procedures belonging to the so-called \emph{Data Depth Concept} (DDC), which are especially useful in robust exploration of socio economic phenomena. The package is available under GPL-2 license on \proglang{CRAN}, \proglang{R-forge}, and \proglang{GitHub} servers.\\
The rest of the paper is organized as follows: Section 2 introduces basic notions related to the DDC, Section 3 presents procedures offered by the package and Section 4 presents illustrative examples of the available procedures applications. The paper ends with some conclusions and references. All empirical datasets and examples studied in the paper are available after installing the package.

 This paper uses the following notation. ${S}^{d-1}$  is the $(d-1)$ dimensional unit sphere in ${\mathbb{R}}^{d}$ , ${S}^{d-1}=\left\{ x\in {{\mathbb{R}}^{d}}:\left\| x \right\|=1 \right\}$ . $\mathcal{B}^{d}$ denotes Borel $\sigma$ algebra in ${\mathbb{R}}^{d}$. The transpose of a vector $x\in \mathbb{R}^{d}$ is written by ${x}^\top$. A sample consisting of $n$ observations is denoted by ${X}^{n}=\{{x}_{1},...,{x}_{n}\},$ $F$ denotes a probability distribution in $\mathbb{R}^{d}$, and ${F}_{n}$ its empirical counterpart.

\section{Data depth concept}
Data depth concept was originally introduced as a method to generalize the concepts of the median and
the quantiles to a multivariate framework. A detailed presentation of the concept may be found in
\cite{Liu:1999}, \cite{Zuo:2000}, \cite{Serfling:2004}, \cite{Serfling:2006}, and \cite{Mosler:2013}. Nowadays the DDC offers various powerful techniques for the exploration and inference of economic phenomena involving robust clustering and classification, robust quality control and streaming data analysis, robust multivariate location, scale, and symmetry tests. Important theoretical aspects of the concept could be found, for example, in \cite{Ruts1999depth}, \cite{Zuo:2000}, \cite{Zuo:2003}, \cite{Dyck:2004}, \cite{Kong:2010}. Recent developments of the computational aspects of especially important multidimensional depths are presented in some studies, for example, \cite{ZuoLai2011},  \cite{Shao:2012}, \cite{LiuZuoWang2013}, \cite{ProjMatlab2015}, \cite{DyckMoz2016}, \cite{Dyck2016}. In recent years, very interesting concepts of depth for functional data have also been proposed (\cite{Nieto} and \cite{Nagy} with an overview and very useful comparative study of several functional depths). Within our package \pkg{DepthProc}, one can find the so-called location depths and their derivatives, i.e., the regression depth and the Student depth, the modified band depth for functional data (see \cite{Lop:2009}), and several examples of integrated functional depth (\cite{Nagy}). The \pkg{DepthProc} also implements the concept of local depth presented in \cite{Pain:2012} and \cite{Pain:2013}. The local versions of depth are available for multivariate as well as functional data. A developer version of the package, which is available on \proglang{R-forge} and \proglang{GitHub} servers, additionally consists of algorithms for performing several depth-induced clustering (\cite{zakopanemediany}), procedures for classification (\cite{zakopaneSVM}), and procedures for the detection of \emph{size} as well as \emph{shape} functional outliers (\cite{Nicolas}, \cite{Func_out}), a \emph{weighted by the local depth kernel estimation of a predictive distribution of data stream procedure}, and procedures dedicated for detecting a structural change in functional time series (\cite{Horvath}, \cite{Structural}).

\subsection{Basic definitions}
Consider the depth of a point, with respect to a probability distribution. Let
$\mathcal{P}_{0}$ be the set of all probability measures on $(\mathbb{R}^{d},\mathcal{B}^{d})$
and $\mathcal{P}$ a subset of $\mathcal{P}_{0}$. To each probability measure $F\in
\mathcal{P},$ a real function $D(\cdot ;F):{\mathbb{R}^{d}}\to {{\mathbb{R}}_{+}}$, the so-called depth function with respect to $F,$ the depth assigns.

The set of all points that have depth at least $\alpha $ is called \textbf{$\alpha-$trimmed region}. The $\alpha-$ trimmed region with respect to $F$ is denoted by ${D}_{\alpha }(F)$, that is,
\begin{equation}
{D}_{\alpha }(F)=\left\{ z\in {\mathbb{R}}^{d}:D(z;F)\ge \alpha  \right\}.
\end{equation}
In the context of a great part of applications, the probability measure is the distribution ${F}^{X}$ of a $d-$variate random vector $X$. In this case, we write shortly $D(z;X)$ instead of $D(z;{F}^{X})$ and ${D}_{\alpha }(X)$ instead of ${D}_{\alpha}({F}^{X})$. The data depth is then defined on the set $\mathcal{X}$ of all random vectors $X$ for which ${F}^{X}$ is in $\mathcal{P}$. In functional data case, theoretical frameworks for the DDC are more complicated, but one can also find successful proposals for them (\cite{Horvath}, \cite{Bosq:2000}, and references therein).
The first mature definition of a depth function is assigned to R. Liu (\cite{Liu:1990}). The most general and commonly used theoretical framework for defining the statistical depth functions has been proposed in \cite{Zuo:2000}. For a certain class of probability distributions on $\mathbb{R}^{d}$, equivalent definitions of the statistical depth function, to a definition proposed in \cite{Zuo:2000}, was proposed in  \cite{Dyck:2004} and \cite{Mosler:2013}. There is an agreement in the literature that every concept of multivariate depth should satisfy some reasonable properties (\cite{Zuo:2000}, \cite{Ruts1999depth}):
\begin{itemize}
\item[ZS1] \emph{Affine invariance}. The depth of a point $x\in {{\mathbb{R}}^{d}}$ should not depend on the underlying coordinate system or, in particular, on the scales of the underlying measurements.
\item[ZS2] \emph{Maximality at center}. For a distribution having a uniquely defined “center” (e.g., the point of symmetry with respect to some notion of symmetry), the depth function should attain maximum value at this center.
\item[ZS3] \emph{Monotonicity relative to the deepest point}. As  a point $x\in {{\mathbb{R}}^{d}}$ moves away from the “deepest” point” (the point at which the depth function attains maximum value; in particular, for the symmetric distributions, the center) along any fixed ray through the center, the depth at $x$ should decrease monotonically.
\item[ZS4] \emph{Vanishing at infinity}. The depth of a point $x$ should approach zero as $\left\| x \right\|$  approaches infinity.
\end{itemize}

In a milestone paper for the DDC, Zuo and Serfling (\cite{Zuo:2000}) considered general notions of depth function on ${{\mathbb{R}}^{d}}$ with respect to arbitrary distribution, which may either be continuous or discrete, precisely sketching the historical background of consecutive steps for particular milestones of the development of the concept.

Let $\mathcal{P}$ denote the class of distributions on Borel sets on ${{\mathbb{R}}^{d}},$ while ${{F}_{X}}$ denote the distribution of a given random vector $X$ belonging to the class of random vectors $\mathcal{X}$
\vskip0.5mm
\textbf{Definition }(Zuo \& Serfling 2000) Let the mapping $D(\cdot ,\cdot ):{{\mathbb{R}}^{d}}\times \mathcal{P}\to {{\mathbb{R}}_{+}}$ satisfy ZS1, ZS2, ZS3, ZS4. That is, assume:
\begin{itemize}
\item[A1] $D(Ax+b;{{F}_{AX+b}})=D(x;{{F}_{X}})$ for any $d\times d$ nonsingular matrix $A$ , and any $d-$vector $b$;
\item[A2] $D(\theta ;F)=\mathop{\sup }_{x\in {{\mathbb{R}}^{d}}}D(x,F)$ for any $F\in \mathcal{P}$ having center $\theta $ ;
\item[A3] for any $F\in \mathcal{P}$ having deepest point $\theta $ , $D(x;F)\le D(\theta +\alpha (x-\theta );F)$ holds for $\alpha \in [0,1];$ and
\item[A4] $D(x;F)\to 0$ as $\left\| x \right\|\to \infty$ , for each $F\in \mathcal{P}$.
\end{itemize}
Then $D(\cdot ;F)$ is called \emph{a statistical depth function.}\\

A sample version of $D(x;F)$ denoted by $D(x;{{F}_{n}})$ or $D(x;{{X}^{n}})$ may be defined by replacing $F$ with a suitable empirical measure ${{F}_{n}},$ calculated from a sample ${{X}^{n}}=\{{{x}_{1}},...,{{x}_{n}}\}$.

\emph{Remark 1:} Postulates A1 to A4 are formulated in terms of the depth itself. In a computational context, it is very useful to notice that these postulates can also be formulated in terms of the trimmed regions (\cite{Dyck:2004}).\\
\emph{Remark 2:} By the above center, Zuo and Serfling understand a point of symmetry. In the multivariate case, popular notions of symmetry are the central symmetry, angular symmetry, and halfspace symmetry. A random vector $X$ in ${{\mathbb{R}}^{d}}$ is centrally symmetric around $\theta $ if $X-\theta \overset{d}{\mathop{=}}\,\theta -X$, where $"\overset{d}{\mathop{=}}\,"$ denotes equality in the distribution; $X$ is centrally symmetric around $\theta $ if $X-\theta /\left\| X-\theta  \right\|$ is centrally symmetric around origin; $X$ is halfspace symmetric around $\theta $ if $\operatorname{Prob}(X\in H)\ge 1/2$ for each closed halfspace containing $\theta $   (for further details see \cite{Symmetry})

\emph{Note:} An extension of the above definition for functional data case may be found in \cite{Nieto}.

The simplest example of the depth is \textbf{the Euclidean depth} defined as
\begin{equation}
 {D}_{EUK}(y;{X}^{n})=\frac{1}{1+{{\left\| y-\bar{x} \right\|}^{2}}},
 \end{equation}
where $\bar{x}$ denotes the mean vector calculated from the sample ${X}^{n}$ (Figure 1).
\begin{figure}
\centering
\begin{minipage}[t]{.45\textwidth}
  \centering
  \includegraphics[width=.95\linewidth]{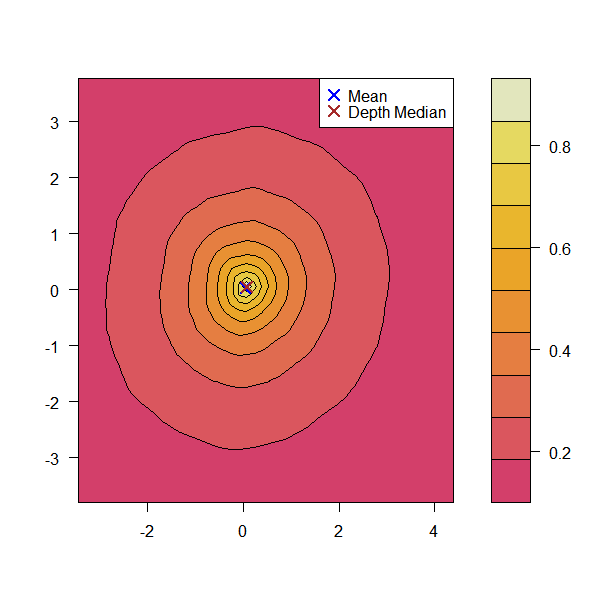}
  \caption{Sample contour plot, the Euclidean depth.}
  \label{fig1}
\end{minipage}
\begin{minipage}[t]{.45\textwidth}
  \centering
  \includegraphics[width=.95\linewidth]{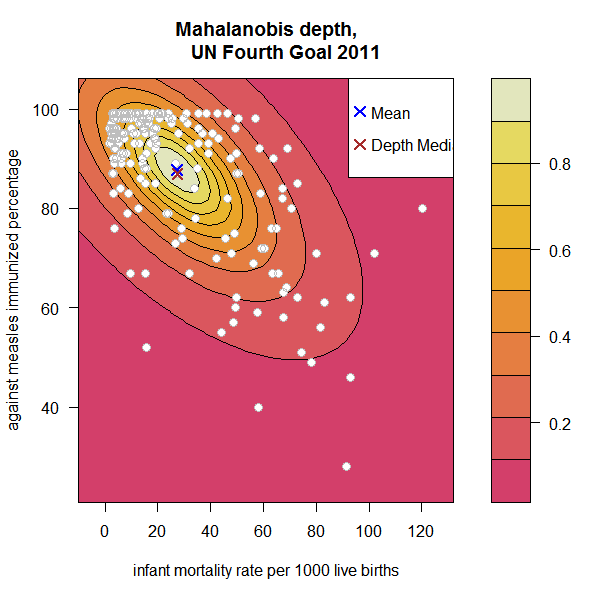}
  \caption{Sample contour plot, Mahalanobis depth.}
  \label{fig2}
\end{minipage}
\end{figure}
As the next example, let us take \textbf{the Mahalanobis depth} (\cite{Mahalanobis:1936})
\begin{equation}
{D}_{MAH}(y;{X}^{n})=\frac{1}{1+{{(y-\bar{x})}^\top}{{S}^{-1}}(y-\bar{x})},
\end{equation}
where $S$ denotes the sample covariance of matrix ${X}^{n}$.\\
\emph{Note:} Putting robust estimators of the covariance matrix and the mean vector, we can obtain its robust version (Figure 2).

The \textbf{symmetric projection depth} $D\left( x;X \right)$ of a point $x\in {{\mathbb{R}}^{d}}$, $d\ge 1$ is defined as
\begin{equation}
                  	D\left( x;X \right)_{PRO}={{\left[ 1+su{{p}_{\left\| u \right\|=1}}\frac{\left| {{u}^{\top}}x-Med\left( {{u}^{\top}}X \right) \right|}{MAD\left( {{u}^{\top}}X \right)} \right]}^{-1}},
\end{equation}
where $Med$ denotes the univariate median, $MAD\left( Z \right)$ = $Med\left( \left| Z-Med\left( Z \right)
\right| \right)$. Its sample version denoted by $D\left( x,{X}^{n} \right)$ or $D\left( x,{X}^{n} \right)$ is
obtained by replacing $F$  with its empirical counterpart ${{F}_{n}}$ calculated from the sample ${X}^{n}$ (Figure 3). This depth, postulated in \cite{Donoho:1992pro} and intensively studied and popularized by Y. Zuo and scientists collaborating with him (see \cite{Zuo:2003}, \cite{Shao:2012}), is one of the best multivariate depths. It is an affine invariant and $D(x,{F}_{n})$ converges uniformly and strongly to $D(x,F)$. The affine invariance ensures that the proposed inference methods are coordinate-free, and the strong convergence of $D(x,{X}^{n})$ to $D(x,X)$ allows us for the approximation of an appropriate quality of $D(x,F)$ by $D(x,{F}^{n}),$ when the $F$ is unknown. Induced by this depth, multivariate location and scatter estimators have very high breakdown points (BP) and Hampel's bounded influence function (IF) (for further details see \cite{Zuo:2003}).

\emph{Note:} It is possible to define an asymmetric projection depth within a theoretical framework proposed by \cite{Dyck:2004} and dedicated for a certain class of depth functions.\\

\begin{figure}
\centering
\begin{minipage}[t]{.45\textwidth}
  \centering
  \includegraphics[width=.95\linewidth]{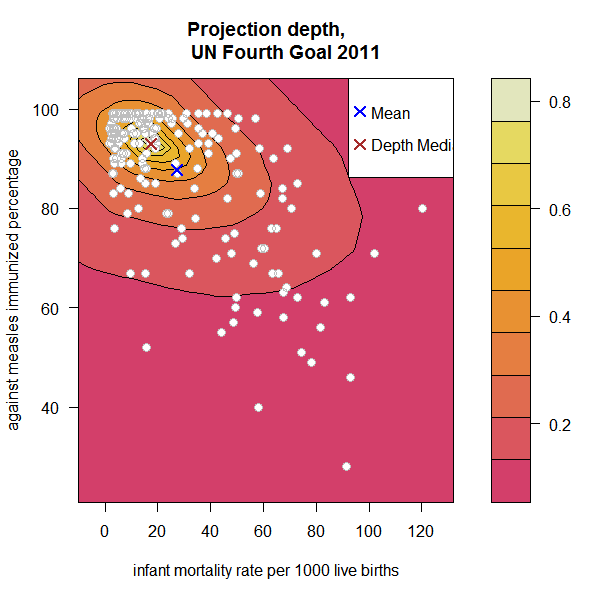}
  \caption{Projection depth contour plot.}
  \label{fig3}
\end{minipage}
\mbox{\hspace{0.1cm}}
\begin{minipage}[t]{.45\textwidth}
  \centering
  \includegraphics[width=.95\linewidth]{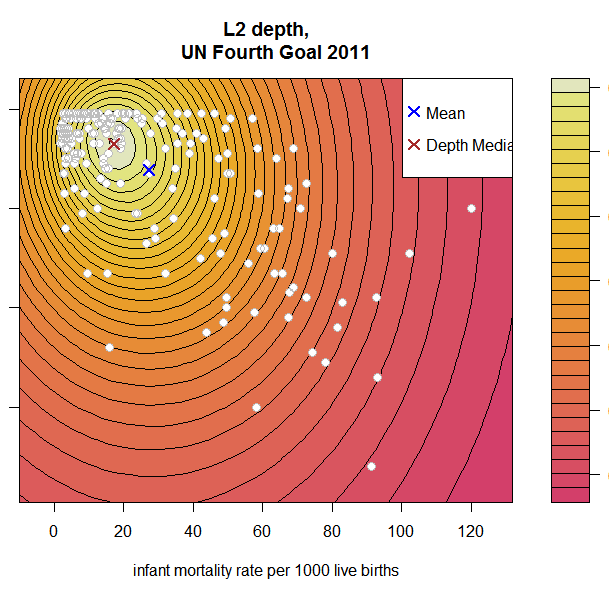}
  \caption{$L^2$ depth contour plot.}
  \label{fig4}
\end{minipage}
\end{figure}

Next very important depth is \textbf{the weighted ${L}^{p}$ depth}. The weighted ${L}^{p}$ depth
$D(\mathbf{x},F)$ of a point $\mathbf{x}\in \mathbb{R}^{d}$, $d\ge 1$ generated by $d$  dimensional random vector $\mathbf{X}$ with distribution $F$,  is defined as (\cite{Zuo:2004}, Figure \ref{fig4})
\begin{equation}
                D(x;F)=\frac{1}{1+\mathbb{E}w({{\left\| x-X \right\|}_{p}})},
\end{equation}
where $w$ is a suitable weight function on $[0,\infty )$, and ${{\left\| \cdot  \right\|}_{p}}$ stands for
the ${L}^{p}$ norm (when $p=2$ we have the usual Euclidean norm). We assume that $w$ is nondecreasing and continuous on $[0,\infty )$ with $w(\infty -)=\infty $, and for $a,b\in {{\mathbb{R}}^{d}}$ satisfying
$w(\left\| a+b \right\|)\le w(\left\| a \right\|)+w(\left\| b \right\|)$. Examples of the weight functions are
$w(x)=a+bx$ , $a,b>0$ or $w(x)={x}^{\alpha }$, $\alpha>0$.
The empirical version of the weighted ${L}^{p}$ depth is obtained by replacing the distribution $F$
of ${X}$ in $\mathbb{E}w({{\left\| {x}-{X} \right\|}_{p}})=\int{w({{\left\| x-t \right\|}_{p}})}dF(t)$ by
its empirical counterpart. The weighted $L^p$ depth from the sample $X^n=\{x_1,...,x_n\}$ is computed as follows:
\begin{equation}
D(x,X^n)=\frac{1}{1+\frac{1}{n}\sum\limits_{i=1}^{n}{w\left( {{\left\| x-{X}_{i} \right\|}_{p}} \right)}},
\end{equation}

The weighted ${L}^{p}$ depth function in a point, has the low BP and unbounded IF (see \cite{Marona:2006}, \cite{Wilcox}, \cite{Genton:2003} for the BP and IF definitions). On the other hand, the weighted ${L}^{p}$ depth-induced medians (multivariate location estimator) are globally robust with the highest BP for any reasonable estimator. The weighted ${L}^{p}$ medians are also locally robust with bounded IFs for suitable weight functions. Unlike other existing depth functions and multivariate medians, the weighted ${L}^{p}$ depth and medians are computationally feasible for online applications and easy to calculate in high dimensions (\cite{Zelias:2014}). The price for this advantage is the lack of affine invariance and equivariance of the weighted ${L}^{p}$ depth and medians, respectively. Theoretical properties of this depth can be found in \cite{Zuo:2004}. This depth is recommendable in a context of \emph{Big Data} analysis (\cite{Zelias:2014}), \cite{JMS}).

\begin{figure}
\centering
\includegraphics[width=.75\linewidth]{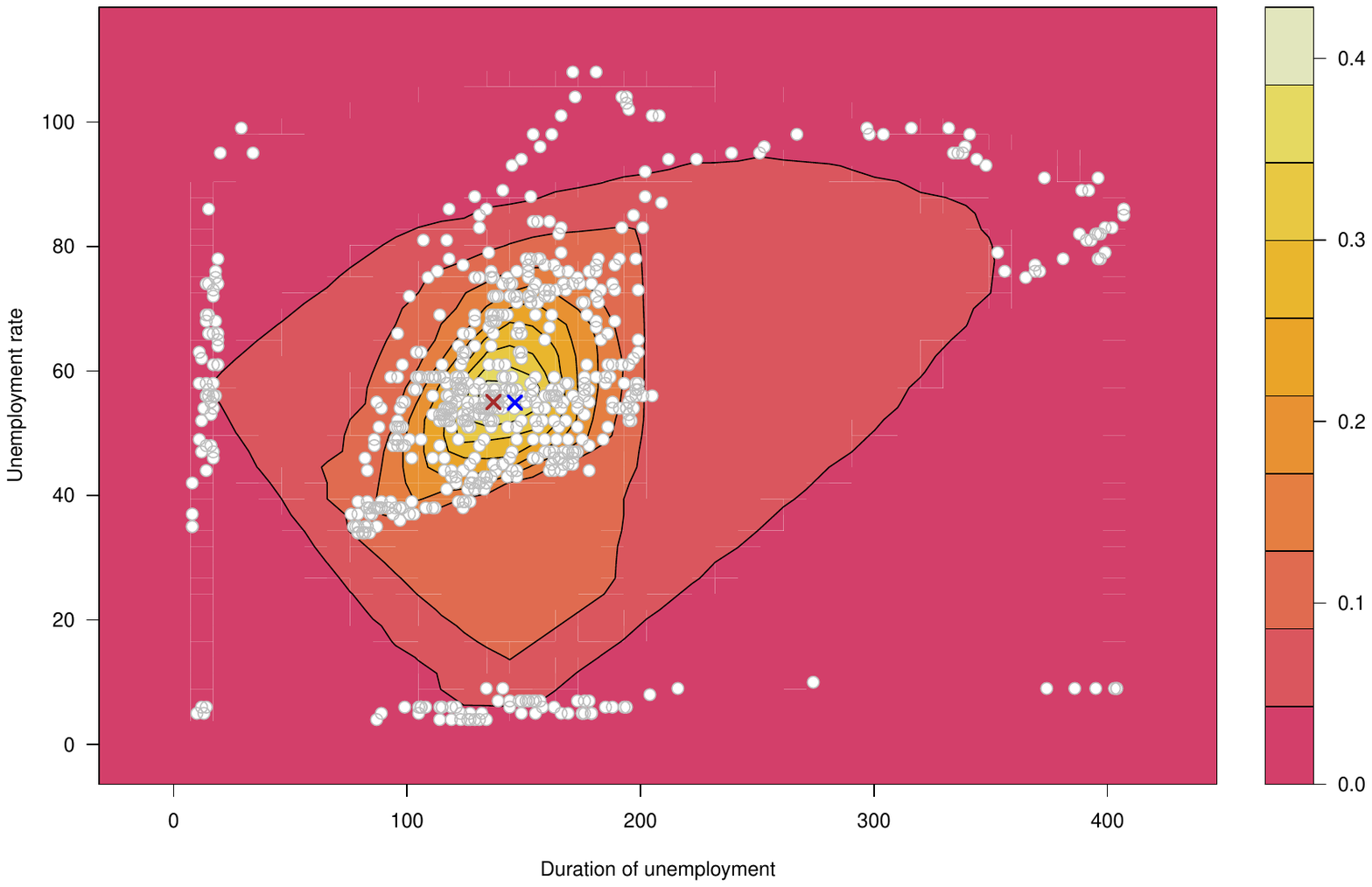}
\includegraphics[width=.75\linewidth]{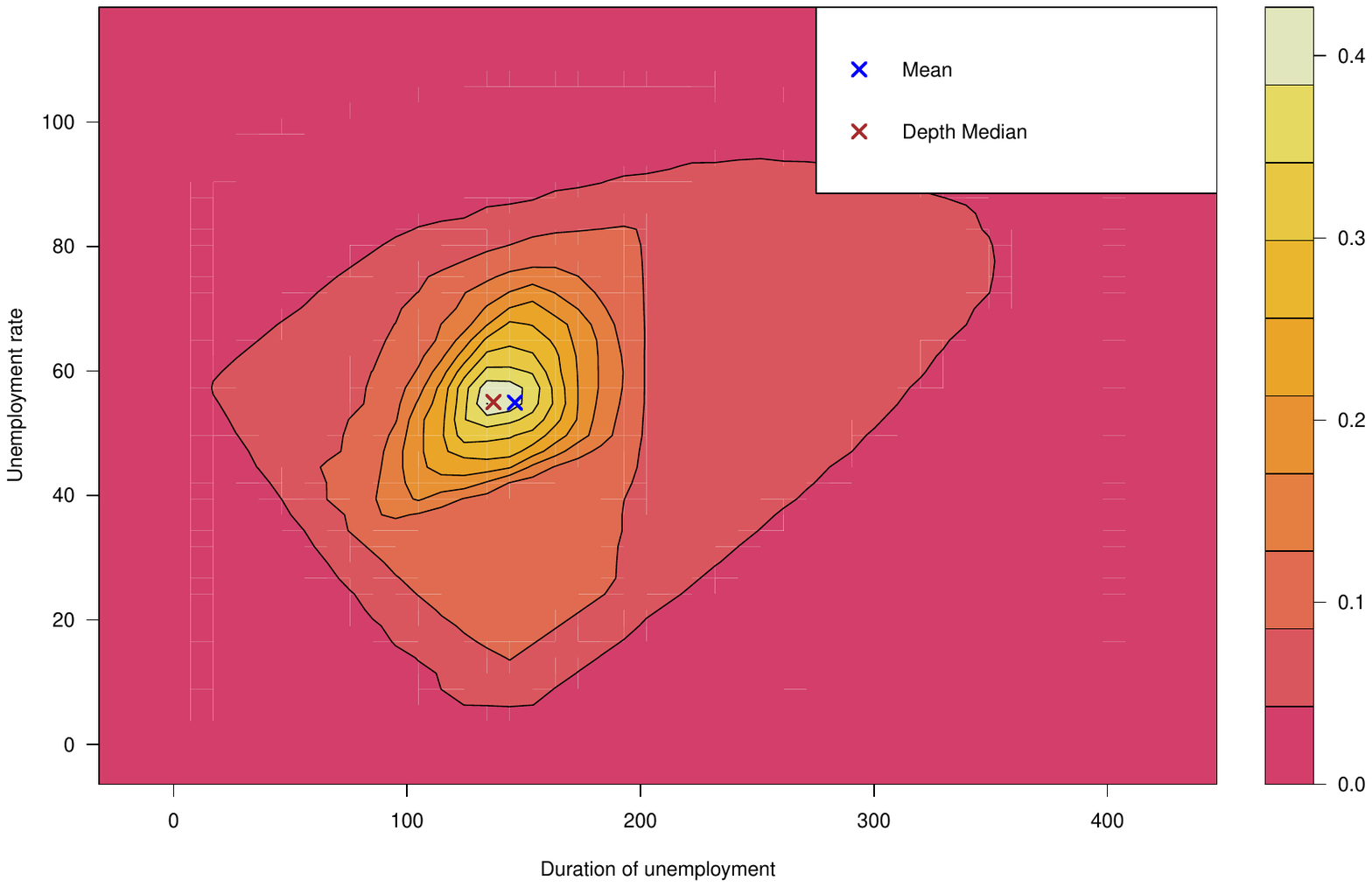}
  \caption{Unemployment rate vs. the duration of unemployment in USA during 1957--2011. Contour plots were prepared using Tukey depth.}
  \label{fig5}
\end{figure}

Next, historically the first and still very important depth is \textbf{the halfspace depth} (\cite{Tukey:1975}, \cite{rousseeuw1998constructing}, \cite{Isodepth}) defined as
\begin{equation}
D(x;F)=\underset{H}{\mathop{\inf }}\,\,\left\{ Prob(H):x\in H\subset {{\mathbb{R}}^{d}},\text{ H is closed subspace} \right\}
\end{equation}
Figure 5 presents sample contour plots prepared using the halfspace depth for a dataset on US economy and mainly regarding the monthly unemployment rate and the duration of unemployment (days) in USA in the period of 1957--2011. Although the difference between mean vector and the Tukey median seems to be insignificant, the shape of the most central regions suggests the relation of the form "bigger duration of unemploynmet, the bigger unemploynment rate" in a more evident way than "classical" data ellipse. Please note that the relation between these economic variables is an open problem in the Economics up to now.
\begin{figure}
\centering
\begin{minipage}[t]{.45\textwidth}
  \centering
  \includegraphics[width=.95\linewidth]{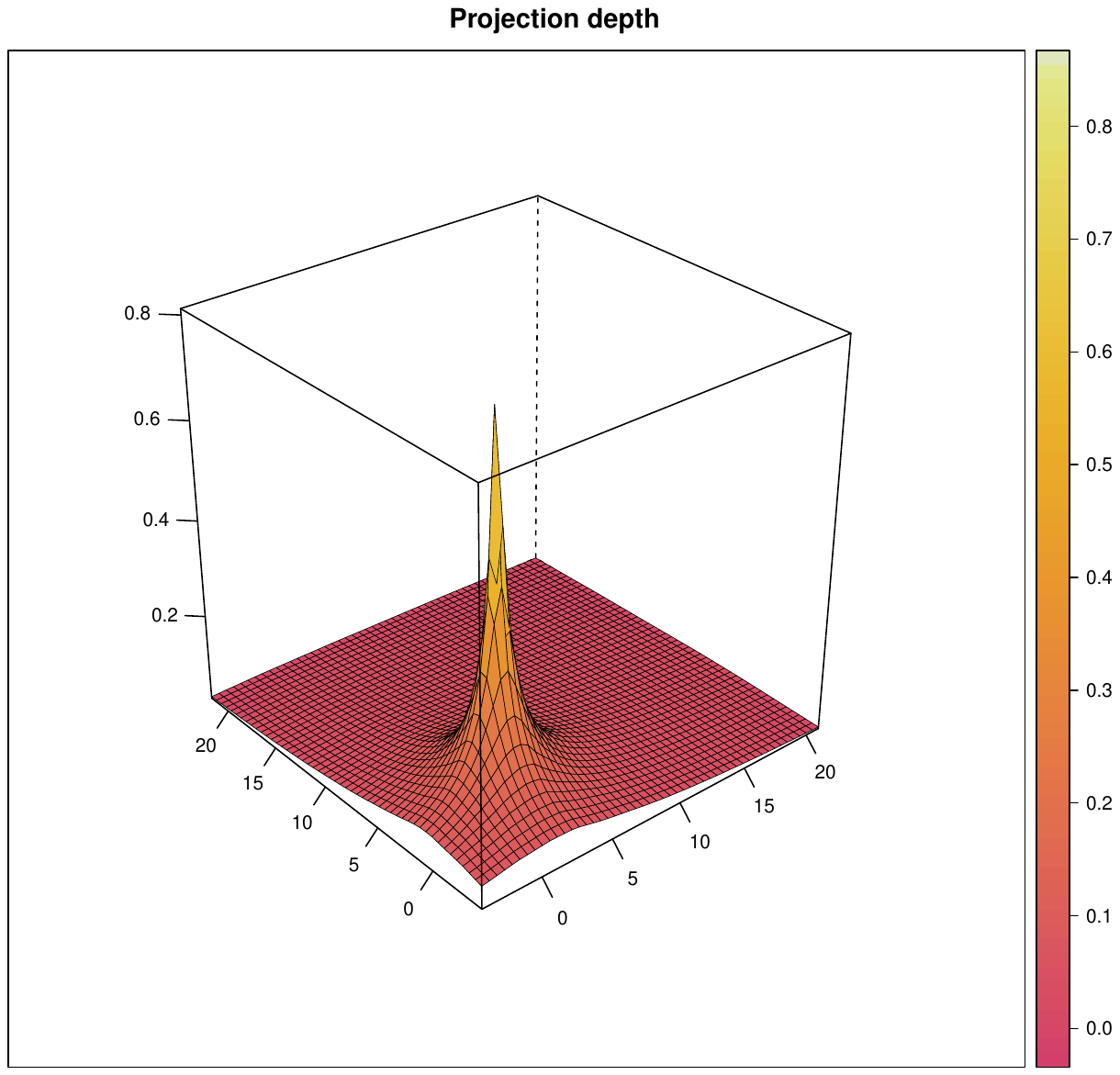}
  \caption{Sample from skewed T(5) distribution; projection depth perspective plot.}
  \label{fig6}
\end{minipage}
\mbox{\hspace{0.1cm}}
\begin{minipage}[t]{.45\textwidth}
  \centering
  \includegraphics[width=.95\linewidth]{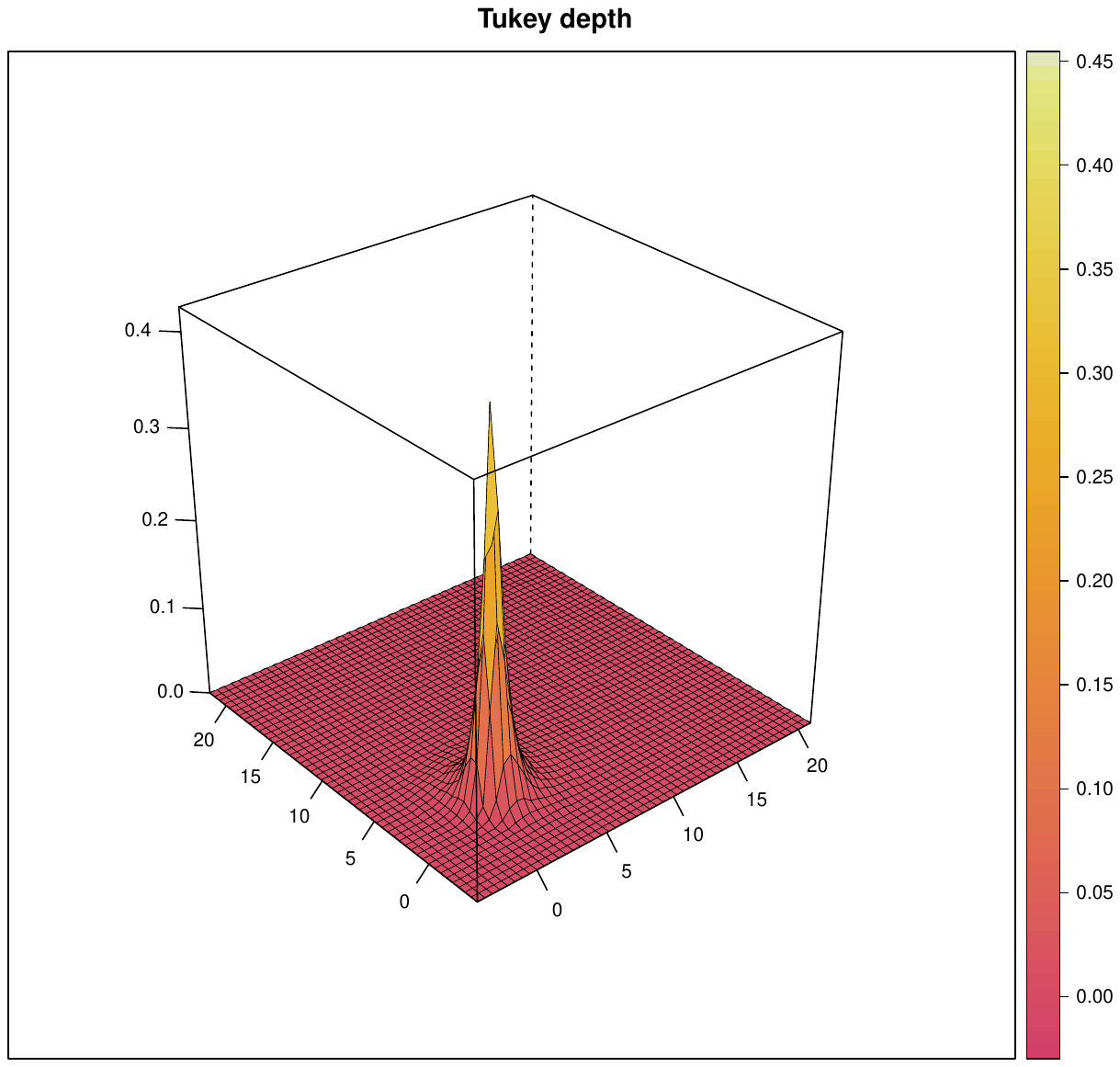}
  \caption{Sample from skewed T(5) distribution; Tukey depth perspective plot.}
  \label{fig7}
\end{minipage}
\end{figure}

\begin{figure}
\centering
\begin{minipage}[t]{.45\textwidth}
  \centering
  \includegraphics[width=.95\linewidth]{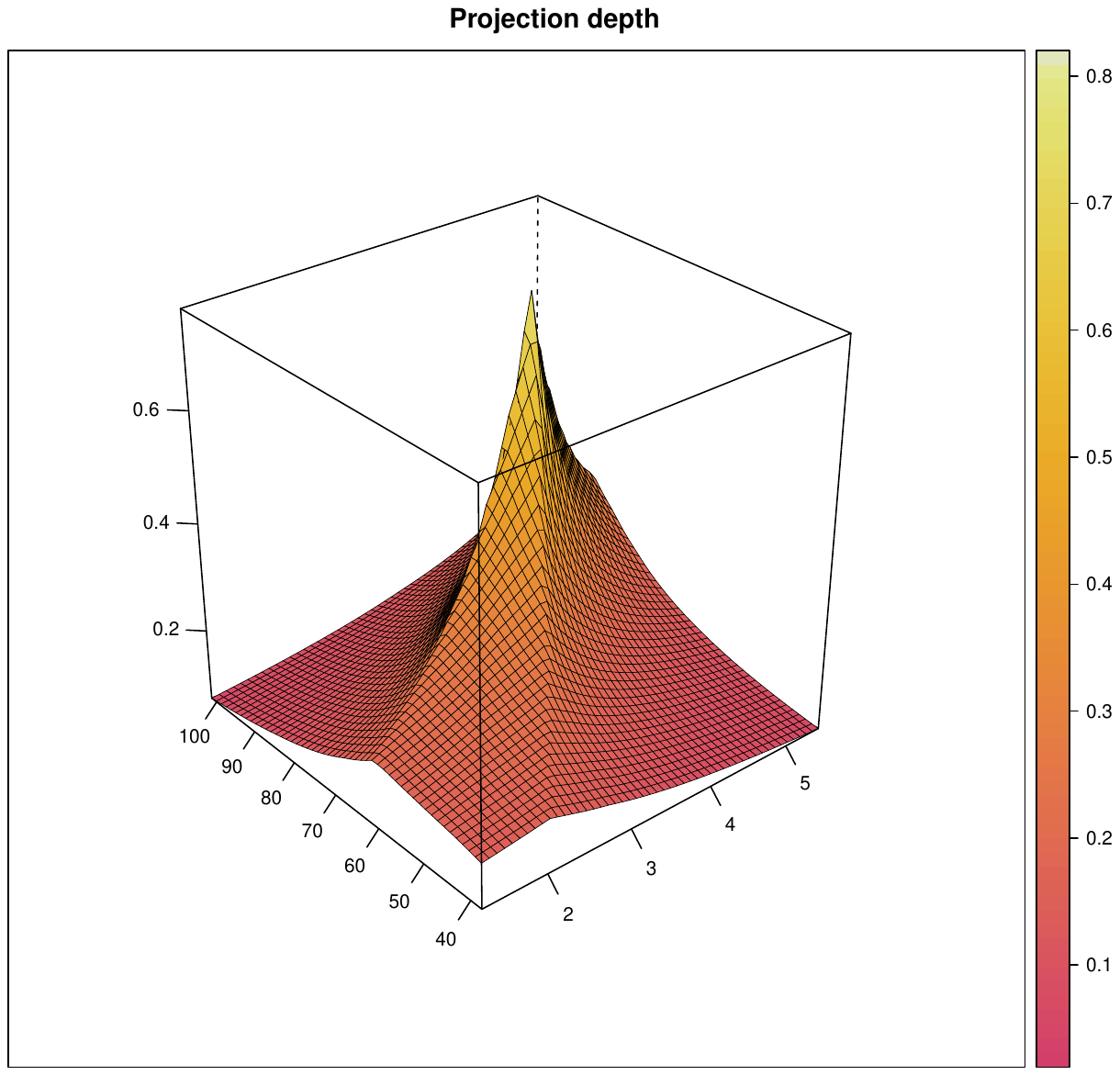}
  \caption{Faithful data in a projection depth perspective plot.}
  \label{fig1}
\end{minipage}
\mbox{\hspace{0.1cm}}
\begin{minipage}[t]{.45\textwidth}
  \centering
  \includegraphics[width=.95\linewidth]{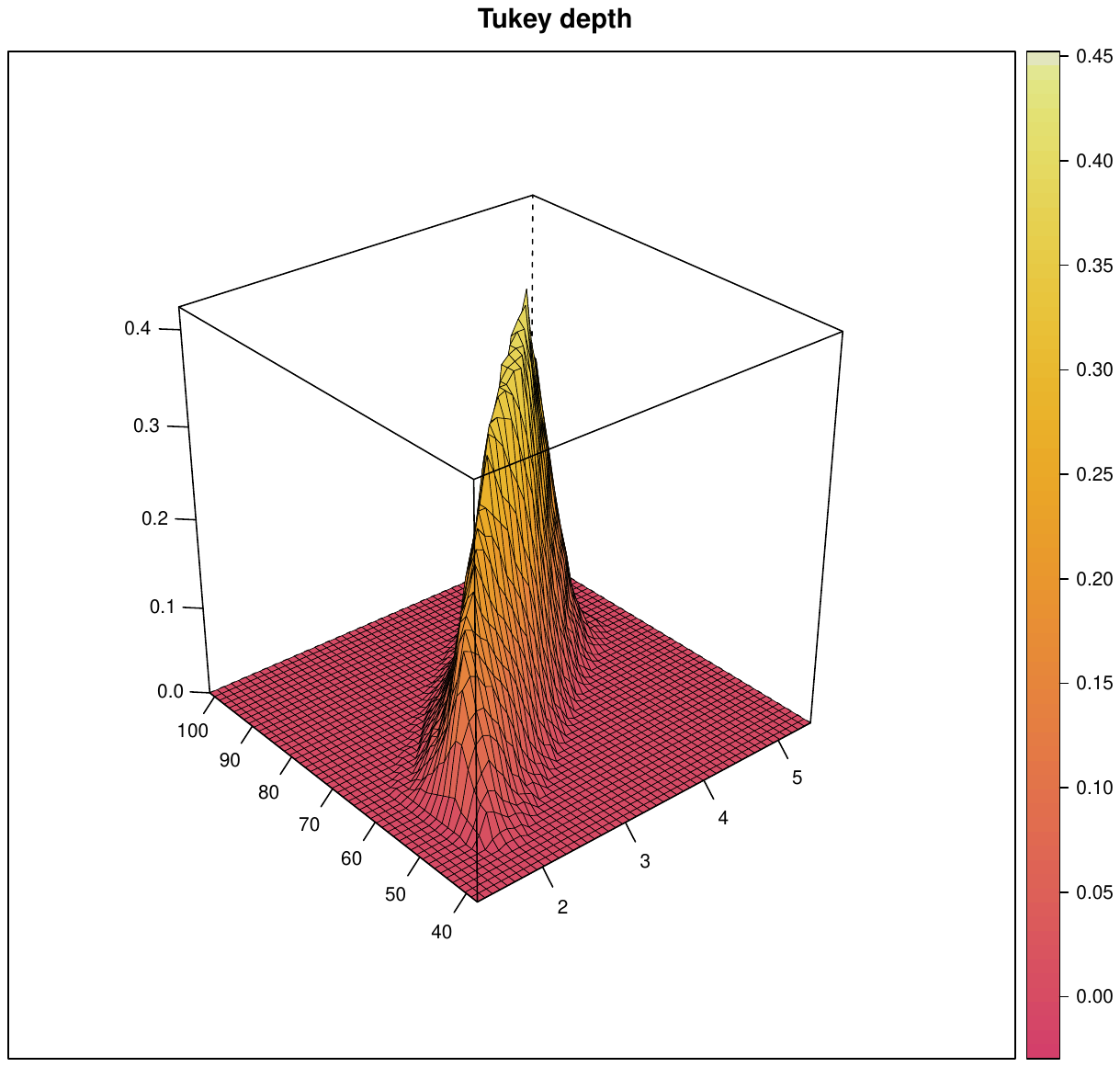}
  \caption{Faithful data in a Tukey depth perspective plot.}
  \label{fig9}
\end{minipage}
\end{figure}
Next important depth is \textbf{the simplicial depth} introduced in \cite{Liu:1990} and defined as
\begin{equation}
SD(x;F)=Prob(x\in S[{{X}_{1}},...,{{X}_{d+1}}]),
 \end{equation}
 where $S[{{X}_{1}},...,{{X}_{d+1}}]$ is the $d$ dimensional simplex in ${{\mathbb{R}}^{d}}$ with vertices ${{x}_{1}},...,{{x}_{d+1}}$ .
 \begin{equation}
SD(x;{{X}^{n}})={{\left( \begin{matrix}
   n  \\
   d+1  \\
\end{matrix} \right)}^{-1}}\sum\limits_{1\le {{i}_{1}}\le \cdots {{i}_{d+1}}\le n}{\mathbf{I}\{x\in S[{{X}_{{{i}_{1}}}},...,{{X}_{{{i}_{d+1}}}}]}\}.
\end{equation}

The depth being very useful for economic application, which originates from Tukey depth, is the \textbf{regression depth} (\cite{Struyf1998computing} \cite{Hubert:1999}). This depth was intensively studied among others in \cite{Van:2000}, \cite{Mizera:2002},  and in the context of its relations to the logistic regression in \cite{overlap}. This concept of depth leads to the deepest regression (DeepReg) estimators of the many important forms of regressions and performs very well in terms of its sensitivity to the choice of the majority of data, which, in general, leads to better merit interpretations of regression in comparison to other very robust regressions (\cite{Visek2002}).

 Let ${Z}^{n}=\left\{ ({x}_{1},{y}_{1}),...,({x}_{n},{y}_{n}) \right\}\subset \mathbb{R}^{d}$ denote a
 sample considered from the following semiparametric model:
\begin{equation}
{{y}_{l}}={{a}_{0}}+{{a}_{1}}{{x}_{1l}}+...+{{a}_{(d-1)l}}{{x}_{(d-1)l}}+{{\varepsilon }_{l}}, l=1,...,n,
\end{equation}
we calculate the depth of a fit $\alpha=(a_{0},...,a_{d-1})$ as
\begin{equation}
RD(\alpha ,{{Z}^{n}})=\underset{u\ne 0}{\mathop{\min }}\left\{\#(\frac{{{r}_{l}}(\alpha
)}{{{u}^{\top}}{{x}_{l}}}<0),l=1,...,n \right\},
\end{equation}
where $r(\cdot )$ denotes the regression residual, $\alpha=(a_{0},...,a_{d-1})$, ${u}^{\top}{x}_{l}\ne 0$.

\textbf{The deepest regression estimator} $DeepRegR(\alpha,{{Z}^{n}})$ is defined as
\begin{equation}
DeepReg(\alpha ,{{Z}^{n}})=\underset{\alpha \ne 0}{\mathop{\arg \max }}\,RD(\alpha ,{{Z}^{n}})
\end{equation}
\begin{figure}
\centering
  \includegraphics[width=.45\linewidth]{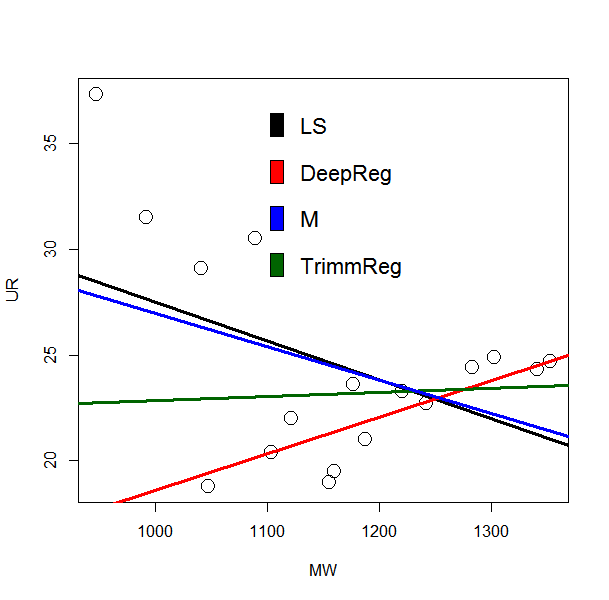}
  \caption{Selected estimators of simple regression expressing relationship between an unemployment rate (UR) and minimal wage (MW) in France in a period 1999-2015. \emph{LS} denotes the least squares estimator, \emph{DeepReg} the deepest regression, \emph{M} denotes the Huber M estimator, and \emph{TrimmReg} denotes least squares estimator for projection depth trimmed data beforehand.}
  \label{fig10}
\end{figure}

Figure 10 presents three estimators of a simple regression applied for  expressing the relationship between the unemployment rate (UR) and the minimal wage (MW) for data on the economy of France in the period of 1999--2015. One can notice differences between the least squares (LS), the deepest regression (DeepReg), Huber M (M) estimators and the least squares estimator for the  dataset trimmed using the projection depth beforehand (TrimmReg), (\cite{ZUOtrimming}). Although the relation between these variables is still not obvious for economists, LS and M estimators show a rather naive point of view on the issue, whereas DeepReg and TrimmReg are much closer to recent empirical findings. The "classical" estimators lead to a recommendation of increasing the minimal wage in order to attain a smaller level of unemployment, whereas "more modern" robust estimators lead to opposite conclusions. Further arguments on the special usefulness of the deepest  regression estimator in this context provide analogous data on economies of Greece, Czech Republic, and Hungary in the same period. The depth-based estimators of regressions lead to recommendations as to economic policy, which are close to recent findings in theoretical economics. \\
The regression depth has its local version, thanks to its relation with the halfspace depth (\cite{Pain:2013}, \cite{Mizera:2002}). The local version of this depth may be easily calculated within the \pkg{DepthProc} package.\\
\vskip1mm
Next is \textbf{the Student depth} which is implemented within the package, originating from \cite{Mizera:2002} and was proposed in \cite{Mizera:2004}. It has been pointed out in \cite{Mizera:2002} that the general halfspace depth can be defined as a measure of the data-analytic admissibility of a fit. Depth of the fit $\theta $ is defined as a proportion of the observations whose omission causes $\theta$ to become \emph{a nonfit}, a fit that can be uniformly dominated by another one.

For a sample ${X}^{n}=\{{x}_{1},...,{x}_{n}\},$ we consider a criterion function ${F}_{i}$, given a fit represented by $\alpha$, the criterion function evaluates the lack of fit of $\alpha $ to the particular observation ${x}_{i}$. It means ${\alpha }^{*}$ is fitting ${x}_{i}$ better than $\alpha $, if ${F}_{i}({\alpha }^{*})<{F}_{i}(\alpha )$.

In \cite{Mizera:2002}, more operational version--the tangent depth of a fit $\alpha $ is defined as
\begin{equation}
d(\alpha )=\underset{\mathbf{u}\ne \mathbf{0}}{\mathop{\inf }}\,\left\{ \#(i/n):{{\mathbf{u}}^{\top }}{{\nabla }_{\alpha }}{{F}_{i}}(\alpha )\ge 0 \right\},
\end{equation}
where $\#(i/n)$ stands for the relative proportion in the index set--its cardinality divided by $n$.

In \cite{Mizera:2004}, the authors make a suggestion by assuming the location-scale model for the data and taking the log-likelihood in the role of the criterion function. They suggest taking the criterion function
\begin{equation}
{F}_{i}(\mu ,\sigma )=-\log f\left( \frac{{{y}_{i}}-\mu }{\sigma } \right)+\log \sigma
\end{equation}
Substituting (14) into (13), we obtain a family of location-scale depths.

\textbf{The Student depth} of $(\mu ,\sigma )\in \mathbb{R}\times [0,\infty )$ is obtained by substituting the density of the $t$ distribution with $v$ degrees of freedom into the above expression.
\begin{equation}
 d(\mu ,\sigma )=\mathop{\inf_{{{u}\ne {0}}}} \left\{\#(i/n) : ({u}_{1},{u}_{2}) \left(
  \begin{array}{c}
                          {\tau }_{i}  \\
                        \frac{v}{v+1}(\tau_{i}^{2}-1) \\
                        \end{array}
                      \right)
      \ge 0 \right\} ,
      \end{equation}
where, by the multiplication we mean the dot product; ${\tau }_{i}$ is a shorthand for $({{y}_{i}}-\mu
)/\sigma $, and we can absorb the constant $v/(v+1)$ into the ${u}$ term (Figures 11--12). It is worth noticing that the Student depth contour plot may be treated as a very powerful graphical tool for the normality assumption inspection in one dimensional case.

\textbf{The Student Median} (SM) is the maximum depth estimator induced by the Student depth. It is a very interesting joint estimator of location and scale in the context of robust time series and data streams analysis. It is robust, but not very robust--its BP is about 33\% and hence is robust to a moderate fraction of outliers, but is sensitive to a regime change of the time series at the same time (\cite{JMS}). It is worth noticing that, by its definition, the SM is not affected by the temporal dependence of the observations \cite{CompstatStud} (for another application of Mizera's idea see \cite{Sparse}). We presuppose an effective application of the SM in the context of candidate attractiveness on a labor market evaluation. The SM and in general the location scale median may be very useful in studies of wages or sex discrimination, where a subjective position of an individual depends on a scatter related to the majority of objects in a group and a distribution describing the group. We also presuppose that ideas of  Mizera and M\"{u}ller may effectively be used for example, in the context of a robust estimation of the gamma regression parameters (\cite{Rydlewski2009}).

 \begin{figure}
\centering
\begin{minipage}[t]{.45\textwidth}
  \centering
  \includegraphics[width=.95\linewidth]{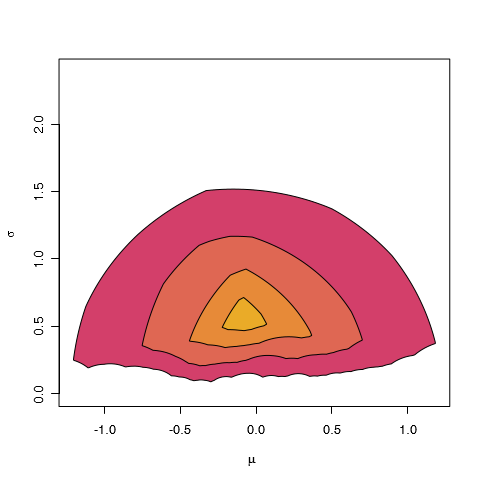}
  \caption{Sample student depth contour plot, data from N(0,1).}
  \label{fig11}
\end{minipage}
\mbox{\hspace{0.1cm}}
\begin{minipage}[t]{.45\textwidth}
  \centering
  \includegraphics[width=.95\linewidth]{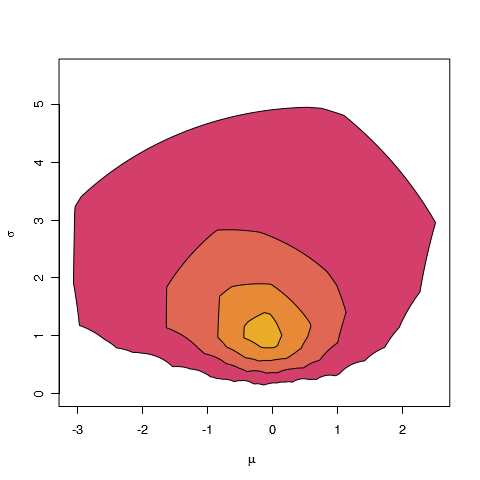}
  \caption{Sample student depth contour plot, data from student t(1).}
  \label{fig12}
\end{minipage}
\end{figure}

\subsection{Local depth}
In an opposition to the density function, the depth function has a global nature, for example, it expresses a centrality of a point with respect to the whole sample. This property is an advantage of depth for some applications, but may be treated as its disadvantage in the context of classification of objects or for k-nearest neighbor rule applications. Depth-based classifier or depth-based k-nearest neighbor density estimators need local version of depths. A successful concept of \textbf{local depth} was proposed in \cite{Pain:2012}. For defining \textbf{the neighbourhood} of a point, authors proposed using an idea of \textbf{symmetrization} of a distribution (a sample) with respect to the point in which the depth is calculated. In their approach, instead of the distribution ${F}^{X}$, a distribution ${{F}_{x}}=1/2{{F}^{X}}+1/2{{F}^{2x-X}}$ is used.

For any $\beta \in (0,1]$, let us introduce the smallest depth region with the probability bigger or equal to $\beta $,
\begin{equation}
{R}^{\beta }(F)=\bigcap\limits_{\alpha \in A(\beta )}{{{D}_{\alpha }}}(F),
\end{equation}
where $A(\beta )=\left\{ \alpha \ge 0:Prob\left[ {{D}_{\alpha }}(F) \right]\ge \beta  \right\}$, $"Prob"$ denotes probability.
Then, for a locality parameter  $\beta \in (0,1]$ we can take the neighborhood of the point $x$ as $R^{\beta }(F_{x})$ (Figures 13--14).
\vskip0.5mm
Formally, let $D(\cdot,F)$ be the depth function. Then the \textbf{local depth }with the locality parameter $\beta \in (0,1]$ with respect to the point $x$ is defined as
\begin{equation}
L{{D}^{\beta }}(z,F):z\to D(z,F_{x}^{\beta }),
\end{equation}
where $F_{x}^{\beta }(\cdot )=F\left( \cdot |R_{x}^{\beta }(F) \right)$ is a conditional distribution of $F,$ conditioned on $R_{x}^{\beta }(F)$.

For $\beta=1,$ the local depth reduces to its global counterpart (no localization).
\begin{figure}
\centering
\begin{minipage}[t]{.45\textwidth}
  \centering
  \includegraphics[width=.95\linewidth]{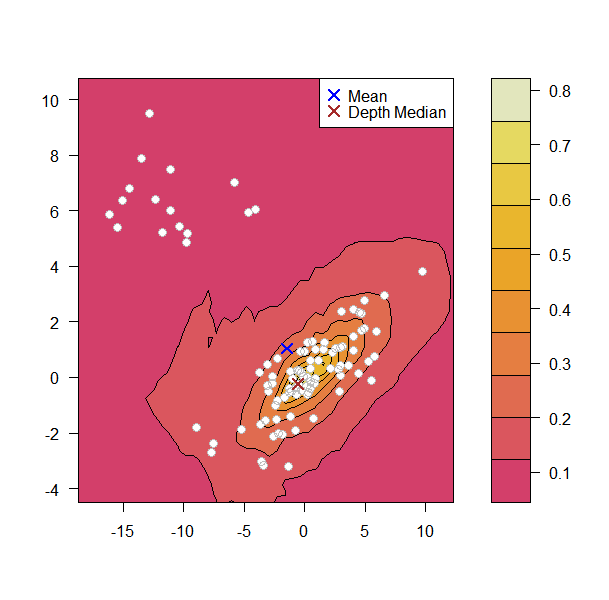}
  \caption{Sample local $L^2$ depth contour plot, $\beta=0.2$.}
  \label{fig9}
\end{minipage}
\mbox{\hspace{0.1cm}}
\begin{minipage}[t]{.45\textwidth}
  \centering
  \includegraphics[width=.95\linewidth]{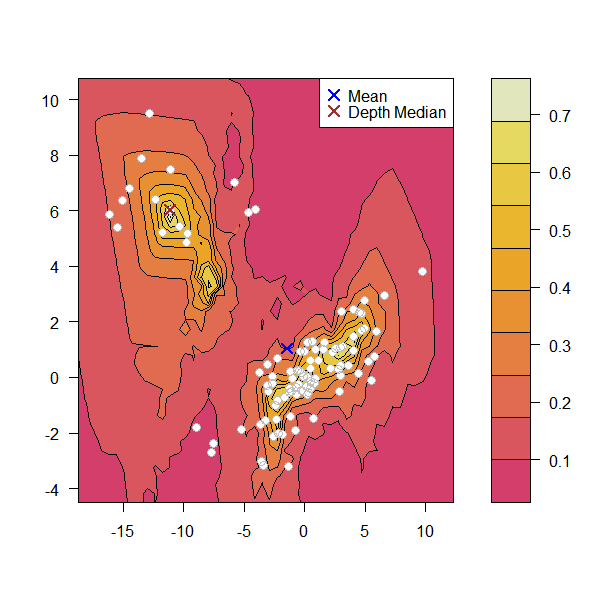}
  \caption{Sample local $L^2$ depth, $\beta=0.6$.}
  \label{fig10}
\end{minipage}
\end{figure}
In a sample case ${{X}^{n}}=\{{{x}_{1}},...,{{x}_{n}}\}$, in the first step, we calculate depth of a point $y$ by adding to the original observations ${{x}_{1}},...,{{x}_{n}}$ their reflections $2y-{{x}_{1}},...,2y-{{x}_{n}}$ with respect to $y,$ let us denote this combined sample, $X_{n}^{y}$ and then calculate the usual depth. Then we order observations from the original sample with respect to $D(\cdot ,X_{n}^{y}),$ the sample depth calculated from the combined sample: $D({{x}_{(1)}},X_{n}^{y})\ge ...\ge D({{x}_{(n)}},X_{n}^{y})$. We choose the locality parameter $\beta \in (0,1],$ determining the size of the depth-based neighborhood of the point $x$. Then we determine ${{n}_{\beta }}(X_{n}^{y})=\max \left\{ l=\left\lceil n\beta  \right\rceil ,...,n \right\}:D({{x}_{(l)}},X_{n}^{y})=D({{x}_{(\left\lceil n\beta  \right\rceil )}},X_{n}^{y})\}$. Finally we calculate $L{{D}^{\beta }}(y,{{X}^{n}})=D(y,X_{n}^{y,\beta })$, where $X_{n}^{y,\beta }$ denotes the subsample ${{x}_{(1)}},...,{{x}_{(n\beta )}}$ of $X_{n}^{y}$. Further theoretical properties involving its weak continuity and almost sure consistency can be found in \cite{Pain:2012} and \cite{Pain:2013}. For an example on the application of this concept of locality in a simple cooperative game, refer \cite{SGH}.

\section{Depths in functional data analysis}
A variety of economic phenomena may be described by means of functions. For instance, consumer utility curves, development paths of companies or countries, day and night electricity consumption, or the concentration of dangerous particles in atmosphere for a week. In recent decades, a very useful statistical methodology has been proposed in this context and is now being intensively developed. The family of statistical methods, named functional data analysis (FDA), enables for functional generalizations of well-known univariate and multivariate statistical techniques, like analysis of variance, kernel regression, or k-nearest neighbor classification techniques (\cite{Ramsay:2009}, \cite{Horvath}, \cite{Ferraty}, \cite{Gorecki2016}).
The FDA is significantly different from one and multivariate statistical analysis, both from an empirical  as well as mathematical point of view. The most important difference relates to the fact that within the FDA we treat observations as realizations of a functional random variable or as trajectories of an appropriate defined stochastic process. Following the above cited authors, we consider a random curve as a real function, whose domain is an interval $[0, T]$, where $T$ is known. We treat these functions as elements of a separable Hilbert space $L^2[0,T]$ of square-integrable functions with a natural inner product. In \cite{Bosq:2000}, one can find proofs on the existence of probability distributions for such objects defined in the Hilbert spaces. Samples are realizations of random functions, i.e., certain random elements of infinite-dimensional real separable Banach or Hilbert space. The separability of a space ensures that a linear combination of random elements belongs to the space.\\
Assume we consider random functions of a form $X:(\Omega ,\mathcal{B},\mathcal{P})\to \mathcal{V}$, where $(\Omega ,\mathcal{B},\mathcal{P})$ is a probabilistic space and $\mathcal{V}$ denotes real and separable Banach or Hilbert space equipped with a norm $\left\| \cdot  \right\|$; in case of the Hilbert space, the norm is induced by the inner product. For all $\omega \in \Omega,$ we have ${{X}_{\omega }}:t\to X(\omega ,t)\in \mathcal{V}$. It is obvious that in practice at our disposal are discrete data, which are transformed to a form of functions (\cite{Horvath}, \cite{Ramsay:2009}).

The FDA offers novel methods for the decomposition of income distributions or yield curves, and for analyzing conditional distributions basing on huge, sparse economic datasets (\cite{Sparse}). The FDA enables us for effective statistical analysis of data, when the number of variables exceeds the number of observations.

\textbf{In economic applications}, we assume that we observe a sample consisting of $N$ curves ${{X}_{1}},{{X}_{2}},...,{{X}_{N}}$ drawn from a certain regular model. For example, let ${{y}_{t}}(x)$ denotes a function such as \textbf{monthly income} for the \textbf{continuous age variable} $x$ in a year $t$. We assume that there is an underlying smooth function ${{f}_{t}}(x)$ that is observed with error at discretized grid points of $x$. In practice, we observe $\{{{x}_{i}},{{y}_{t}}({{x}_{i}})\},$ for $t=1,2,...,n$ and $i=1,2,...,p$ , from which we extract a smooth function ${{f}_{t}}(x)$, given by
$${{y}_{t}}({{x}_{i}})={{f}_{t}}({{x}_{i}})+{{\sigma }_{t}}({{x}_{i}}){{\varepsilon }_{t,i}},$$
where ${{\varepsilon }_{t,i}}$ is usually i.i.d. standard normal variable, ${{\sigma }_{t}}({{x}_{i}})$ allows the amount of noise to vary with ${{x}_{i}}$, and $\{{{x}_{1}},{{x}_{2}},...,{{x}_{p}}\}$ is a set of discrete data points.
A special case of ${{\left\{ {{y}_{t}}(x) \right\}}_{t\in \mathbb{N}}}$ is when the continuous variable $x$ is also a time variable. Let $\{{{Z}_{w}},w\in [1,N]\}$ be a seasonal time series that has been observed at $N$ equiv-spaced time points.
We divide the time series into $n$ trajectories, and then consider each trajectory of length $p$ as a curve rather than $p$ distinct data points. \textbf{The functional time series} (FTS) is given by
$${{y}_{t}}(x)=\{{{Z}_{w}},w\in (p(t-1),pt]\} , t=1,2,...,n.$$
The FTS is series of functions indexed by a while the observation is done.
The FDA enables us also for an effective analysis of \textbf{economic data streams}, i.e., an analysis of unequally spaced observed time series, for which the classical Box-Jenkins methodology is not applicable (e.g., future contracts analysis). The time series techniques proposed within the FDA enables for prediction of whole future trajectory of a phenomenon instead of predicting single consecutive observation (\cite{JMS}, \cite{KosRyMie:2017computational}).
\subsection{Global and local depths for functional data}
Depths describe certain global properties of data cloud or the underlying distribution in terms of degree of outlyingness of a point from a center--the median. However, in many  situations local properties of data are of prime importance. To these situations belong the clustering issues, the probability distribution estimation, or pattern recognition problems. In this context, several local extensions of depths have been proposed (\cite{Pain:2013}). Local versions of functional depths, which are available within the \pkg{DepthProc}, are based on Paindaveine and Van Bever concept of the locality and are appropriately adjusted for the functional case (\cite{Hierarchical}). For other possibilities in this context refer, for instance, \cite{Sguera2016}.\\
Let us concentrate on  the most popular Lopez-Pintado and Romo concepts of depths for functional data, which are implemented within the \pkg{DepthProc} package. Formal definition of the functional depth and inspiring discussions of their theoretical properties may be found in \cite{Nieto} and \cite{Nagy}. Let ${{x}_{1}}(t),...,{{x}_{n}}(t)$ denote a set of real functions, for simplicity let us assume that they belong to $C[0,1]$, a space of continuous functions defined on an interval $[0,1]$. A graph of a function $x$ is a subset ${{\mathbb{R}}^{2}}$
\begin{equation}
     G(x)=\{(t, t(x)):t\in [0,1]\}.
\end{equation}
A band in ${{\mathbb{R}}^{2}}$ determined by $k$ functions from a sample ${{x}_{1}},...,{{x}_{n}}$ is defined as following:
            $$V({{x}_{{{i}_{1}}}},{{x}_{{{i}_{2}}}},...,{{x}_{{{i}_{k}}}})=\left\{ (t,y):t\in [0,1],\underset{r=1,...,k}{\mathop{\min }}\,{{x}_{{{i}_{r}}}}(t)\le y\le \underset{r=1,...,k}{\mathop{\max }}\,{{x}_{{{i}_{r}}}}(t) \right\}$$
$$=\left\{ (t,y):t\in [0,1],y={{\alpha }_{t}}\underset{r=1,...,k}{\mathop{\min }}\,{{x}_{{{i}_{r}}}}(t)+(1-{{\alpha }_{t}})\underset{r=1,...,k}{\mathop{\max }}\,{{x}_{{{i}_{r}}}}(t),{{\alpha }_{t}}\in [0,1] \right\}.$$

For any function $x$ and set of functions $\{{{x}_{1}},...,{{x}_{n}}\},$ an index of $j$ functions,
             $$S_{n}^{(j)}(x)={{\left( \begin{matrix}
   n  \\
   j  \\
\end{matrix} \right)}^{-1}}\sum\limits_{1\le {{i}_{1}}<{{i}_{2}}<\cdots <{{i}_{j}}\le n}{I\left\{ G(x)\subset V({{x}_{{{i}_{1}}}},{{x}_{{{i}_{2}}}},...,{{x}_{{{i}_{j}}}}) \right\}},$$ $j\ge 2$,
expresses a fraction of bands $V({{x}_{{{i}_{1}}}},{{x}_{{{i}_{2}}}},...,{{x}_{{{i}_{j}}}})$ determined by $j$ different functions ${{x}_{{{i}_{1}}}},{{x}_{{{i}_{2}}}},...,{{x}_{{{i}_{j}}}},$ covering a graph of $x$.\\
\textbf{Definition 2} (Lopez-Pintado \& Romo 2009): For functions ${{x}_{1}},...,{{x}_{n}},$ the band depth (BD) of a function $x$ equals
                               $${{S}_{n,J}}(x)=\sum\limits_{j=2}^{J}{S_{n}^{(j)}(x)},$$
                               $J\ge 2$.
In case, when ${{X}_{1}},...,{{X}_{n}}$ are independent copies of stochastic process $X$, which generates ${{x}_{1}},...,{{x}_{n}}$, population versions of depth indices are defined:
                                 $${{S}^{(j)}}(x)=P\left\{ G(x)\subset V({{X}_{{{i}_{1}}}},{{X}_{{{i}_{2}}}},...,{{X}_{{{i}_{j}}}}) \right\},$$
                                 $${{S}_{J}}(x)=\sum\limits_{j=2}^{J}{{{S}^{(j)}}(x)=\sum\limits_{j=2}^{J}{P\left\{ G(x)\subset V({{X}_{{{i}_{1}}}},{{X}_{{{i}_{2}}}},{{X}_{{{i}_{j}}}}) \right\}}}.$$

A function being a sample median with respect to the sample ${{\hat{m}}_{n,J}}$ is a curve, which maximizes the sample depth:
                                                 $${{\hat{m}}_{n,J}}=\underset{x\in \{{{x}_{1}},...,{{x}_{n}}\}}{\mathop{\arg \max }}\,{{S}_{n,J}}(x).$$
In a population case, as the median, we take a curve ${{m}_{J}}$ in $C[0,1]$ that maximizes ${{S}_{J}}(\cdot )$.
	Unfortunately, there are great difficulties in applying the above BD concept of functional depth in the case of economic time series. Trajectories of economic objects are crossing for many times, which makes the band depth rather useless. Lopez-Pintado and Romo have proposed a much better concept of functional depth for economic applications in \cite{Lop:2009} and have named it \emph{the modified band depth} (MBD).

For each function $x$ from a sample of functions ${X}^{n}=\left\{{x}_{1},...,{x}_{n}\right\}$ and for any $j=1,2,…,n,$ let
\begin{equation}
{{A}_{j}}(x)\equiv A(x;{{x}_{{{i}_{1}}}},{{x}_{{{i}_{2}}}},...,{{x}_{{{i}_{j}}}})\equiv \left\{ t\in I:\underset{r={{i}_{1}},...,{{i}_{j}}}{\mathop{\min }}\,{{x}_{r}}(t)\le x(t)\le \underset{r={{i}_{1}},...,{{i}_{j}}}{\mathop{\max }}\,{{x}_{r}}(t) \right\}
\end{equation}
denote a subset of an interval $I$ (on which the function $x$ is defined ), on which the function $x$ lies inside a band determined by observations ${{{x}_{{{i}_{1}}}},{{x}_{{{i}_{2}}}},...,{{x}_{{{i}_{j}}}}}$. Let  $\lambda$ denote the Lebesgue measure. Then, for $j=2,3,…,n,$ we define a quantity
\begin{equation}
MBD_{n}^{(j)}(x)={{\left( \begin{matrix}
   n  \\
   j  \\
\end{matrix} \right)}^{-1}}\sum\limits_{1\le {{i}_{1}}<...<{{i}_{j}}\le n}^{{}}{\frac{\lambda ({{A}_{j}}(x))}{\lambda (I)}}
\end{equation}
 measuring ”how frequent” (assuming that $I$ denotes an interval of time) a given observation $x$ is inside the band. If we fix $J=2,3,…,n$, then the modified band depth of a function $x$ with respect to a sample ${{x}_{{{i}_{1}}}},{{x}_{{{i}_{2}}}},...,{{x}_{{{i}_{j}}}}$ is equal to
\begin{equation}
MBD_{n,J}^{{}}(x)=\sum\limits_{j=2}^{J}{MBD_{n}^{(j)}(x)}.
\end{equation}
In applications, one usually assumes $J=2$, hence only considers bands determined by each pair of observations. A population version of the MBD takes a form
\begin{equation}
MBD_{J}^{{}}(x)=\sum\limits_{j=2}^{J}{MBD_{{}}^{(j)}(x)},
\end{equation}
where
$$ MBD_{{}}^{(j)}(x)=\mathbb{E}\sum\limits_{1\le {{i}_{1}}<...<{{i}_{j}}\le n}^{{}}{\frac{\lambda ({{A}_{j}}(x;{{X}_{1}},{{X}_{2}},...,{{X}_{j}}))}{\lambda (I)}}.$$
It is worth noticing that the BD takes into account a shape of curves in a higher degree than the MBD, whereas the last one is more concentrated on amplitudes of curves. Curves being almost always in the center and taking extremal values on short intervals shall take high value of the MBD and small value of BD (\cite{Kos_Ry_Mie:2017canadian}, \cite{Func_out}).

Figure 15 presents functional boxplots showing trajectories of dangerous substances in air concentration during a day and night in Cracow city of Poland in December of 2016. The pollution with dust particles PM10 and PM2.5 relate to activities of heating systems and they influence on allergy issues. The pollution with NO relate to a traffic intensity and the increased nuisance due to an incident of smog. One can notice a smaller degree of NO and NOx pollution between 10.00 and 16.00, pollution of PM2.5 about midnight, which may be especially dangerous for infants and small children.  The functional boxplots of this kind may be used in the context of optimization of municipal health and ecological politics including the designing of smog alert system and pro ecological taxation system.\\
Figures 16 to 19 present the Internet user activities in a certain Internet service measured by means of the number of users and the number of "clicks". The boxplots may be used in the context of intrusion into computer systems detection. The boxplots were prepared using the modified band depth (MBD) and Frainman and Muniz depth (FM). Departures from the median trajectories may signal events, which need the attention of the administrator of the service. It is worth noticing that boxplots presentingthe number of users do not directly correspond to boxplots presenting the number of "cliks". This fact suggests the differences in types of users in particular intervals of a day and night. Considering an year scale of the phenomenon, a danger related to the trajectory may be expressed in terms of its closeness to the appropriate functional median (\cite{zakopanemediany}). An automatic alerting system may be formulated in terms of the classifier for functional data (\cite{zakopaneSVM}). Figures 20--21 present the comparison of two services considered with respect to the number of users obtained using the Depth vs. depth plots (\cite{Liu:1999}). Shapes of patterns of points on the figures indicate differences in the location between the considered services, which roughly speaking, denote different prices of advertisement spaces in the services. The pattern should lead an analyst to further study the nature of the differences (multivariate skewness, kurtosis \cite{Liu:1999}).

\begin{figure}
\centering

  \includegraphics[width=.95\linewidth]{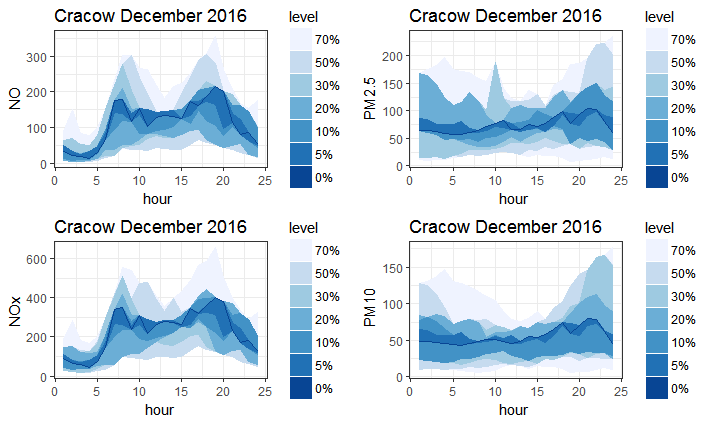}
  \caption{The functional boxplots: Air pollution in Cracow during a day and night in December 2016.}
  \label{fig15}
\end{figure}

\begin{figure}
\centering
\begin{minipage}[t]{.45\textwidth}
  \centering
  \includegraphics[width=.95\linewidth]{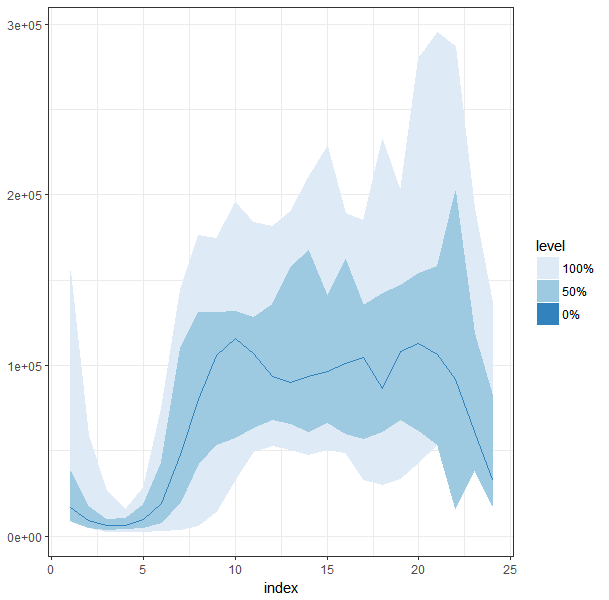}
  \caption{The functional boxplot: number of users in service 1, the MBD depth.}
  \label{fig16}
\end{minipage}
\mbox{\hspace{0.1cm}}
\begin{minipage}[t]{.45\textwidth}
  \centering
  \includegraphics[width=.95\linewidth]{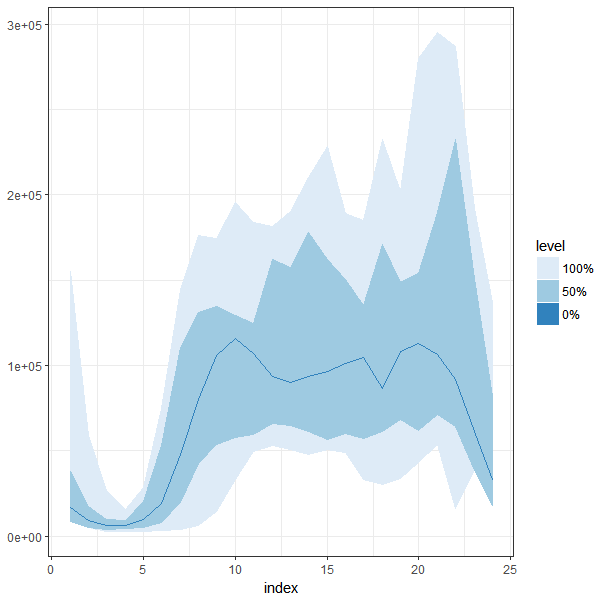}
  \caption{The functional boxplot: number of users in service 1, the FM depth.}
  \label{fig17}
\end{minipage}
\end{figure}

\begin{figure}
\centering
\begin{minipage}[t]{.45\textwidth}
  \centering
  \includegraphics[width=.95\linewidth]{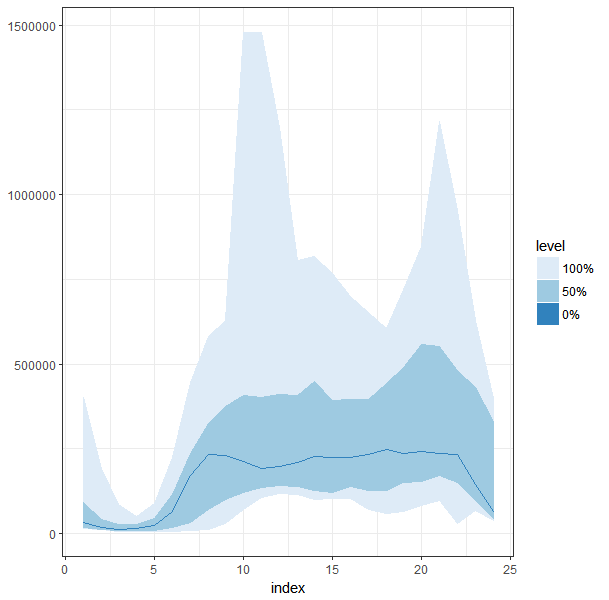}
  \caption{The functional boxplot: number of page views in service 1, the MBD depth.}
  \label{fig18}
\end{minipage}
\mbox{\hspace{0.1cm}}
\begin{minipage}[t]{.45\textwidth}
  \centering
  \includegraphics[width=.95\linewidth]{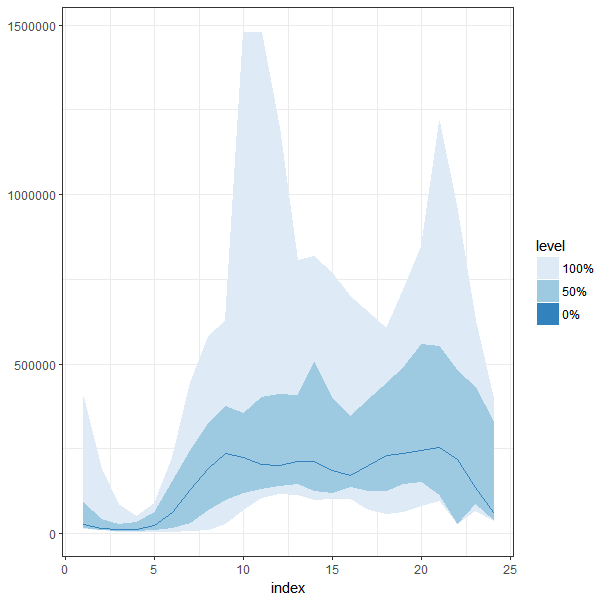}
  \caption{The functional boxplot: number of page views in service 1, the FM depth.}
  \label{fig19}
\end{minipage}
\end{figure}

\begin{figure}
\centering
\begin{minipage}[t]{.45\textwidth}
  \centering
  \includegraphics[width=.95\linewidth]{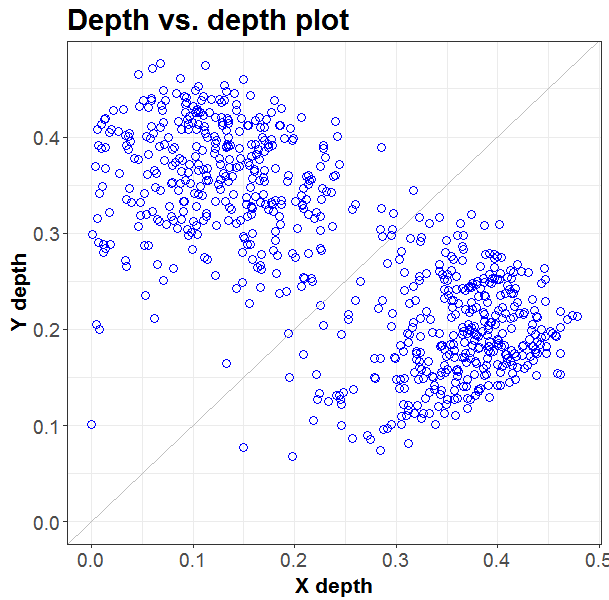}
  \caption{The DD-plot: number of users in service 1 vs. number of users in service 2; local MBD depth, $\beta=0.25$.}
  \label{fig20}
\end{minipage}
\mbox{\hspace{0.1cm}}
\begin{minipage}[t]{.45\textwidth}
  \centering
  \includegraphics[width=.95\linewidth]{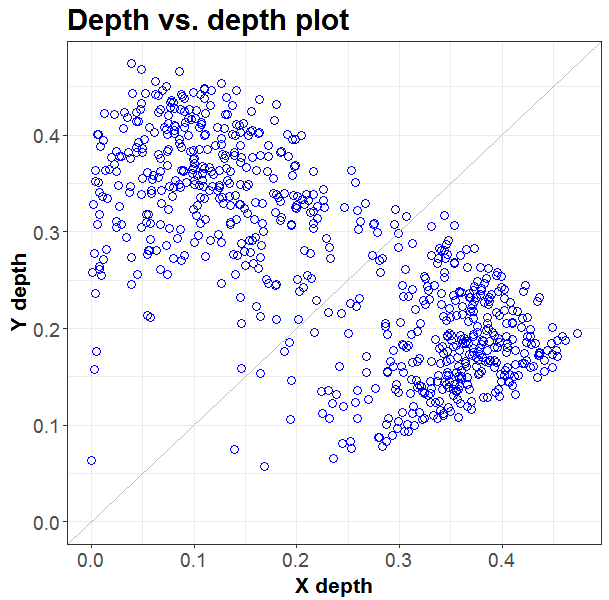}
  \caption{The DD-plot: number of users in service 1 vs. number of users in service 2; local MBD depth, $\beta=0.45$.}
  \label{fig21}
\end{minipage}
\end{figure}
For further generalization of the modified band depth and their theoretical properties see \cite{Nieto}.

It should be stressed that depths for functional data lead to robust functional data analysis and interesting methods of functional outliers detection (\cite{KosRyMie:2017computational} \cite{Nicolas}, \cite{Kos_Ry_Mie:2017canadian})

\subsection{Approximate depth calculation}
Despite certain very important results obtained in last years (\cite{LiuZuoWang2013}, \cite{ProjMatlab2015}, \cite{DyckMoz2016}), a direct calculation of many statistical depth functions is still a very challenging computational issue. On the other hand, a computational tractability of depths and procedures induced by them is especially important for an on line economy management and monitoring, such as studying of high frequency financial data, social networks monitoring, and the Internet shopping center management (\cite{JMS}, \cite{Gab:2012}).

Within the \pkg{DepthProc} package, we use an approximate algorithm proposed in \cite{Dyck:2004} for the calculation of a certain class of location depth functions (depth functions possessing the so-called weak projection property) and dedicated for the centrally symmetrical distributions on $\mathbb{R}^{d}$. Theoretical background of the approach may be found in \cite{Cuesta:2008} and references therein. We use our own fast algorithm for the MBD calculation, an algorithm proposed in \cite{Struyfregdepth} for the deepest regression calculation, and the direct algorithm \pkg{lsdepth} for the Student depth calculation proposed in \cite{lsdepth}. For the calculation local depths, we use a direct method described in \cite{Pain:2012}.\\
In the \pkg{DepthProc}, in order to decrease the computational burden related to sample depth calculation, we use proposition 11 from \cite{Dyck:2004}. By default we use 1000 random projections from the uniform distribution on a sphere of an appropriate dimension. If the number 1000 seems to lead to "too sparse" randomization for a researcher it is easy to increase the number of projections.\\
\emph{Note:} Having at disposal a family of one dimensional depths, we can define and easily calculate the corresponding family of integrated functional depths \cite{Nagy}. We use this idea within the \pkg{DepthProc} package for the FM depth calculation and for the derivative procedures.

\subsection{Existing software for depth calculation}
Currently there are several packages available on  \pkg{CRAN} servers, which are directly dedicated for certain aspects of depth calculation that are especially important for procedures proposed by the authors of these packages (\pkg{depth} of \cite{depthPack}, \pkg{depthTools} of \cite{depthTools}, \pkg{localdepth} of \cite{localdepth} and \pkg{ddalpha} of \cite{Mosler:2014}). Additionally, the three packages \pkg{fda.usc} by \cite{fda.usc}, \pkg{fda} by \cite{Ramsay:2009}, and \pkg{rainbow} by \cite{rainbow} comprise of a very rich and useful family of tools related to the calculation and visual presentation of depths for functional data, including the preparation of functional boxplots. For instance, within the \pkg{fda.usc}, one can find several classifiers as well as clustering procedures for functional data induced by popular functional depths.

The \pkg{depth} package allows for exact and approximate calculation of Tukey, Liu, and Oja depths. It also provides tools for vizualisation contour plots and perspective plots of depth functions, and functions for depth median calculation. It is wort noting that the commands \code{depthContour} and \code{depthPersp}, which are available within the \pkg{DepthProc}, were patterned on these \pkg{depth} commands.

The \pkg{depthTools} is focused on the MBD for functional data (\cite{Lop:2009}). It provides scale curve, rank test based on the MBD, and two techniques of supervised classification--the  distance to the trimmed mean classification method (DS) and the weighted trimmed mean distance classification method (TAD).

The \pkg{localdepth} package enables us for the calculation of local version of "simplicial", "ellipsoid", "halfspace" (Tukey's depth), "Mahalanobis" and "hyperspheresimplicial" depth functions. The \pkg{localdepth} also has a function for the depth-vs-depth plot, which differs from the function that is available within the \pkg{DepthProc}. In the \pkg{localdepth}, the DD-plot is a plot of normalized localdepth versus normalized depth. We should also note that the version of the local depth, which is available within the \pkg{localdepth}, differs from the more general version proposed in \cite{Pain:2013} that is available within the \pkg{DepthProc}.

The \pkg{ddalpha} package originally concentrated around a new method for the classification basing on the DD-plot prepared using the random Tukey depth and the zonoid depth--now offers implementations of recent computational developments (e.g., \cite{DyckMoz2016}) in the DDC area. It is worth noticing the package \pkg{WRS2} being the part of a very important book on general aspects of robust statistics from \cite{Wilcox}, while it also consists of a selection of multivariate depths. It is also worth noticing the recently (after the submission of this paper) appeared package associated with \cite{Nicolas}, that enables for certain kind of the "magnitude" as well as "shape" functional outliers detection.\\
Our package, however, seems to be an user-friendly selection of tools dedicated for "robust economic analysis". (\cite{Kos:2012mon}). Its properties seem to be a reasonable choice in the context of a trade-off between the precision, speed, price, and the offered visualization opportunities. For example, the Student median, which is rather not well known even in a statistical community, seems to be especially interesting in the context of attractiveness of a candidate on a labor market evaluation where the closeness to the center is taken into account regarding the dispersion and the shape of a distribution. For using a very good \pkg{CompPD}, package (\cite{ProjMatlab2015}) we need a rather expensive \pkg{MATLAB} program and by using it we can analyze up to eight variables (using free \pkg{Octave} program we can analyze only 2 variables, because of the fact that many \pkg{MATLAB} build-in-functions do not have free counterparts). We would like to stress that a significant part of the procedures implemented within our package has a local version, which is especially interesting from an economic point of view, where the locality concept being taken from \cite{Pain:2013}, \cite{SGH}.
\vskip1mm
\begin{center}
\includegraphics[width=0.85\textwidth]{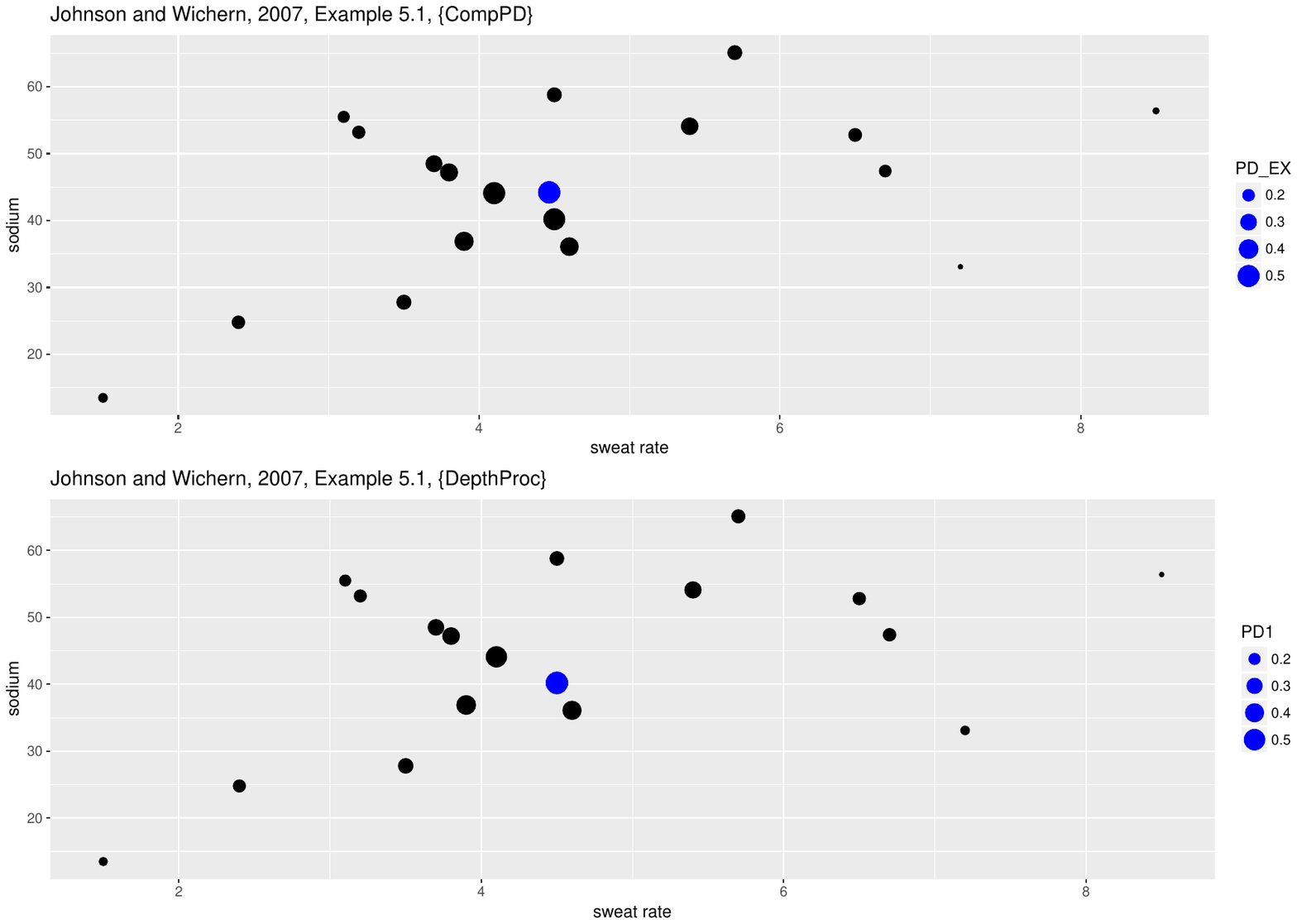}
\captionof{figure}{Exact sample projection depth vs. calculated using \pkg{DepthProc}}
\label{fig:22}
\includegraphics[width=0.85\textwidth]{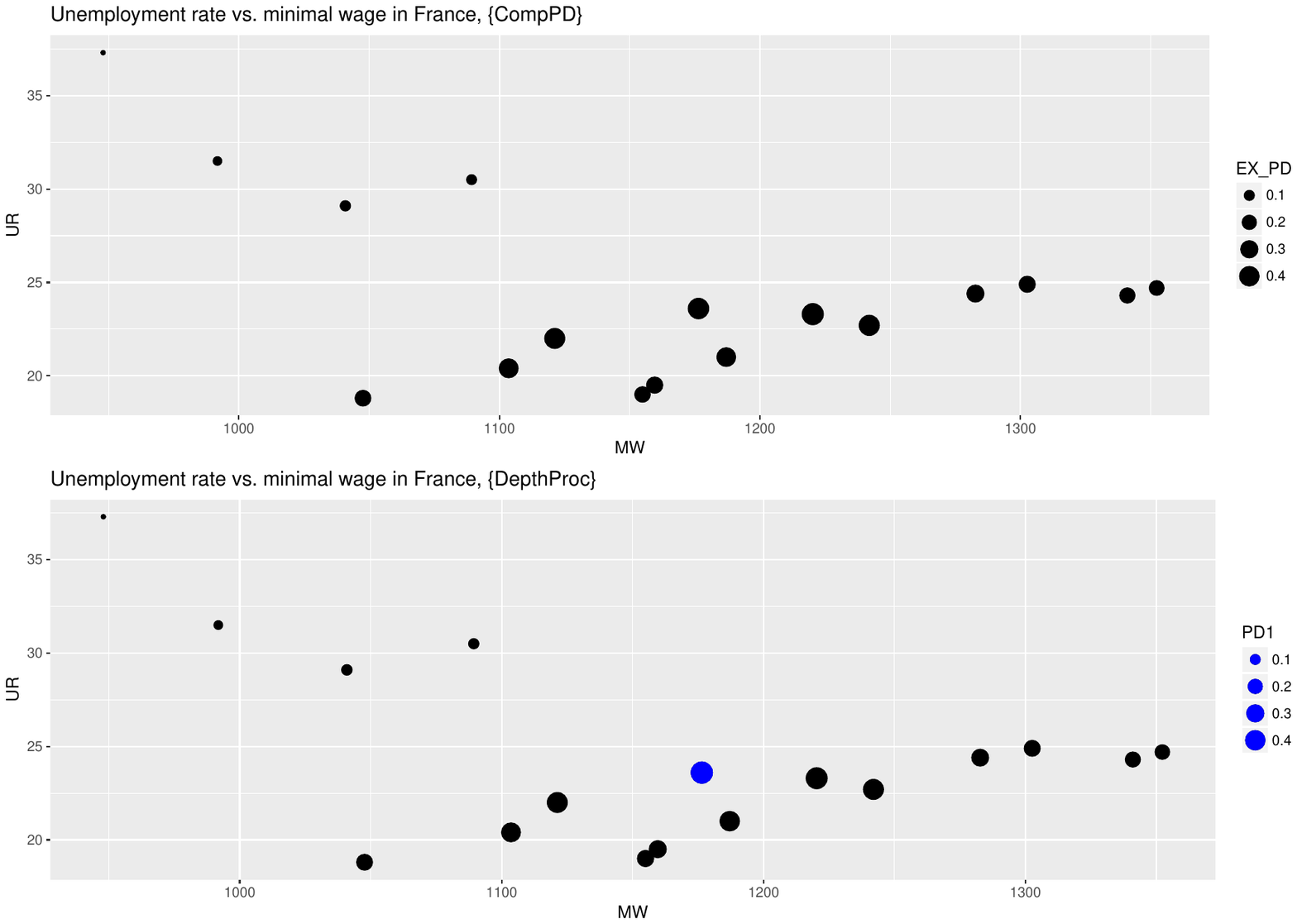}
\captionof{figure}{Exact sample projection depth vs. calculated using \pkg{DepthProc}}
\label{fig:R23}
\includegraphics[width=0.85\textwidth]{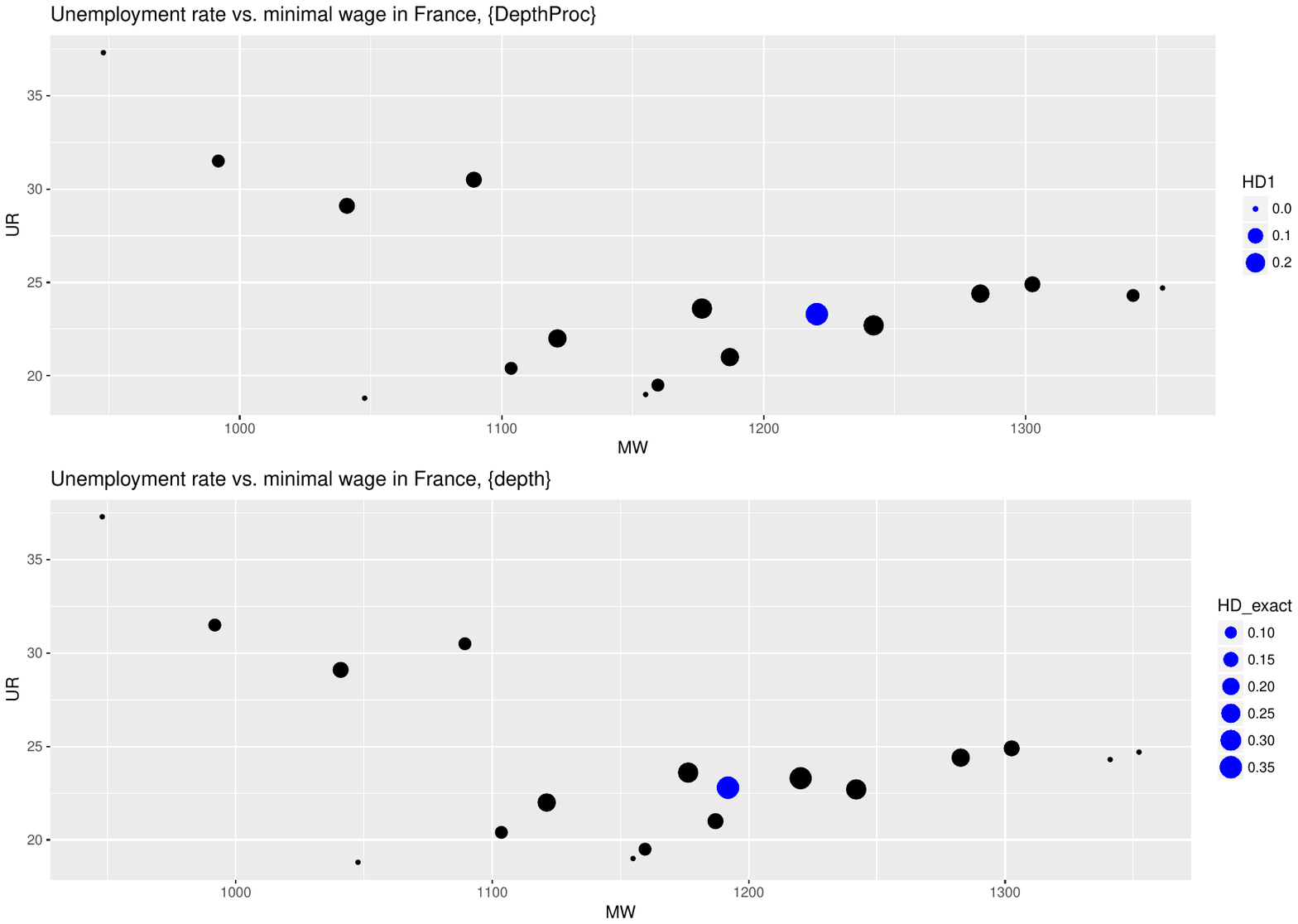}
\captionof{figure}{HD exact \emph{depth} vs. HD \emph{Depthproc}}
\label{fig:R24}
\includegraphics[width=0.85\textwidth]{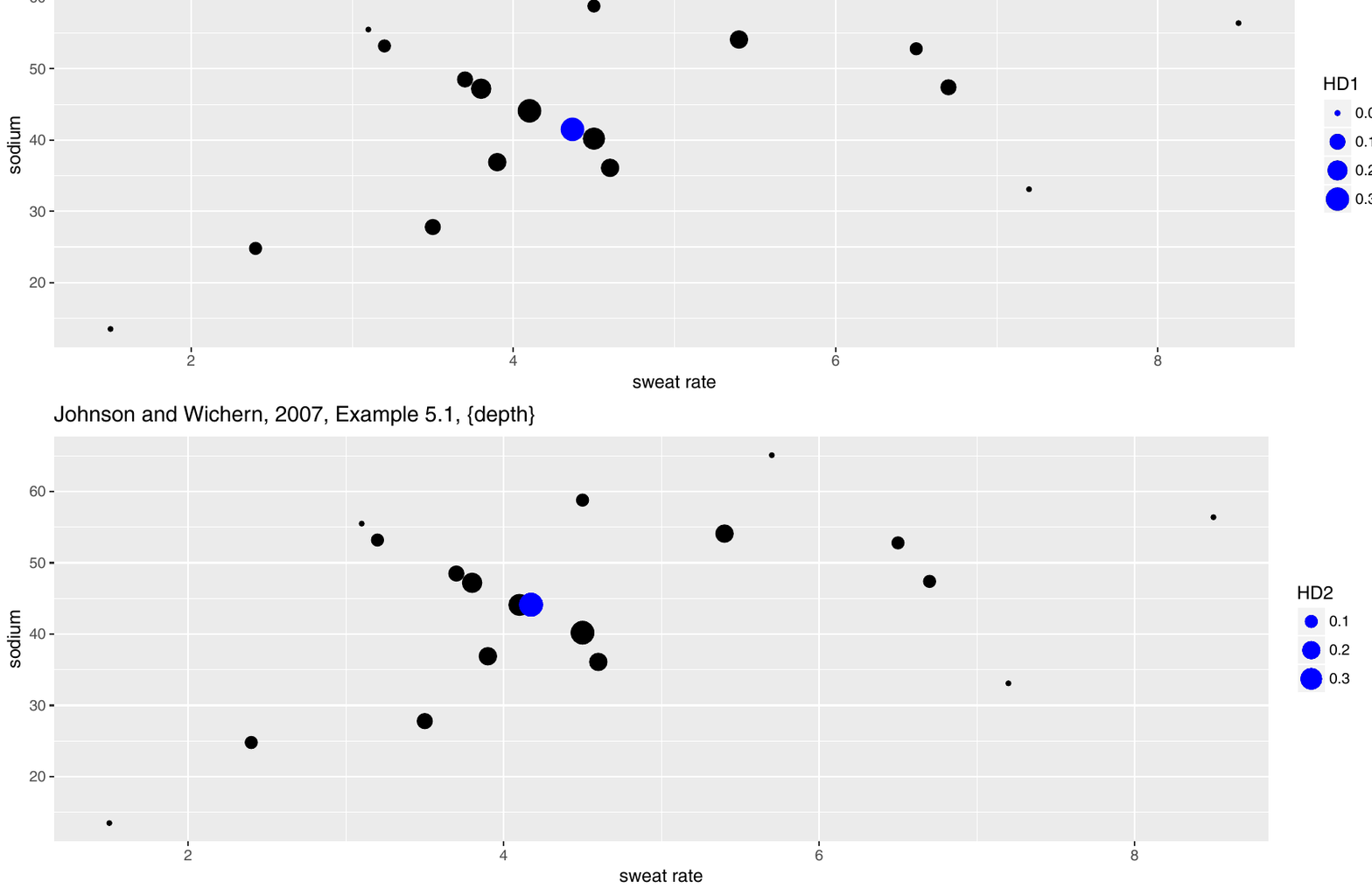}
\captionof{figure}{HD approximate via \emph{depth} vs. HD via \emph{Depthproc}}
\label{fig:25}

\begin{minipage}[t]{0.45\textwidth}
\includegraphics[width=0.85\textwidth]{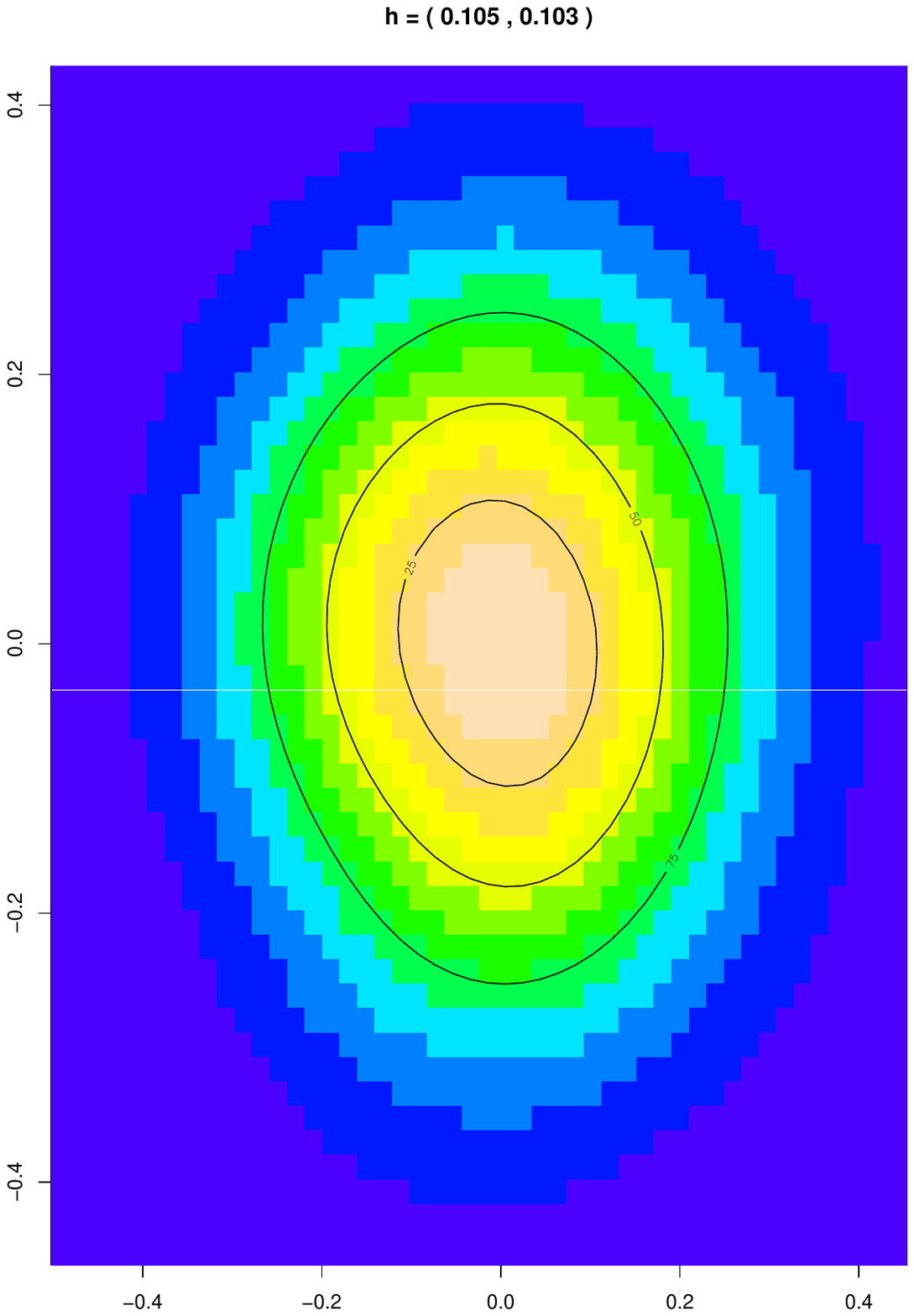}
\captionof{figure}{Kernel density estimate of sample projection median, 100 element sample from 2D normal distribution.}
\label{fig:26}
\end{minipage}
\begin{minipage}[t]{0.45\textwidth}
\includegraphics[width=0.85\textwidth]{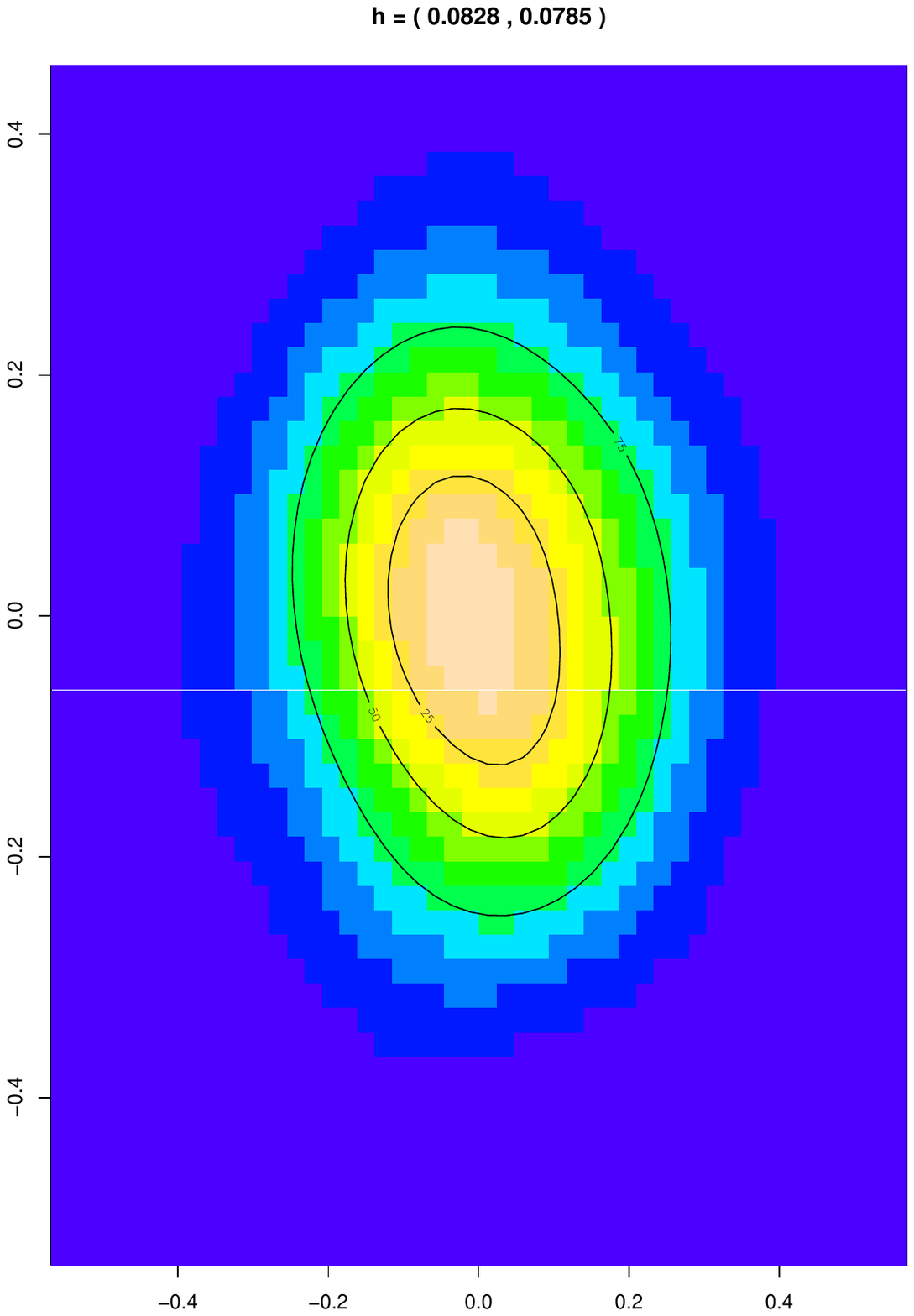}
\captionof{figure}{Kernel density estimate of sample Tukey median, 100 element sample from 2D normal distribution.}
\label{fig:27}
\end{minipage}
\begin{minipage}[t]{0.45\textwidth}
\includegraphics[width=0.85\textwidth]{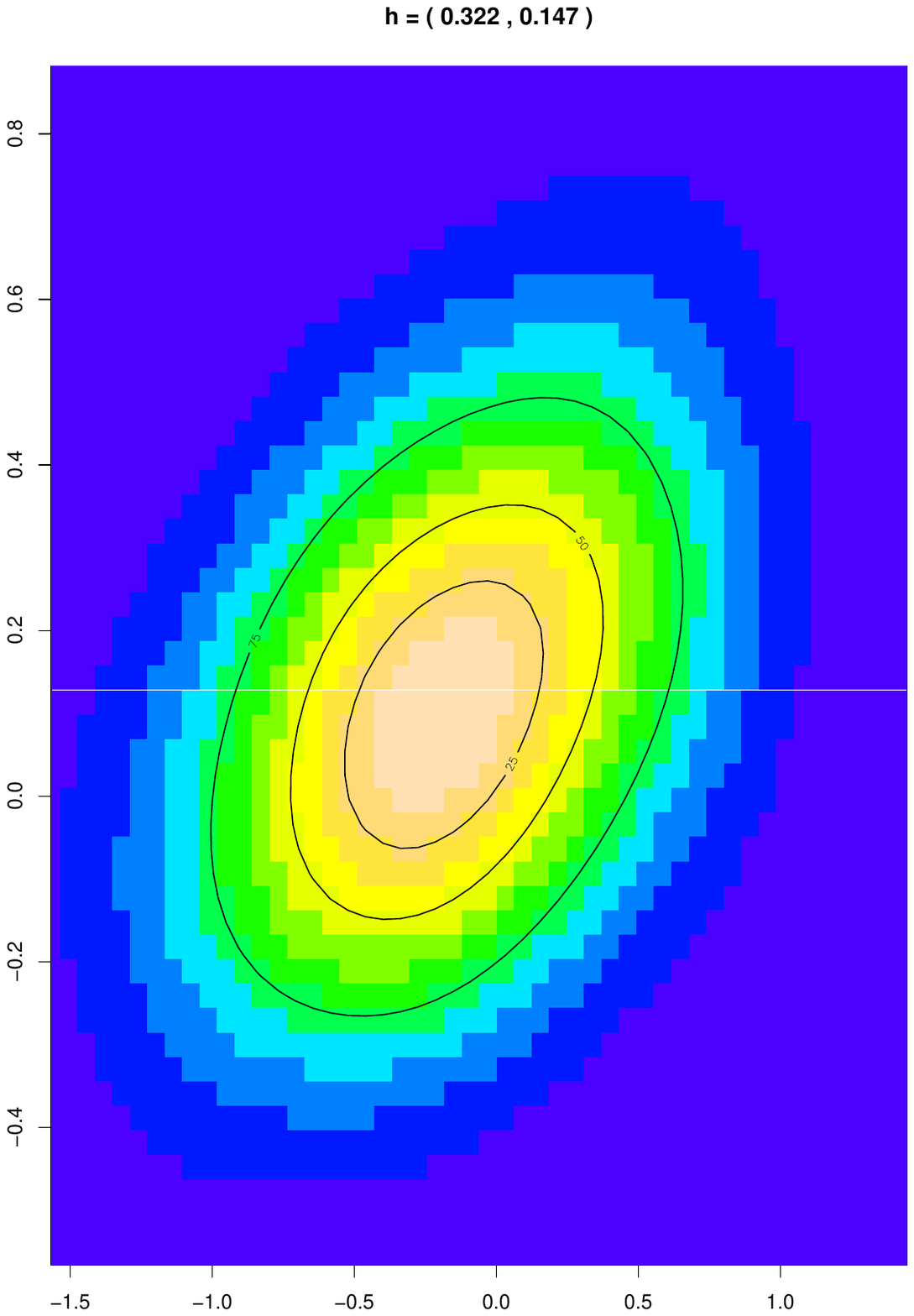}
\captionof{figure}{Kernel density estimate of sample projection median, 100 element sample from a mixture of two 2D normal distributions.}
\label{fig:28}
\end{minipage}
\begin{minipage}[t]{0.45\textwidth}
\includegraphics[width=0.85\textwidth]{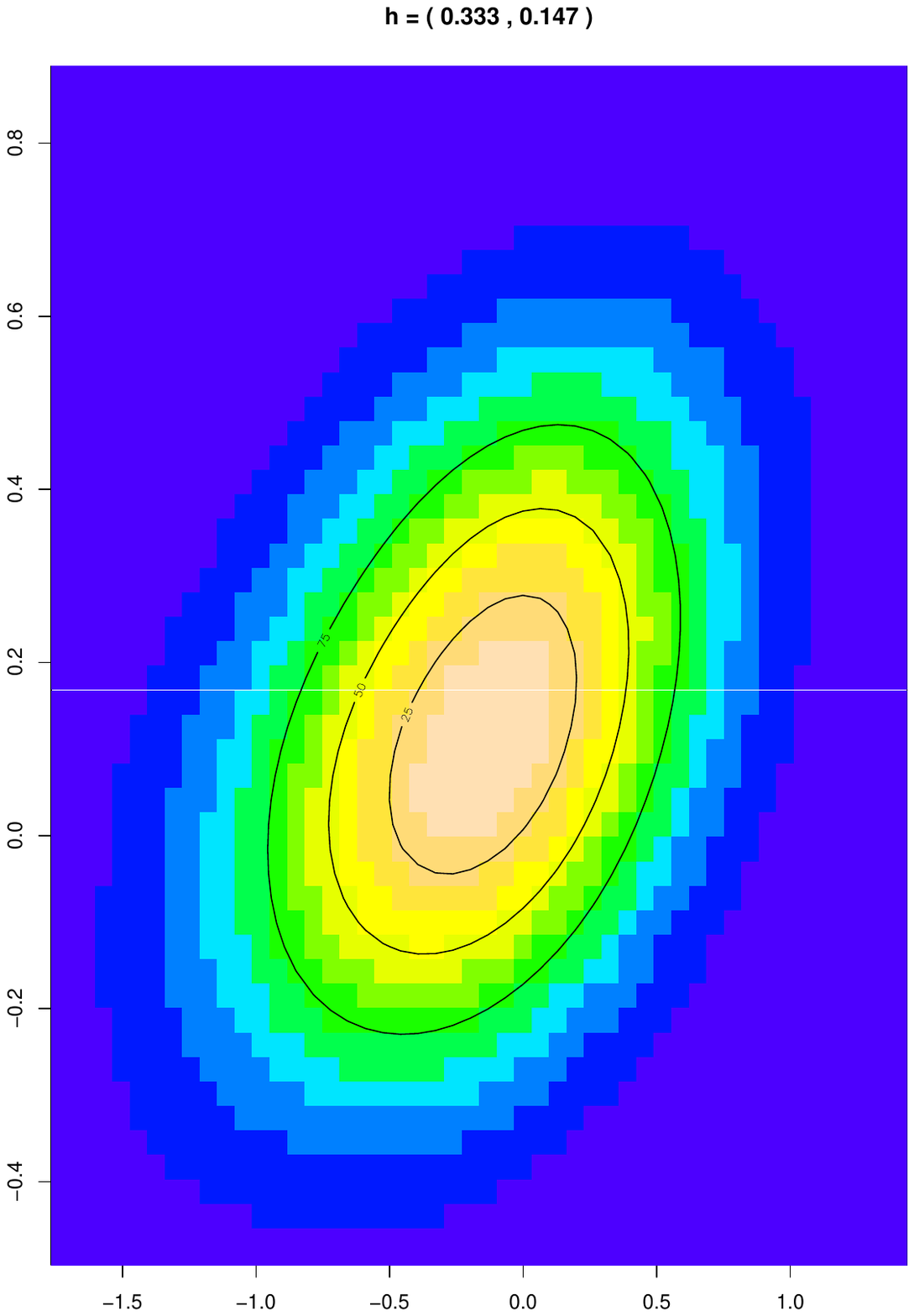}
\captionof{figure}{Kernel density estimate of sample Tukey median, 100 element sample from a mixture of two 2D normal distributions.}
\label{fig:29}
\end{minipage}
\begin{minipage}[t]{0.45\textwidth}
\includegraphics[width=0.85\textwidth]{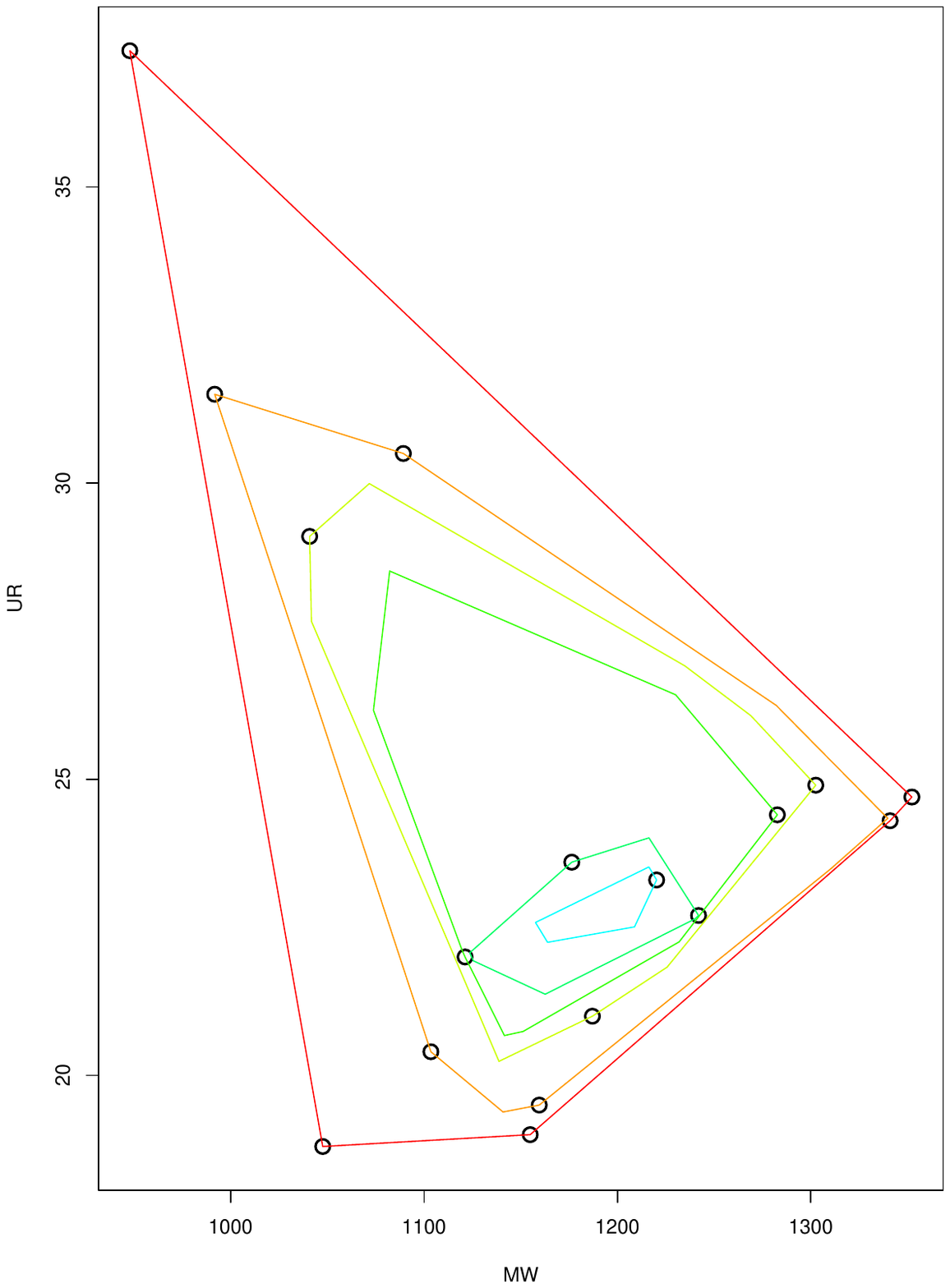}
\captionof{figure}{Tukey depth contour plot using depth of dataset "France".}
\label{fig:30}
\end{minipage}
\begin{minipage}[t]{0.45\textwidth}
\includegraphics[width=0.85\textwidth]{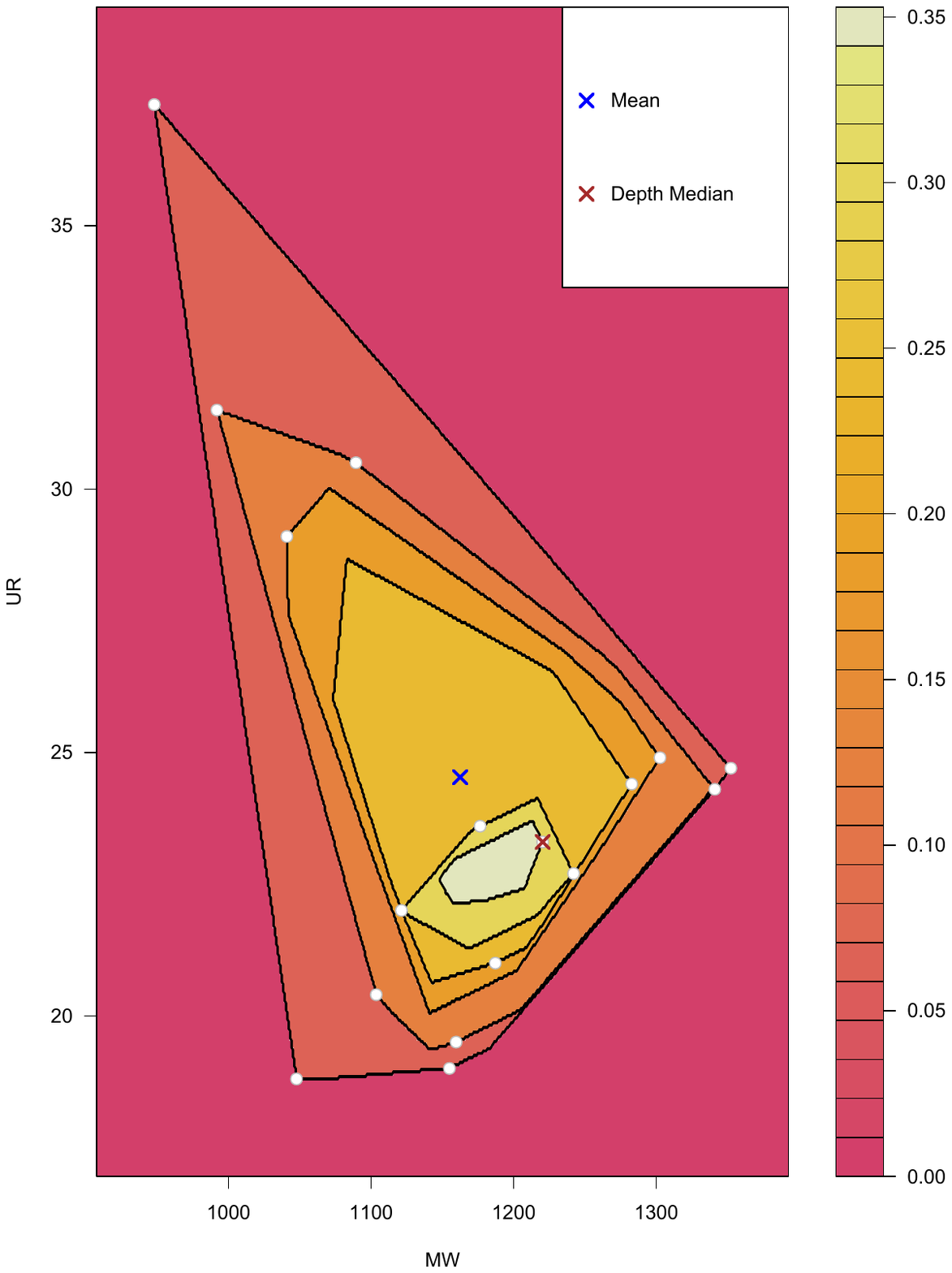}
\captionof{figure}{Tukey depth contour plot using \pkg{DepthProc} of dataset "France".}
\label{fig:31}
\end{minipage}
\end{center}
\vskip2mm
Figures 22--25 present results of comparisons of projection and Tukey depths calculated using the exact and approximate algorithms by means of \pkg{depth}, \pkg{DepthProc}, and \pkg{ddalpha} and ideas taken from \cite{LiuZuoWang2013}. For the comparisons, two empirical datasets were used--Sweat Data (Johnson and Wichern, 2017, example 5.1) and dataset France, which is available within the \pkg{DepthProc}. Generally speaking, we do not observe significant differences between results obtained by the above-listed packages, both depths in points as well as values of the depth-induced medians.\\
Figures 26--27 present kernel density estimates for the corresponding projection and Tukey medians calculated from 100 element sample drawn from a 2D skewed  Student's t- distribution with 5 degrees of freedom.
Figures 28--29 present kernel density estimates for the corresponding projection and Tukey medians calculated from 100 - element sample drawn from a mixture of two 2D normal distributions differing with respect to their location and scale parameters. Figures 30--31 present a comparison of Tukey depth contour plots for dataset France prepared using exact algorithm available within the \pkg{depth} package and an approximate available within the \pkg{DepthProc}. Structurally, the plots look very similar.
\vskip2mm
\begin{table}[ht]
\centering
\begin{tabular}{rlrrrr}
  \hline
 & Package & 1st Qu. & Mean & Median & 3Qu. \\
  \hline
1 & \pkg{ddalpha} (1000 directions) & 10.38 & 10.47 & 10.43 & 10.49 \\
  2 & \pkg{DepthProc} (1000 directions) & 10.52 & 12.04 & 10.64 & 10.73 \\
  3 & \pkg{ddalpha} (5000 directions) & 51.88 & 52.15 & 52.00 & 52.18 \\
  4 & \pkg{DepthProc} (5000 directions) & 47.12 & 47.43 & 47.34 & 47.69 \\
  5 & \pkg{ddalpha} (10000 directions) & 104.16 & 104.69 & 104.39 & 104.89 \\
  6 & \pkg{DepthProc} (10000 directions) & 93.23 & 93.56 & 93.41 & 93.81 \\
   \hline
\end{tabular}
\caption{A comparison of time (in milliseconds) of calculation of projection depth for a sample of 1000 observations for \pkg{DepthProc} and \pkg{ddalpha} for selected numbers of directions. The experiment was repeated 100 times.}
\label{tab1}
\end{table}

\begin{table}[ht]
\centering
\begin{tabular}{rlrrrr}
  \hline
 & Package & 1st Qu. & Mean & Median & 3rd Qu. \\
  \hline
1 & \pkg{depth} & 48.70 & 49.63 & 49.10 & 49.75 \\
  2 & \pkg{DepthProc} & 5.37 & 6.30 & 5.60 & 5.73 \\
   \hline
\end{tabular}
\caption{A comparison of time  (in milliseconds) of calculation of 2D Tukey depth for 1000 observations for \pkg{DepthProc} and \pkg{depth} and samples of 500 observations. The experiment was repeated 100 times.}
\label{tab2}
\end{table}
\begin{table}[ht]
\centering
\begin{tabular}{rlrrrr}
  \hline
 & Package & 1st Qu. & Mean & Median & 3rd Qu. \\
  \hline
1 & \pkg{depth} & 163.17 & 165.29 & 163.95 & 165.48 \\
  2 & \pkg{DepthProc} & 177.50 & 181.93 & 181.96 & 184.22 \\
   \hline
\end{tabular}
\caption{A comparison of time of calculation of 5D Tukey depth for for \pkg{DepthProc} and \pkg{depth} in case of and samples of 200 observations. The experiment was repeated 100 times.}
\label{tab3}
\end{table}
Tables 1-3 presents of results of comparisons of time in milliseconds of sample depth calculation for packages \pkg{ddalpha}, \pkg{DepthProc}, and \pkg{depth}. Samples consisted of 1000, 500 and 200 observations and were generated from normal (table 1), Student 2D with 3 degrees of freedom (table 2), Student 5D with 3 degrees of freedom (table 3) distributions.

Tables 4-6 present results of comparisons of time in milliseconds of sample Tukey median calculation for packages \pkg{DepthProc} and \pkg{depth}. Samples consisted of 500 and 1000 observations and were generated from Student 2D with 3 degrees of freedom (table 4), Student 5D with 3 degrees of freedom (table 5), and Student 7D with 3 degrees of freedom distributions. By "depth (exact)" we denote time of calculation in case of exact algorithm offered by the \pkg{depth} package, by "depth (1000 directions)" and analogously for the \pkg{DepthProc 1000 directions}" we denote an approximate algorithm for Tukey depth calculation using 1000 directions, which is available in the corresponding package. Table 7 presents a comparison of precision of a sample projection depth in a point calculation for the \pkg{ddalpha} and the \pkg{DepthProc} using 1000 and 10000 directions,  Student 4D t distribution with 3 degrees of freedom and samples of 1000 observations. The comparisons were conducted using the \pkg{microbenchmark} package \cite{microbenchmark}. The experiments were repeated for a fixed number of times. The tables consist of basic summary statistics of the experiments (lower quartile (1st Qu.), upper quartile (3rd Qu.)).

\begin{table}[ht]
\centering
\begin{tabular}{rlrrrr}
  \hline
 & Package & 1st Qu. & Mean & Median & 3rd Qu. \\
  \hline
1 & \pkg{depth} (exact) & 12685.47 & 12728.72 & 12721.20 & 12773.32 \\
  2 & \pkg{DepthProc} (1000 directions) & 328.37 & 391.03 & 340.01 & 463.89 \\
   \hline
\end{tabular}
\caption{A comparison of time (in milliseconds) of calculation of Tukey median for \pkg{DepthProc} and \pkg{depth} in case of samples of 500 observations from Student T 2D distribution with 3 degrees of freedom.}
\label{tab4}
\end{table}

\begin{table}[ht]
\centering
\begin{tabular}{rlrrrr}
  \hline
 & Package & 1st Qu. & Mean & Median & 3rd Qu. \\
  \hline
1 & \pkg{depth} (1000 directions) & 75.41 & 76.35 & 76.09 & 76.93 \\
  2 & \pkg{DepthProc} (1000 directions) & 331.92 & 389.55 & 338.30 & 465.02 \\
   \hline
\end{tabular}
\caption{A comparison of time of calculation of Tukey median for \pkg{DepthProc} and \pkg{depth} in case of samples of 500 observations from Student T 5D distribution with 3 degrees of freedom.}
\label{tab5}
\end{table}
\begin{table}[ht]
\centering
\begin{tabular}{rlrrrr}
  \hline
 & Package & 1st Qu. & Mean & Median & 3 Qu. \\
  \hline
1 & \pkg{depth} (1000 directions) & 200.52 & 203.33 & 201.38 & 205.87 \\
  2 & \pkg{DepthProc} (1000 directions) & 492.28 & 571.49 & 614.24 & 625.72 \\
   \hline
\end{tabular}
\caption{A comparison of time of calculation of Tukey median for \pkg{DepthProc} and \pkg{depth} in case of samples of 1000 observations from Student T 7D distribution with 3 degrees of freedom.}
\label{tab6}
\end{table}

\begin{table}[ht]
\centering
\begin{tabular}{rrrrrrr}
  \hline
 Package & SD & 1st Qu. & Median & Mean & 3rd Qu. & MAD \\
  \hline
\pkg{DepthProc}1000 & 0.0089 & 0.45 & 0.46 & 0.46 & 0.47 & 0.0092 \\
  \pkg{ddalpha}1000 & 0.0092 & 0.46 & 0.47 & 0.47 & 0.47 & 0.0083 \\
  \pkg{DepthProc}10000 & 0.0056 & 0.44 & 0.45 & 0.45 & 0.45 & 0.0057 \\
  \pkg{ddalpha}10000 & 0.0061 & 0.45 & 0.45 & 0.45 & 0.46 & 0.0061 \\
   \hline
\end{tabular}
\caption{A comparison of a precision of an approximate estimation of the sample projection depth in a point for  \pkg{DepthProc} and \pkg{ddalpha}. Coordinates of 200 5D observations were generated independently from Student t distribution with 1 degree of freedom. The experiment was repeated times.}
\label{tab7}
\end{table}
Table 7 presents a comparison of a precision of an approximate projection depth calculation in a point conducted using the \pkg{ddalpha} and \pkg{DepthProc} packages. Samples of 200 5-dimensional observations were combined from independent one dimensional Student t distributions with one degree of freedom.
The experiments were repeated for a fixed number of times. The table consist of basic summary statistics of the repetitions (SD denotes standard deviation, MAD denotes median of absolute deviations from the median).
Results of the comparisons lead to a general conclusion, that \pkg{DepthProc} exhibits significant advantages over alternative packages, which offer exact algorithms of depth calculation, in terms speed of calculation, and offers similar or better properties in a comparison to implementations of alternative approximate algorithms, in terms of their speed and precision.
\section{Package description and illustrative examples}
The package comprises commands listed in Table 8.
\begin{table}
\center
\begin{tabular}{|c|c|}
  \hline
  \textbf{Command} & \textbf{Short description} \\
  \hline \hline
  \proglang{asymmetryCurve} & multivariate asymmetry functional \\
  \hline
  \proglang{binningDepth2d} & depth-based simple binning of 2D data \\
  \hline
  \proglang{CovLP}& $L^p$ depth-weighted location and scatter estimator \\
  \hline
  \proglang{ddmvnorm} & multivariate quantile-quantile normality plot \\
  \hline
  \proglang{deepReg2d} & deepest regression estimator for simple regression \\
  \hline
  \proglang{depth} & depth calculation \\
  \hline
  \proglang{depthContour} & depth contour plot \\
  \hline
  \proglang{depthDensity} & depth-weighted density estimator \\
  \hline
  \proglang{depthMBD} & fast modified band depth calculation\\
  \hline
  \proglang{depthmedian} & multivariate median calculation \\
  \hline
  \proglang{depthPersp} & depth perspective plot \\
  \hline
  \proglang{depthLocal} & local depth calculation \\
  \hline
  \proglang{fncBoxPlot} & functional boxplot\\
  \hline
 \proglang{lsdSampleMaxDepth} & Student median calculation \\
  \hline
 \proglang{medianDepthConfinterval} & bootstrap region for a multivariate median \\
  \hline
  \proglang{mWilcoxonTest} & multivariate and functional global and local Wilcoxon test \\
  \hline
  \proglang{ScaleCurve} & multivariate scatter functional \\
  \hline
  \proglang{trimmReg2d} & projection depth-trimmed regression 2D \\
  \hline
  \proglang{kMedian} & k-local multivariate and functional medians clustering\\
  \hline
  \proglang{fncClass} & robust classifier for functional data\\
  \hline
  \proglang{fncPred} & robust predictor for functional time series\\
  \hline
\end{tabular}
\caption{Main commands available within the \pkg{DepthProc}.}
\label{tab8}
\end{table}
The \proglang{depthDensity}, \proglang{kMedian}, \proglang{FuncClass}, \proglang{FuncPred} commands corresponding to nonparametric, weighted by the local depth conditional probability density estimator, for k- local medians clustering for multivariate and functional data, for classification of functional objects and for robust prediction of functional time series are under development. These commands indicate directions of further development of the package.
\subsection{Available depth functions}
A basic command for depth calculation is\\
\vskip 0.5mm
\code{depth(u, X, depth\_params = list(method="Projection"), threads = -1, ...)}
\vskip 0.5mm
\textbf{Arguments}\\
\vskip 0.5mm
\textbf{u}: Numerical vector or matrix, whose depth is to be calculated. The dimension has to be the same as that of the observations.\\
\textbf{X}: The data as a matrix, a data frame, or a list. If it is a matrix or data frame, then each row is treated as one multivariate observation. If it is a list, all components must be numerical vectors of equal length (coordinates of the observations).\\
\textbf{depth\_params}: list of parameters for the depth function\\

\textbf{method}: name of the desired depth function. Can be one of "Projection", "Tukey", "Mahalanobis", "Euclidean", "LP", "MBD", "FM" or local.\\
\textbf{threads}: number of threads used in parallel computations. Default value -1 means that all possible cores will be used.
...: other parameters specific to the selected depth function.

\subsection{Maximal depth estimators}
The \pkg{DepthProc} enables for calculating multivariate medians induced by depth functions.
\vskip0.5mm
\code{depthMedian(x, depth\_params = list())}
\vskip0.5mm
\textbf{Arguments}:
\vskip0.5mm
\textbf{x}: The data as a $k\times n$ matrix or data frame.\\
\textbf{depth\_params}: list of parameters passed to the depth function.
\vskip1mm
\emph{Note: by default we use a definition of the depth median taken from Liu Parelius and Sighn (1999): "Given a notion of data depth, there is a natural choice of location parameter for the underlying distribution, namely the deepest point or the average of the deepest points if there is more than one. however as pointed one of the Reviewer  this is an incorrect definition of the median in general. For instance, it is typical, even in two dimensions that the maximum Tukey depth will be higher than the maximum reached at any point. therefore, the median is should be  defined as the centroid of the convex set with maximal depth. Due to this remark, the \pkg{DepthProc} offers an option "centroid=FALSE or TRUE".}
\vskip1mm
Table 9 presents a comparison of selected location characteristics calculated for empirical dataset on Polish stock branch indices WIG TELECOMMUNICATION, WIG FOOD, WIG MEDIA, WIG FUELS, WIG CONSTRUCTION, WIG BANKING, and WIG CHEMISTRY for a period from 14.04.1991 to 07.03.2018.
\vskip0.5mm
The dataset consists of $6279\times7$ observations. The \emph{ProDepthProc} denotes the projection median calculated via approximate algorithm provided by the \pkg{DepthProc} package, The \emph{Tukddalpha} denotes the Tukey median calculated via approximate algorithm provided by the \pkg{ddalpha} package, the \emph{TukDepthProc} denotes the Tukey median calculated via approximate algorithm provided by the \pkg{DepthProc} package, \emph{MeanVec} denotes the sample mean vector, and \emph{Med1D} denotes a vector of one-dimensional medians. The measures obtained via \pkg{DepthProc} are in fact averages from 1000 repetitions of the approximate median calculations.\\
\emph{Interpretation:} the sample projection median as well as sample Tukey median provide a better insight into the general market tendency within the considered period. Note  that the considered period involved financial crashes, which produced rather atypical observations. Note also that an application of exact algorithms for multivariate medians for this dataset were practically impossible in the case of using an average class laptop available in 2016 (Intel I7, 16GB). Therefore, in similar cases, we recommend using approximate algorithms of depth calculations.
\begin{table}
\begin{tabular}{|c|c|c|c|c|c|c|c|}
  \hline
  Measure/Index & TELECOM & FOOD& MEDIA & FUELS & CONST& BANKING& CHEM \\
  \hline
  Exact ProMed& 13900 & 21426& 23929& 24397 & 22931 & 30168 & 24397 \\
  DepthProc ProMed& 1252.848 & 1166.272 & 1401.83 & 1401.83 & 1316.2 & 1535.8 & 1401.83 \\
  Tukddalpha & 1292.6 & 1254.6 & 1438.44 & 1438.44& 1374.92 &1515.28 & 1438.44 \\
  TukDepthProc & 1261.01 & 1307.11 & 1422.53 & 1420.33 & 1399.6 & 1567.54 & 1511.74 \\
  MeanVec & 1123.13 & 2317.14 & 2573.24 & 2567.7 & 2762.75 & 4130.4 & 4981.4 \\
  Med1D & 1110.31 & 2290.69 & 2439.89 & 2356.98 & 1917.81 & 3602.57 & 2630.37 \\
  \hline
\end{tabular}
\caption{A comparison of location characteristics calculated using the \pkg{DepthProc}, the \pkg{ddalpha} packages for  7 Polish stocks branch indices in the period of 1991--2018.}
\label{tab9}
\end{table}
\subsection{depthContour and depthPersp}
Basic statistical plots offered by \pkg{DepthProc} are \textbf{the contour plot} and \textbf{the perspective plot} (Figures 1--8).
\vskip1mm
\code{depthContour(x, xlim = extendrange(x[, 1], f = 0.1), ylim = extendrange(x[,
  2], f = 0.1), n = 50, pmean = TRUE, mcol = "blue", pdmedian = TRUE,
  mecol = "brown", legend = TRUE, points = FALSE, colors = heat_hcl,\\
  levels = 10, depth_params = list(),  graph_params = list())} \\
\code{depthPersp(x, plot_method = "lattice", xlim = extendrange(x[, 1], f = 0.1),
  ylim = extendrange(x[, 2], f = 0.1), n = 50, xlab = "x", ylab = "y",
  plot_title = NULL, colors = heat_hcl, depth_params = list(),
  graph_params = list(), contour_method = c("auto", "convexhull", "contour"))}\\
\vskip 0.5mm
\textbf{Arguments}
\vskip1mm
\textbf{x}: The data as a $k\times 2$ matrix or data frame.\\
\textbf{plot\_method}: {There are two options "lattice" and "rgl" --- see details. \emph{Note:} \pkg{rgl} can cause some problems with installation on clusters without OpenGL.\\
\textbf{pmean}: {Logical. If TRUE mean will be marked.}\\
\textbf{mcol}: {Determines the color of lines describing the mean.}\\
\textbf{pdmedian}: {Logical. If TRUE depth median will be marked.}\\
\textbf{mecol}: {Determines the color of lines describing the depth median.}\\
\textbf{legend}: {Logical. If TRUE legend for mean and depth median will be drawn.}\\
\textbf{points}: {Logical. If TRUE points from matrix x will be drawn.}\\
\textbf{xlim}: {Limits for x-axis.}\\
\textbf{ylim}: {Limits for y-axis.}\\
\textbf{n}: {Number of points in each coordinate direction to be used in the plot.}\\
\textbf{xlab}: {Description of x-axis.}\\
\textbf{ylab}: {Description of y-axis.}\\
\textbf{colors}: {Function for colors pallete (e.g., gray.colors).}\\
\textbf{depth\_params}: {List of parameters for the depth function.}\\
\textbf{graph\_params}: {List of graphical parameters for functions rgl::persp3d and lattice::wireframe.}\\
\textbf{contour\_method}: {determines the method used to draw the contour lines. The default value ('auto') tries to determine the best method for given depth function. 'convexhull' uses a convex hull algorithm to determine boundaries. 'contour' uses the algorithm from filled.contour.}\\

\subsection{DD-plots}
For two probability distributions $F$ and $G$, both distributions in $\mathbb{R}^{d}$, we can define \textbf{the DD-plot} as being a very useful generalization of the one-dimensional quantile-quantile plot:
\begin{equation}
DD(F,G)=\left\{ \left( D({z},F),D({z},G) \right),{z}\in {{\mathbb{R}}^{d}} \right\}.
\end{equation}
Its sample counterpart calculated for two samples ${{{X}}^{n}}=\{{{X}_{1}},...,{{X}_{n}}\}$ from $F$, and
${{Y}^{m}}=\{{{Y}_{1}},...,{{Y}_{m}}\}$ from $G$ is defined as
\begin{equation}
DD({{F}_{n}},{{G}_{m}})=\text{  }\left\{ \left( D({z},{{F}_{n}}),D({z},{{G}_{m}}) \right),{z}\in \text{  }\!\!\{\!\!\text{
}{{{X}}^{n}}\cup {{{Y}}^{m}}\} \right\}.
\end{equation}
A detailed presentation of the DD-plot can be found in \cite{Liu:1999}. Figure 32 presents a DD-plot with heart-shaped pattern in the case of differences in a location between two samples, whereas Figure 33 presents moon-shaped pattern in the case of scale differences between the samples. Applications of the DD-plot and theoretical properties of statistical procedures using this plot can be found in \cite{Li:2004}, \cite{Liu:1993}, \cite{Jure:2012}, \cite{Zuo:2006}, \cite{Liu:1999}. In \cite{Mosler:2014}, an application of the DD-plot for the classification can be found.
\vskip1mm
In order to investigate differences in the location and the scale and in order to inspect the "normality of a sample", within the \pkg{DepthProc}, one can use DD-plot in the following way:\\
\code{ddPlot(x, y, scale = FALSE, location = FALSE, name = "X", name_y = "Y",
  title = "Depth vs. depth plot", depth_params = list())}\\
\code{ddMvnorm(x, size = nrow(x), robust = FALSE,
alpha = 0.05,\\
title ="ddMvnorm", ...)} \\
\vskip1mm
\textbf{Arguments}\\
\textbf{x}: {The data as a matrix or a data frame.}\\
\textbf{y}: {The second data sample. \code{x} and \code{y} must be of the same number of columns.}\\
\textbf{scale}: {Logical. Determines whether the dispersion is to be aligned.}\\
\textbf{location}: {Determines whether the location is to be aligned to 0 vector with the depth median.}\\
\textbf{name}: {Name for data set x. It will be passed to the drawing function.}\\
\textbf{name\_y}: {As above for y}\\
\textbf{title}: {Title of the plot.}\\
\textbf{depth\_params}: {List of parameters for the depth function.}\\
\textbf{size}: Size of a theoretical set.\\
\textbf{robust}: Logical, the default is FALSE. If TRUE, robust measures are used to estimate the parameters of the theoretical distribution.\\
\textbf{alpha}: Cutoff point for the robust measure of covariance.
\begin{figure}
\centering
\begin{minipage}[t]{.45\textwidth}
  \centering
  \includegraphics[width=.95\linewidth]{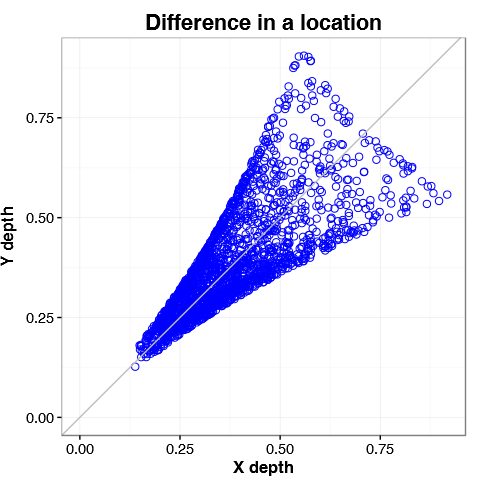}
  \caption{DD-plot, a difference in a location.}
  \label{fig20}
\end{minipage}
\begin{minipage}[t]{.45\textwidth}
  \centering
  \includegraphics[width=.95\linewidth]{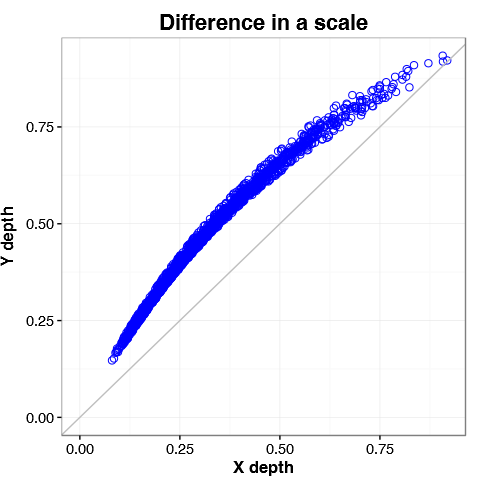}
  \caption{DD-plot, a difference in a scale.}
  \label{fig21}
\end{minipage}
\end{figure}
\vskip0.5mm
\subsection{Multivariate Wilcoxon test}
Having two samples $\mathbf{X}^{n}$ and $\mathbf{Y}^{m},$ using any depth function, we can compute depth values in a combined sample $\mathbf{Z}^{n+m}$ = $\mathbf{X}^{n}\cup \mathbf{Y}^{m}$, assuming the empirical distribution calculated basing on all observations or only on observations belonging to one of the samples $\mathbf{X}^{n}$ or $\mathbf{Y}^{m}.$

For example, if we observe that ${X}_{l}'s$ depths are more likely to cluster tightly around the center of the combined sample, while ${Y}_{l}'s$ depths are more likely to scatter at outlying positions, then we conclude that $\mathbf{Y}^{m}$ was drawn from a distribution with a larger scale.

Properties of the DD-plot-based statistics in the i.i.d. setting were studied in \cite{Li:2004}. Authors proposed several DD-plot based statistics and presented bootstrap arguments for their consistency and good effectiveness in comparison to Hotelling $T^2$ and multivariate analogs of Ansari-Bradley and Tukey-Siegel statistics. Asymptotic distributions of a depth-based multivariate Wilcoxon rank-sum test statistic under the null and general alternative hypotheses were obtained in \cite{Zuo:2006}. Several properties of the depth-based rank test involving its unbiasedness was critically discussed in \cite{Jure:2012}.

Basing on the DD-plot object, which is available within the \pkg{DepthProc}, it is possible to calculate other multivariate and functional  generalizations of one-dimensional rank and order statistics.

The depth-based multivariate Wilcoxon rank sum test is especially useful for the multivariate scale changes detection and it was introduced, among other, by \cite{Liu:1993}\\
For the samples ${{\mathbf{X}}^{m}}=\{{{\mathbf{X}}_{1}},...,{{\mathbf{X}}_{m}}\}$ , ${{\mathbf{Y}}^{n}}=\{{{\mathbf{Y}}_{1}},...,{{\mathbf{Y}}_{n}}\}$ and a combined sample ${\mathbf{Z}}={{\mathbf{X}}^{n}}\cup {{\mathbf{Y}}^{m}},$ the \textbf{Wilcoxon statistic} is defined as
\begin{equation}
S=\sum\limits_{i=1}^{m}{{{R}_{i}}},
\label{eq4}
\end{equation}
where ${R}_{i}$ denotes the rank of the i-th observation, $i=1,...,m$ in the combined
sample
 $R({{\mathbf{x}}_{l}})=  \#(\left\{ {{\mathbf{z}}_{j}}\in {{\mathbf{Z}}}:D({{\mathbf{z}}_{j}},{\mathbf{Z}})\le D({{\mathbf{x}}_{l}},{\mathbf{Z}}) \right\}), l=1,...,m.$
\vskip0.5mm
The distribution of $S$ is symmetric about  $E(S)=1/2m\text{(}m\text{+}n\text{+1)}$.
Its variance equals  ${{D}^{2}}(S)={1}/{12}\;mn(m+n+1).$
For theoretical properties statistic refer \cite{Li:2004} and \cite{Zuo:2006}.\\
The Wilcoxon statistic induced by a depth function may be used for multivariate as well as functional time series monitoring (\cite{JMS}, \cite{Structural}).
\vskip0.5mm
\code{mWilcoxonTest(x, y, alternative = "two.sided", depth_params = list())}\\
\textbf{Arguments}\\
\textbf{x, y}: Data matrices or data frames of the same dimension.\\
\textbf{alternative}: Character string determining the alternative, as in one-dimensional Wilcoxon test.\\
\textbf{depth\_params}: {List of parameters for the depth function.}\\
\vskip0.5mm
The example  1 shows an application of the multivariate Wilcoxon test for two samples drawn from normal distributions differing with respect to a scale.
 \vskip1mm
EXAMPLE 1
\begin{Sinput}
R> library("MASS")
R> x <- mvrnorm(100, c(0,0), diag(2))
R> y <- mvrnorm(100, c(0,0), diag(2)*1.4)
R> mWilcoxonTest(x, y)
\end{Sinput}
\begin{Soutput}
Multivariate Wilcoxon test for equality of distributions
data:  dep_x and dep_y
W = 6034, p-value = 0.01156
alternative hypothesis: distributions are not equal
\end{Soutput}
\subsection{Scale and asymmetry curves }
For sample depth function $D({x};{{{Z}}^{n}})$, ${x}\in {{\mathbb{R}}^{d}}$, $d\ge 2$,
${Z}^{n}=\{{{{z}}_{1}},...,{{{z}}_{n}}\}\subset {{\mathbb{R}}^{d}}$ and ${{D}_{\alpha }}({{{Z}}^{n}})$, denoting $\alpha-$central region, we can define \textbf{the scale curve} (Figure 34) as
\begin{equation}
SC(\alpha )=\left( \alpha ,vol({{D}_{\alpha }}({{{Z}}^{n}}) \right)\subset {{\mathbb{R}}^{2}},\hskip2mm   for \hskip2mm \alpha \in [0,1],
\end{equation}
and \textbf{the asymmetry curve} as (\cite{Serfling:2004}, Figure 35)
\begin{equation}
AC(\alpha )=\left( \alpha ,\left\| {{c}^{-1}}(\{{\bar{z}}-med|{{D}_{\alpha
                       }}({{{Z}}^{n}})\}) \right\| \right)\subset {{\mathbb{R}}^{2}}, \hskip2mm for \hskip2mm \alpha \in [0,1]
\end{equation}
being a nonparametric scale and a asymmetry functional, where $c--$denotes a "consistency" constant, ${\bar{z}}-$denotes a mean vector, and $med$ denotes a median induced by a prefixed depth function and $vol$ denotes the volume.
Further information on the scale and the asymmetry curves can be found in \cite{Liu:1999}, \cite{Serfling:2006}, \cite{Serfling:2004}, \cite{Serfling:2006a}, \cite{Mosler:2013}.
\vskip0.5mm
\code{scaleCurve(x, y = NULL, alpha = seq(0, 1, 0.01),\\
method = "Projection", name = "X", name_y = "Y", title = "Scale Curve", ...)}
\vskip0.5mm
\textbf{Arguments}\\
\textbf{x}: {Multivariate data in the form of a matrix.}\\
\textbf{y}: {Additional matrix with multivariate data.}\\
\textbf{alpha}: {Vector with values of central area to be used in the computation.}\\
\textbf{name\_x}: {Name of the X matrix used in the legend.}\\
\textbf{name\_y}: {Name of the Y matrix used in the legend.}\\
\textbf{title}: {Title of the plot.}\\
\textbf{depth\_params}: {List of parameters for the depth function.}\\

\code{asymmetryCurve(x, y = NULL, alpha = seq(0, 1, 0.01), movingmedian = FALSE,
  name = "X", name_y = "Y", depth_params = list(method = "Projection"))}
\vskip0.5mm
\textbf{Arguments}\\

\textbf{x}: {The data as a matrix or a data frame. If it is a matrix or a data frame, then each row is viewed as one multivariate observation.}\\
\textbf{y}: {Additional matrix of multivariate data.}\\
\textbf{alpha}: {An ordered vector containing indices of central regions used for asymmetry curve calculation.}\\
\textbf{movingmedian}: {Logical. For default FALSE only one depth median is used to compute asymmetry norm. If TRUE, for every central area, a new depth median will be used and this approach needs much more time.}\\
\textbf{name}: {Name of set X, used in plot legend.}\\
\textbf{name\_y}: {Name of set Y, used in plot legend.}\\
\textbf{depth\_params}: {List of parameters for the depth function.}\\

\begin{figure}
\centering
\begin{minipage}[t]{.45\textwidth}
  \centering
  \includegraphics[width=.95\linewidth]{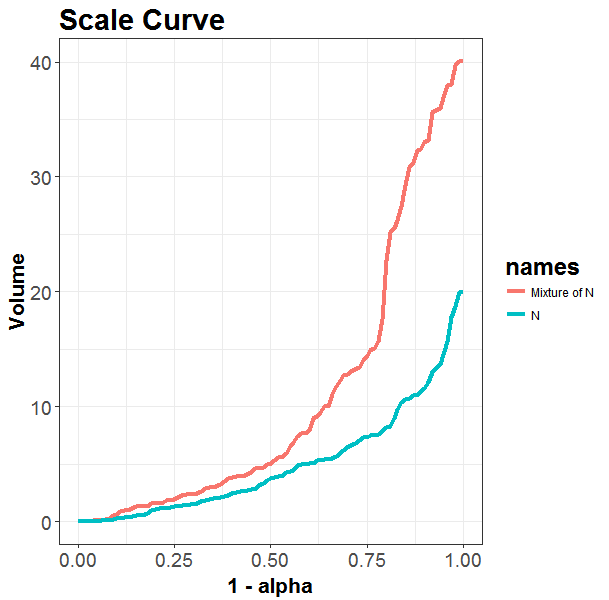}
  \caption{Scale curves.}
  \label{fig34}
\end{minipage}
\begin{minipage}[t]{.45\textwidth}
  \centering
  \includegraphics[width=.95\linewidth]{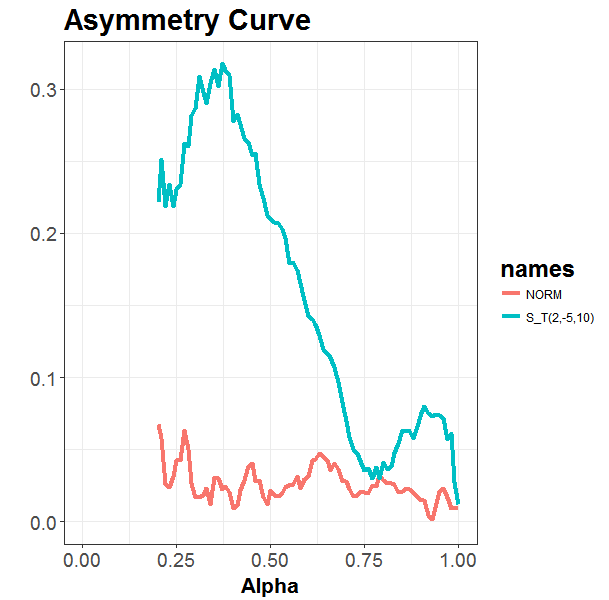}
  \caption{Asymmetry curves.}
  \label{fig35}
\end{minipage}
\end{figure}
 The example 2 shows the comparison of samples obtained by means of scale curves and asymmetry curves\\
 \vskip1mm
 EXAMPLE 2
\begin{Sinput}
R> x  <- mvrnorm(1000, c(0,0), diag(2))
R> s1 <- scaleCurve(x, name = "Curve 1")
R> s2 <- scaleCurve(x*2, x*3, name = "Curve 2", name_y = "Curve 3")
R> w  <- getPlot(combineDepthCurves(s1, s2)) + ggtitle("Plot")
R> w + theme(text = element_text(size = 25))
R> xx <- mvrnorm(1000, c(0,0), diag(2))
R> yy <- mvrnorm(1000, c(0,0), diag(2))
R> p  <- asymmetryCurve(xx, yy)
R> getPlot(p) + ggtitle("Plot")
\end{Sinput}
\vskip0.5mm
\begin{Sinput}
R> xx <- mvrnorm(1000, c(0, 0), diag(2))
R> yy <- mvrnorm(1000, c(0, 0), diag(2))
R> p <- asymmetryCurve(xx, yy)
R> getPlot(p) + ggtitle("Plot")
\end{Sinput}

\subsection{Simple robust regressions}
Within the package, two simple (two dimensional) robust regressions are available: \textbf{the deepest regression},   and \textbf{the least squares regression for projection depth-trimmed sample (TrimReg)} (Figures 10, 49, and 50).\\
\vskip1mm
\proglang{deepReg2d(x, y)}\\
\proglang{trimProjReg2d(x, y, alpha = 0.1)}
\vskip0.5mm
\textbf{Arguments}
\vskip0.5mm
\textbf{x,y}: Data vectors\\
\textbf{alpha}: Trimming parameter\\
 The example 3 shows a comparison of the deepest regression and least squares estimators of the simple regression for a "reference dataset" considered in the context of robust regression \emph{starsCYG} (Rousseeuw and Leroy 1987).
\vskip0.5mm
EXAMPLE 3
\begin{CodeInput}
R> plot(starsCYG, cex=1.4)
R> deepreg  <- deepReg2d(starsCYG$log.Te, starsCYG$log.light)
R> trimreg  <- trimProjReg2d(starsCYG$log.Te, starsCYG$log.light)
R> least.sq <- lm(starsCYG$log.Te ~ starsCYG$log.light)
R> abline(deepreg,  lwd = 3, col = "red")
R> abline(trimreg,  lwd = 3, col = "brown")
R> abline(least.sq, lwd = 3, col = "blue")
\end{CodeInput}

\begin{CodeOutput}
R> # coefficients:
R> deepreg@coef
  # -7.903043  2.913043
R> trimreg@coef
  # -7.403531    2.802837
\end{CodeOutput}
The example 4 shows a comparison of the deepest regression, the least median of squares, and least squares estimators of the simple regression with a least squares regression applied to the projection depth-trimmed data. The data relate to a relationship between a minimal wage (MW) and an unemployment rate (UR) in France in a period 1999-2015.
\vskip1mm
\emph{Interpretation: Simple regression belongs to the most popular tools of economic analysis. Although by regression we generally mean a function expressing the relationship between a conditional mean of one variable and a condition related to other variables, which linearity is very rare in a practice of economic analysis, the simple regression is commonly used as a rough indicator of general direction of the relation. The deepest regression estimator and the least squares estimator for trimmed data by the projection depth seem to indicate the relation taking into account its consistency with a theory of economics the best. We should note, that there does not exist a simple justification application of robust regression in economics. It should be recommended when an influential majority of data is of a prime importance, but pointing out that "influential majority" may posses many meanings.}
\vskip0.5mm
EXAMPLE 4
\begin{Sinput}
R> library("MASS")
R> library("quantreg")
R> library("DepthProc")
R> data("france")
R> attach("france")
R> plot(MW, UR, cex=2)
R> RES1 <- lm(UR ~ MW)
R> abline(RES1, lwd=2, cex=3, col='red')
R> summary(RES1)
R> RES2 <- rlm(UR ~ MW)
R> summary(RES2)
R> abline(RES2, lwd=5, col="blue")
R> deviance(RES2)
R> (RES4 <- lqs(UR ~ MW,method = "lms"))
R> (RES5 <- lqs(UR ~ MW, method = "lts"))
R> abline(RES4, lwd = 2, col = "green")
R> abline(RES5, lwd = 2, col = "pink")
R> lines(lowess(MW, UR, f=0.5, iter = 0), lwd = 2)
R> RES6 <- trimProjReg2d(MW, UR)
R> abline(RES6, lwd = 3, col = "darkgreen")
\end{Sinput}

\subsection{Weighted estimators of location and scatter}
Using the depth function, one can define a depth-weighted multivariate location and scatter estimators possessing high breakdown points that are computationally tractable (\cite{Zuo:2005}). In the case of location, the estimator is defined as
\begin{equation}
  L(F)={\int{{x}{{w}_{1}}(D({x},F))dF({x})}}/\int{{{w}_{1}}(D({x},F))dF({x})},
\end{equation}
Subsequently, a depth-weighted scatter estimator is defined as
\begin{equation}
S(F)=\frac{\int{({x}-L(F)){{({x}-L(F))}^{\top}}{{w}_{2}}(D({x},F))dF({x})}}{\int{{{w}_{2}}(D({x},F))dF({x})}},
\end{equation}
where ${{w}_{2}}(\cdot )$ is a suitable weight function that can be different from ${{w}_{1}}(\cdot )$.
\vskip0.5mm
The \pkg{DepthProc} package offers these estimators in the case of computationally feasible weighted ${L}^{p}$ depth. Note that $L(\cdot )$ and $S(\cdot )$ include multivariate versions of trimmed means and covariance matrices. Sample counterparts of (28) and (29) take the forms
\begin{equation}
{{T}_{WD}}({{{X}}^{n}})={\sum\limits_{i=1}^{n}{{w({d}_{i})}{{X}_{i}}}}/{\sum\limits_{i=1}^{n}{{w({d}_{i})}}} ,
\end{equation}
\begin{equation}
DIS({{{X}}^{n}})=\frac{\sum\limits_{i=1}^{n}{{w({d}_{i})}\left(
{{{X}}_{i}}-{{T}_{WD}}({{{X}}^{n}}) \right){{\left( {{{X}}_{i}}-{{T}_{WD}}({{{X}}^{n}})
\right)}^{T}}}}{\sum\limits_{i=1}^{n}{{w({d}_{i})}}},
\end{equation}
where ${{d}_{i}}$ are sample depth weights,  ${{w}_{1}}(x)={{w}_{2}}(x)=a\cdot x +b$, $a, b \in \mathbb{R}$. \\

Computational complexity of the scatter estimator crucially depend on the complexity of the depth used.
For the weighted ${L}^{p}$ depth, we have $O({{d}^{2}}n+{{n}^{2}}d)$ complexity and a good perspective its distributed calculation \cite{Zuo:2004}, \cite{Zelias:2014}. These facts are of prime importance, for example, in context of a robust on line portfolio optimization.
\vskip0.5mm
\code{CovLP(x, pdim = 2, la = 1, lb = 1)}\\
 The example 5 shows an application of the weighted by the $L^p$ depth multivariate location and the scatter estimator for a sample drawn from the mixture of two normal distributions.
 \vskip1mm
EXAMPLE 5
\begin{CodeInput}
R> require("MASS")
R> Sigma1 <- matrix(c(10, 3, 3, 2), 2, 2)
R> X1 <- mvrnorm(n = 8500, mu = c(0, 0), Sigma1)
R> Sigma2 <- matrix(c(10, 0, 0, 2), 2, 2)
R> X2 <- mvrnorm(n = 1500, mu = c(-10, 6), Sigma2)
R> BALLOT <- rbind(X1,X2)
R> train  <- sample(1:10000, 500)
R> data   <- BALLOT[train, ]
R> cov_x  <- CovLP(data, 1, 1, 1)
R> cov_x
\end{CodeInput}
\begin{CodeOutput}
Call:
Method:  Depth Weighted Estimator
Robust Estimate of Location:
[1]  -1.6980   0.8844
Robust Estimate of Covariance:
        [,1]      [,2]
[1,]  15.249  -2.352
[2,]  -2.352   4.863
\end{CodeOutput}
\subsection{Depth-induced binning}
Let us recall that binning is a popular method, which allows for faster computation by reducing the continuous sample space to a discrete grid (\cite{Hall:1996}). It is useful, for example, in the case of a predictive distribution estimation by means of kernel methods.  To bin a window of $n$ points ${W}_{i,n}=\left\{{X}_{i-n+1},...,{X}_{i} \right\}$ to a grid ${X}'_{1},...,{X}'_{m},$ we simply assign each sample point ${X}_{i}$ to the nearest grid point ${X}'_{j}$. When binning is completed, each grid point ${X}'_{j}$ has an associated number ${c}_{i}$, which is the sum of all the points that have been assigned to ${X}'_{j}$. This procedure replaces the data ${W}_{i,n}=\left\{ {X}_{i-n+1},...,{X}_{i} \right\}$ with the smaller set ${W}'_{j,m}=\left\{ {X}'_{j-m+1},...,{X}'_{j} \right\}$. Although simple binning can speed up the computation, it is criticized for the lack of a precise control over the accuracy of the approximation. Robust binning, however, stresses properties of the majority of the data and decreases the computational complexity of the data stream analysis at the same time (\cite{Operational}).\\
For a 1D window ${W}_{i,n}$, let ${Z}_{i,n-k}$ denote a 2D window created basing on ${W}_{i,n}$ and it consisted of $n-k$ pairs of observations and the $k$ lagged observations ${Z}_{i,n-k}$=$\left\{ ({X}_{i-n-k},{X}_{i-n+1})\right\}$, $1\le i\le n-k.$ Robust 2D binning of the ${Z}_{i,n-p}$ is a very useful technique in the context of robust estimation of a predictive distribution of a time series (\cite{Operational}) or robust monitoring of a data stream (\cite{JMS}).

Assume we analyze a data stream $\{{X}_{t}\}$ using a moving window of a fixed length $n$, i.e., ${W}_{i,n}$ and the derivative window ${Z}_{i,n-1}$. In the first step, we calculate the weighted sample $L^p$ depth for ${W}_{i,n}$. Next, we choose equally spaced grid of points ${l}_{1},...,{l}_{m}$; in this way of $[{{l}_{1}},{{l}_{m}}]\times [{{l}_{1}},{{l}_{m}}]$ covers a fraction of the  $\beta$ central points of ${Z}_{i,n-1}$ with respect to the calculated $L^p$ depth,it covers ${R}^{\beta }({Z}_{i,n-1})$ for certain prefixed threshold $\beta \in (0,1)$. For both ${X}_{t}$ and ${X}_{t-1},$ we perform a simple binning using following bins: $(-\infty ,{l}_{1})$, $({l}_{1},{l}_{2})$,..., $({l}_{m},\infty )$.

For robust binning, we reject "border" classes and further use only midpoints and binned frequencies for classes $({l}_{1},{l}_{2})$, $({l}_{2},{l}_{3})$,...,  $({l}_{m-1},{l}_{m})$.

Figures 36--37 present the idea of a simple $L^{p}$ binning in the case of data generated from a mixture of two two-dimensional normal distributions. The midpoints are represented by triangles.\\
\begin{figure}
\centering
\begin{minipage}{.47\textwidth}
  \centering
  \includegraphics[width=.99\linewidth]{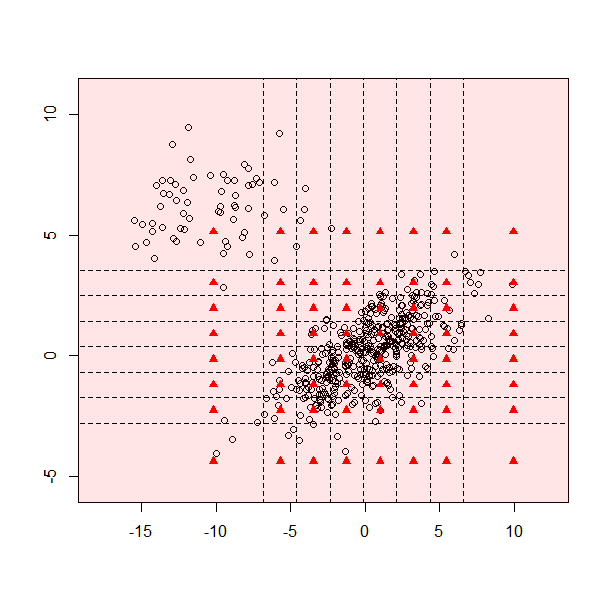}
  \caption{The first step in $L^p$ depth binning.}
  \label{fig36}
\end{minipage}%
\mbox{\hspace{0.1cm}} 
\begin{minipage}{.47\textwidth}
  \centering
  \includegraphics[width=.99\linewidth]{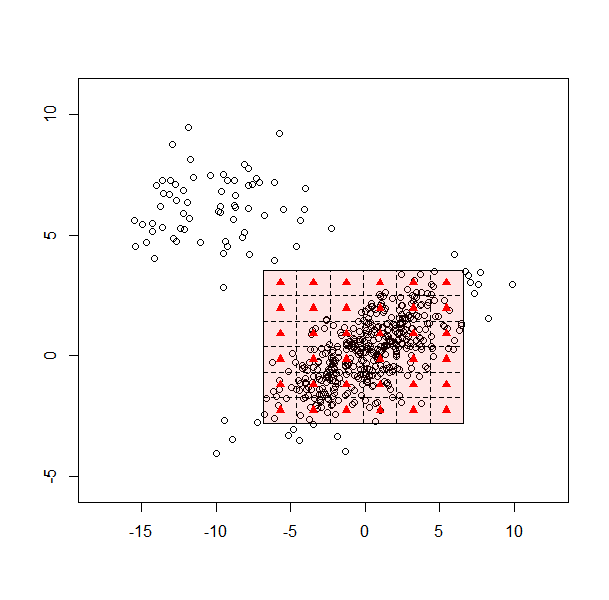}
  \caption{The second step in $L^p$ depth binning.}
  \label{fig37}
\end{minipage}
\end{figure}
The example 6 shows a general idea of the $L^p$ binning.
\vskip 0.5mm
EXAMPLE 6
\begin{Sinput}
R> require("MASS")
R> Sigma1 <- matrix(c(10, 3, 3, 2), 2, 2)
R> X1     <- mvrnorm(n = 8500, mu= c(0, 0), Sigma1)
R> Sigma2 <- matrix(c(10,0,0,2), 2, 2)
R> X2     <- mvrnorm(n = 1500, mu = c(-10, 6), Sigma2)
R> BALLOT <- rbind(X1, X2)
R> train  <- sample(1:10000, 500)
R> data   <- BALLOT[train, ]
R> plot(data)
R>
R> b1 <- binningDepth2D(data, remove_borders = FALSE, nbins = 12, k = 1)
R> b2 <- binningDepth2D(data, nbins = 12, k = 1, remove_borders = TRUE)
R> plot(b1)
R> plot(b2)
\end{Sinput}
 The example 7 shows an application of the $L^p$ binning for data concerning an evaluation of the Fourth Millennium Development Goal of The United Nations.
\vskip0.5mm
EXAMPLE 7
\begin{Sinput}
R> data("under5.mort")
R> data("maesles.imm")
R> data2011 <- cbind(under5.mort[, 22], maesles.imm[, 22])
R> plot(binningDepth2D(data2011, nbins = 8, k = 0.5,
+    remove_borders = TRUE ))
\end{Sinput}
\section{Empirical applications of functional depths}
 The example 8 shows the application of functional boxplots (\cite{Ramsay:2009} for a definition of the functional boxplot) of data  on airpollution with suspended dangerous particles in air dust of PM10 on a day and night in cities of Katowice and Cracow in Poland in the selected periods of 2016 and 2017. These boxplots may be useful in designing a smog alert system and optimal proecological regional taxation system.
Examples 9--13 present an evaluation of typical behavior of Internet users obtained by means of functional boxplots. Examples 13--15 are dedicated for robust prediction of a hierarchical functional time series using the local moving median (\cite{Hierarchical} and \cite{Shang:hierarchic} for an alternative generalized least squares predictor, which is not robust). Note that the functional time series methodology enables for the successful prediction of unequally spaced economic time series (\cite{JMS}). Example 16 shows an application of the local Wilcoxon statistic for the detection of a structural change in a functional time series \cite{Structural}. The test may be used to detect atypical behavior of the Internet users in a scale of day and night, week, month as well as the change in the shape of an yield curve of a country, which may signal an approaching crisis.
\vskip0.5mm
EXAMPLE 8 "Air pollution in Katowice and Cracow in Poland in 2016 and 2017"
\begin{Sinput}
R> data("katowice.airpollution")
R> katowice.raw <- as.matrix(katowice.airpollution)
R> matplot(t(katowice.raw), type = "l",
+   col  = terrain.colors(181), main = 'KATOWICE',
+   xlab = 'hour', xlim = c(0, 24), ylab = 'pollution')
R> w1 <- fncBoxPlot(katowice.airpollution,
+    bands = c(0, 0.05, 0.10, 0.5, 0.90, 0.95), method = "MBD")
R> print(w1 + ggtitle("Air pollution in Katowice 2016 - 2017") +
+    labs(y = "pollution ", x = "hour "))
R> data("cracow.airpollution")
R> cracow.pm10 <- matrix(cracow.airpollution[,"PM10"], ncol = 24, byrow = TRUE)
R> w1 <- fncBoxPlot(cracow.pm10,
+    bands = c(0, 0.05, 0.10, 0.5, 0.90, 0.95), method = "MBD")
R> print(w1 + ggtitle("Air pollution in December 2016") +
+    labs(y = "pollution ", x = "hour "))
\end{Sinput}
\vskip1mm
\emph{Interpretation:} By preparing boxplots for trajectories of day and night air pollution, one may obtain an insight into the problem, and learn how it is typical and what is anomalous. This insight may be effectively used in the process of creation of local ecological policy involving traffic and tax regulations. Depth-induced methods may also effectively be used in robust pollution forecasting (\cite{CEJEME:2018})\\
\vskip0.5mm
Examples 9--12 show elements of analysis of dataset \emph{internet.users}, containing 1728 working days of two Internet services considered on the basis of number of unique users (users) and the number of page views (views). One-dimensional time series are divided into 24-hour periods for obtaining functional time series (\cite{Structural}).
\vskip0.5mm
EXAMPLE 9
\begin{Sinput}
R> data("internet.users")
R> users<-internet.users[1:17280,5]
R> views<-internet.users[1:17280,6]
R> library("zoo")
R> window <- function(x) { x }
R> users.m <- rollapply(users, width = 24, by = 24, window,
+                  by.column = FALSE)
R> views.m <- rollapply(views, width = 24, by = 24, window,
+                  by.column = FALSE)
R> depths_1 <- depthLocal(users.m, beta=0.45, depth_params1 = list(method = "MBD"))
R> depths_2 <- depthLocal(views.m, beta=0.45, depth_params1 = list(method = "MBD"))
R> par(mfrow = c(1, 2))
R> plot(depths_1, xlab = "hour", ylab = "users", main = "local depth, beta = 0.45")
R> plot(depths_2, xlab = "hour", ylab = "views", main = "local depth, beta = 0.45")
\end{Sinput}
\begin{figure}
\centering
  \includegraphics[width=.99\linewidth]{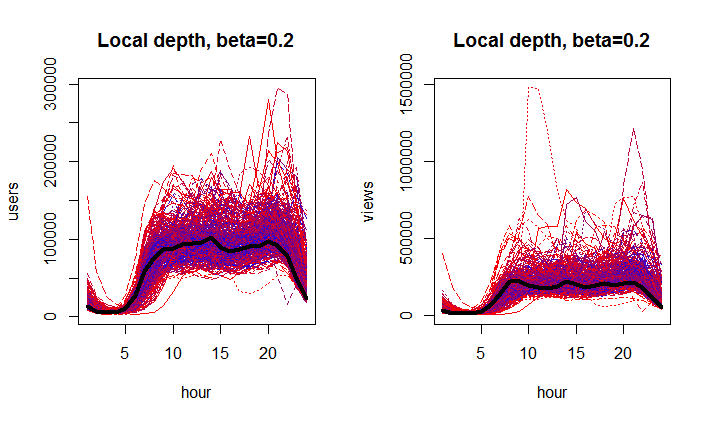}
  \caption{Local functional depth for Internet users data.}
  \label{fig38}
\end{figure}
EXAMPLE 10
\begin{Sinput}
R> data("internet.users")
R> ind_1 <- which(internet.users[, 1] == 1)
R> DATA_1 <- internet.users[ind_1, ] # the first Internet service
R> ind_2  <- which(internet.users[, 1] == 2)
R> DATA_2 <- internet.users[ind_2, ] # the second Internet service
R> users_1 <- DATA_1[1:8759, 5]
R> # the number of unique users in the service 1
R> views_1 <- DATA_1[1:8759, 6]
R> # the number of page views in the service 1
R> users_2 <- DATA_2[1:8759, 5]
R> # the number of unique users in the service 2
R> views_2 <- DATA_2[1:8759, 6]
R> # the number of page views in the service 2
\end{Sinput}
\vskip0.5mm
\begin{Sinput}
R> library("zoo")
R> window<-function(x){x}
R> users.m.1 <- rollapply(users_1, width=24, by=24, window, by.column=FALSE)
R> views.m.1 <- rollapply(views_1, width=24, by=24, window, by.column=FALSE)
R> users.m.2 <- rollapply(users_2, width=24, by=24, window, by.column=FALSE)
R> views.m.2 <- rollapply(views_2, width=24, by=24, window, by.column=FALSE)
\end{Sinput}
\vskip0.5mm
EXAMPLE 11: "Functional boxplots"
\begin{Sinput}
R> fncBoxPlot(users.m.1, bands = c(0, 0.05, 0.5, 0.95,1), method = "MBD")
R> fncBoxPlot(users.m.1, bands = c(0, 0.05, 0.5, 0.95,1), method = "FM")
R> fncBoxPlot(views.m.1, bands = c(0, 0.05, 0.5, 0.95,1), method = "MBD")
R> fncBoxPlot(views.m.1, bands = c(0, 0.05, 0.5, 0.95,1), method = "FM")
R> fncBoxPlot(users.m.2, bands = c(0, 0.05, 0.5, 0.95,1), method = "MBD")
R> fncBoxPlot(users.m.2, bands = c(0, 0.05, 0.5, 0.95,1), method = "FM")
R> fncBoxPlot(views.m.2, bands = c(0, 0.05, 0.5, 0.95,1), method = "MBD")
R> fncBoxPlot(views.m.2, bands = c(0, 0.05, 0.5, 0.95,1), method = "FM")
\end{Sinput}
\emph{Interpretation:} Functional boxplots may be interpreted analogous to their one dimensional counterparts. Inside the box one may find more and more central trajectories with the most central trajectory---the functional median of a day and night behaviors of the service users. Using the locality parameter $\beta$ one may take into account the multi modality of data (differences between a holiday trajectory and a working day trajectory) or choose "a resolution" at which the phenomenon is observed. A significant (relative to a size of the box) departure from the central trajectory should incline to the service administrator for further investigations.
\vskip0.5mm
EXAMPLE 12: "Comparison of two services with respect to number of unique users and page views"
\begin{Sinput}
R> ddPlot(x = users.m.1, y = users.m.2, depth_params = list(method = "Local",
+     beta = 0.45, depth_params1 = list(method = "MBD")))
R> ddPlot(x = views.m.1, y = users.m.2, depth_params = list(method = "Local",
+     beta = 0.45, depth_params1 = list(method = "MBD")))
R> ddPlot(x = views.m.1, y = users.m.2, depth_params = list(method = "Local",
+     beta = 0.25, depth_params1 = list(method = "MBD")))
R> par(mfrow=c(1, 1))
\end{Sinput}
\vskip0.5mm
\emph{Interpretation: Shapes of patterns on vs. depth plots indicate the differences in underlying distributions. Activities of Internet users differ between services. Further investigation of the issue may lead to an effective valuation of advertisement time and the space of the service and allocation of communicates between the services.}
\vskip0.5mm
EXAMPLE 13: "Functional time series prediction using a moving functional median"
\begin{Sinput}
R> wrapMBD = function(x) {
R>  depthMedian(x, depth_parms = list(method = "Local",
R>        beta = 0.45, depth_params1 = list(method = "MBD")))
R> }
R>
R> SV  <- function(n, gamma, phi, sigma, delta) {
R>   epsilon <- rnorm(n)
R>   eta <- rnorm(2 * n, 0, delta)
R>   h <- rnorm(1)
R>   for(t in 2:(2 * n)) {
R>     h[t] <- exp(gamma + phi * (h[t - 1] - gamma) + sigma * eta[t])
R>   }
R>   Z <- sqrt(tail(h, n)) * epsilon
R>   return(Z)
R> }
R> example <- SV(100, 0, 0.2, 0.5, 0.1)
R> plot(ts(example))
\end{Sinput}
\vskip0.5mm
\emph{Interpretation:} The function \code{SV} is designed for generating trajectories from a popular stochastic volatility model in econometrics. The function \code{wrapMBD} is designed for calculating a functional median specified by a list of parameters
\vskip0.5mm
EXAMPLE 14: "Simple functional time series simulator"
\begin{Sinput}
R> m.data1 <- function(n,a,b) {
R>   M <- matrix(nrow = n, ncol = 120)
R>   for(i in 1:n) M[i,]<- a*SV(120, 0, 0.3, 0.5, 0.1) + b
R>   M
R> }
R>
R> m.data.out1 <- function(eps, m, n, a, b, c, d) {
R>   H <- rbind(m.data1(m, a, b), m.data1(n, c, d))
R>   ind <- sample((m+n),eps)
R>   H1  <- H[ind,]
R>   H1
R> }
\end{Sinput}
\vskip0.5mm
\emph{Interpretation: Basing on the function \code{SV}, two simple functions designed for generating functional time series are designed. The function \code{m.data1} generates a matrix of dimension $n\times 120$, consisting of $n$ free from outliers trajectories observed in 120 time points, whereas the function \code{m.data.out.1} produces a matrix of functional observations, which consists of a $(m+n)/eps$ fraction of outliers considered in a certain simple but rather naive way on the FDA ground (\cite{Nicolas}).}
\vskip0.5mm
EXAMPLE 15: "Simple R script, the example showing how to calculate base forecasts for three
hierarchical FTS levels using a moving functional median implemented within the \pkg{DepthProc}."
\begin{Sinput}
R> require("DepthProc")
R> require("RColorBrewer")
R> require("zoo")
R> m <- matrix(c(1, 0, 1, 3, 2, 3, 2, 0), nrow = 2, ncol = 4)
R> m[2,] <- c(2,2,3,3)
R> m[1,] <- c(0,1,1,0)
R> M2A <- m.data.out1(150, 3000, 7000, 5, 0, 1, 25)
R> M2B <- m.data.out1(150, 3000, 7000, 2, 0, 1, 15)
R> M2C <- m.data.out1(150, 3000, 7000, 3, 0, 1, 10)
R> matplot(t(M2A), type="l", col = topo.colors(151), xlab = "time", main = "FTS")
R> matplot(t(M2B), type="l", col = topo.colors(151), xlab = "time", main = "FTS ")
R> matplot(t(M2C), type="l", col = topo.colors(151), xlab = "time", main = "FTS")
\end{Sinput}
Below are the moving local medians applied to the above series, window lengths = 15 obs.,
locality parameters beta = 0.45
\begin{Sinput}
R> result4A = rollapply(t(M2A),width = 15, wrapMBD, by.column = FALSE)
R> result4B = rollapply(t(M2B),width = 15,wrapMBD, by.column = FALSE)
R> result4C = rollapply(t(M2C),width = 15, wrapMBD, by.column = FALSE)
R> matplot(result4A, type="l", col = topo.colors(87), xlab = "time",
+          main="15-obs mov. med.")
R> matplot(result4B, type="l", col = topo.colors(87), xlab = "time",
+          main="15-obs mov. med.")
R> matplot(result4C, type="l", col = topo.colors(87), xlab = "time",
+          main="15-obs mov. med.")
\end{Sinput}
\vskip0.5mm
\emph{Interpretation: Moving functional median may effectively be used in the forecasting of phenomena exhibiting a natural hierarchy, i.e., e.g., a turnover of a company with regard to product lines and/or client target groups (see \cite{Shang:hierarchic}, \cite{CEJEME:2018}).}\\
\vskip0.5mm
EXAMPLE 16: "Global and local Wilcoxon test"
\begin{Sinput}
R> Md1 = m.data1(100, 1, 2)
R> Md2 = m.data1(100, 1, 7)
R> mWilcoxonTest (t(Md1), t(Md2), depth_params = list(method = "MBD"))
R> mWilcoxonTest(t(Md1), t(Md2),
+    depth_params  = list(method="Local", beta=0.25,
+    depth_params1 = list(method = "MBD")))
\end{Sinput}
\vskip0.5mm
\emph{Interpretation: nonparametric Wilcoxon test for functional data may be a reasonable alternative for statistical procedures assuming the normality of data generating processes or resampling procedures, especially in the functional case, where straightforward generalizations of one dimensional statistical techniques does not exist (\cite{Nicolas} and references therein). The $\beta$ parameter may be interpreted in terms of a resolution at which populations are compared.}
\vskip1mm
EXAMPLE 17: "Procedure of a structural change detection in a functional time series using moving Wilcoxon statistic"
\begin{Sinput}
R> movwilcox <- function(x) {
R>   res <- mWilcoxonTest(x, t(ref), depth_params = list(method = "Local",
+                        beta = 0.25, depth_params1 = list(method = "MBD")))
R>   as.numeric(res[1])
R> }
R> ref <- m.data1(50,1,2)
R> trajectory <- function(n, m) {
R>   ref <- m.data1(50, 1, 2)
R>   # ref is a reference sample, here a 120 x 50 matrix,
R>   # MAA is a data frame, to which we apply a moving window.
R>   # The MAA has dimension 120 x (m+n).
R>   M_1A <- m.data1(n, 1, 2)
R>   M_1B <- m.data1(m, 1, 7)
R>   MAA <- cbind(M_1A, M_1B)
R>   results <- c()
R>   for(i in 50:150) { results[i] <- movwilcox(MAA[, i:(i + 49)]) }
R>   na.omit(results)
R> }
R> example <- trajectory(100,100)
R> plot(example)
\end{Sinput}
\emph{Interpretation: The functional median calculated from a moving window may effectively be used for nonparametric monitoring of economic systems in the context of detecting a regime, a structural change, which may signal a need of adjusting of portfolio and the marketing or investment strategy to new conditions (\cite{Structural}, \cite{Hierarchical}).}

\section{The package architecture}

\subsection{Nomenclature conventions}

There is no agreed naming convention within the \proglang{R} project. In our package we use the following coding style:

\begin{itemize}
  \item \emph{Class} names start with an uppercase letter (e.g., \proglang{DepthCurve}).
  \item For \emph{methods} and \emph{functions} we use lower camel case convention (e.g., \proglang{depthTukey})
  \item All functions related to the location-scale depth starts with an 'lsd' prefix (e.g., \proglang{lsdSampleDepthContours}).
  \item Sometimes we depart from these rules, whenever, to preserve the compatibility with other packages (e.g., \proglang{CovLP}-- t is a function from \pkg{DepthProc} that follows \pkg{rrcov} naming convention).
\end{itemize}

\subsection{Dependencies}

Algorithms for depth functions are written in \proglang{C++}, and they are entirely independent of \proglang{R}. This approach brings more flexibility for the users because she is not locked in R, and can easily use our implementations with any system with \proglang{C++} bindings.  However, to take advantage of the full potential of the depth functions R package should be employed, because it contains a comprehensive set of the graphical procedures.

For matrix operations  \proglang{C++}, we use \pkg{Armadillo Linear Algebra Library} of \cite{Armadillo} and \pkg{OpenMP} library of \cite{openmp2013} for the parallel computing. The communication between \proglang{R} and \proglang{C++} is performed by \pkg{RcppArmadillo} package of \cite{RcppArmadillo}.
For plotting, we use \pkg{base} \proglang{R} graphic (contours plots), \pkg{lattice} package of \cite{lattice} (perspective plot), and \pkg{ggplot2} of \cite{ggplot2} (other plots). We also use functions from \pkg{rrcov} of \cite{rrcov}, \pkg{np} of \cite{np}, \pkg{geometry} of \cite{geometry}.

\subsection{Parallel computing}

By default, the \pkg{DepthProc} uses multithreading and tries to utilize all the available processors. User can control this behavior with \emph{threads} parameter:

EXAMPLE 18: Tested on Intel(R) Core(TM) i5-2500K CPU @ 3.30GHz

\begin{Sinput}
R> x = matrix(rnorm(200000), ncol = 5)
R> system.time(depth(x))
\end{Sinput}
\begin{Soutput}
  user system elapsed
  1.484 0.060 0.420
\end{Soutput}

EXAMPLE 19: Only one thread (approximately 3 times slower)
\begin{Sinput}
R> system.time(depth(x, threads = 1))
\end{Sinput}
\begin{Soutput}
  user system elapsed
  1.368 0.000 1.371
\end{Soutput}
EXAMPLE 20: Any value <1 means "use all possible cores"
\begin{Sinput}
R> system.time(depth(x, threads = -10))
\end{Sinput}
\begin{Soutput}
  user system elapsed
  1.472 0.076 0.416
\end{Soutput}
\subsection{Classes}

Below, we describe only the \code{Depth}, \code{DepthCurve}, and \code{DDPlot} classes in detail, because only they have nonstandard behavior. Other classes are very simple.

\code{CovDepthWeighted} is a class for \code{CovLP} function. It inherits the behavior from \code{CovRobust} class from \pkg{rrcov} package. Description of this class can be found in \cite{rrcov}.

\subsection{UML diagrams and classes}

In this paper, we have exploited the UML class diagrams to describe the behavior of the main \pkg{DepthProc} structures. The UML abbreviation stands for \emph{Unified Modeling Language}, a system of notation for describing object-oriented programs.

In the UML, class is denoted by a box with three compartments, which contain the name, the attributes (slots), and operations (methods) of the class. Each attribute is followed by its type, and each method by its return value. Inheritance relation between the classes are depicted by arrowheads pointing to the base class.

\subsection{Depth class}
\begin{figure}
\centering
\includegraphics[width=0.5\textwidth]{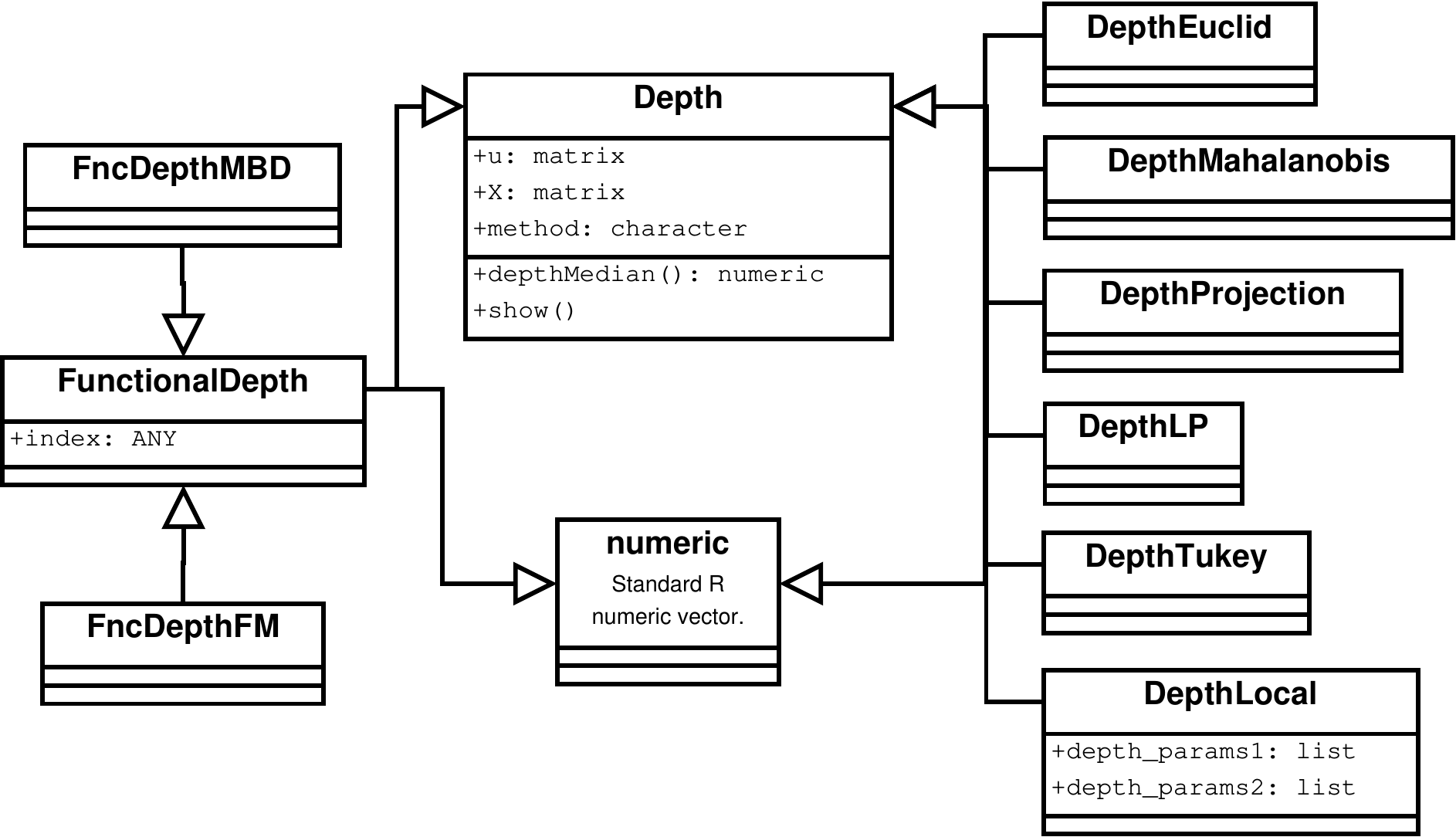}
\caption{Object structure for classes related to depth functions.}
\label{fig27}
\end{figure}

Figure 39 shows an object structure for classes related to depth functions. Each depth class inherits \emph{Depth} and standard \emph{Numeric}. Through inheritance after Numeric these classes are treated as a standard vector, and one can use them with all functions that are appropriate for vectors (e.g., max, min). Depth class is mainly used in internal package operations, but it can also be used for extracting a depth median without recomputing the depth values. This mechanism is shown in the following example:
\vskip0.5mm
EXAMPLE 21: Function for numeric vector
\begin{Sinput}
R> x = matrix(rnorm(1e5), ncol = 2)
R> dep = depth(x)
R> max(dep)
\end{Sinput}
\vskip0.5mm
\begin{Soutput}
  [1] 0.9860889
\end{Soutput}
EXAMPLE 22: Function for raw matrix, all depths must be recomputed
\begin{Sinput}
R> system.time(dx <- depthMedian(x))
\end{Sinput}
\begin{Soutput}
  user  system elapsed
  1.609   0.072   0.451
\end{Soutput}
EXAMPLE 23: Function for depth class, result is immediate
\begin{Sinput}
R> system.time(dm <- depthMedian(dep))
\end{Sinput}
\begin{Soutput}
  user  system elapsed
  0.000   0.000   0.001
\end{Soutput}
\begin{Sinput}
R> # In order to check the equality
R> all.equal(dm, dx)
\end{Sinput}
\begin{Soutput}
  [1] TRUE
\end{Soutput}

\subsection{DepthCurve and DDplot classes}

The \code{DepthCurve} is a main class for storing results from \code{scaleCurve} and the \code{asymmetryCurve} functions, describing their behavior (Figure 28). The \code{DDPlot} stores results from the \code{ddPlot} and \code{ddMvrnorm} functions.

Both classes, \code{DepthCurve} and \code{DDPlot} can be converted into \pkg{ggplot} object for further appearance modifications via \code{getPlot()} function.\\
\vskip0.5mm
EXAMPLE 24:
\begin{Sinput}
R> x <- matrix(rnorm(1e2), ncol = 2)
R> y <- matrix(rnorm(1e2), ncol = 2)
R> ddplot <-  ddPlot(x,y)
R> p <- getPlot(ddplot)
R> # In order to modify a title
R> p + ggtitle("X vs Y")
R> scplot <- scaleCurve(x,y)
R> p <- getPlot(scplot)
R> # In order to change a color palette
R> p + scale_color_brewer(palette = "Set1")
\end{Sinput}

Figure 40 shows class structure for \code{DepthCurve}. Class \code{ScaleCurveList} is a container for storing multiple curves for charting them on one plot. It inherits the behavior from a standard \proglang{R} list, but can also be also converted into \pkg{ggplot} object with \code{getPlot} method.

We have introduced $combineDepthCurves$ operator for combining \code{DepthCurves} into \code{DepthCurveList}. This operator is presented in the following example:\\
\vskip0.5mm
EXAMPLE 25
\begin{Sinput}
R> data("under5.mort")
R> data("maesles.imm")
R> data2011 <- cbind(under5.mort[,"2011"],maesles.imm[,"2011"])
R> data2000 <- cbind(under5.mort[,"2000"],maesles.imm[,"2000"])
R> data1995 <- cbind(under5.mort[,"1995"],maesles.imm[,"1995"])
R> sc2011 <- scaleCurve(data2011, name = "2011")
R> sc2000 <- scaleCurve(data2000, name = "2000")
R> # In order to create ScaleCurveList
R> sclist <- combineDepthCurves(sc2000,sc2011)
R> sclist
R> # In order to add another Curve
R> sc1995 <- scaleCurve(data1995, name = "1995")
R> combineDepthCurves(sclist, sc1995)
\end{Sinput}
\vskip0.5mm
\emph{Interpretation: We compare dispersions of countries regarding the infant mortality rate and children (one-year old) immunized against measles in the period of 1995--2011. Curves representing years closer and closer to 2011 are placed lower and lower. One can therefore conclude that the dispersion (differences) between countries decreased in this period.}
\vskip0.5mm
\begin{figure}
\centering

\includegraphics[width=0.8\textwidth]{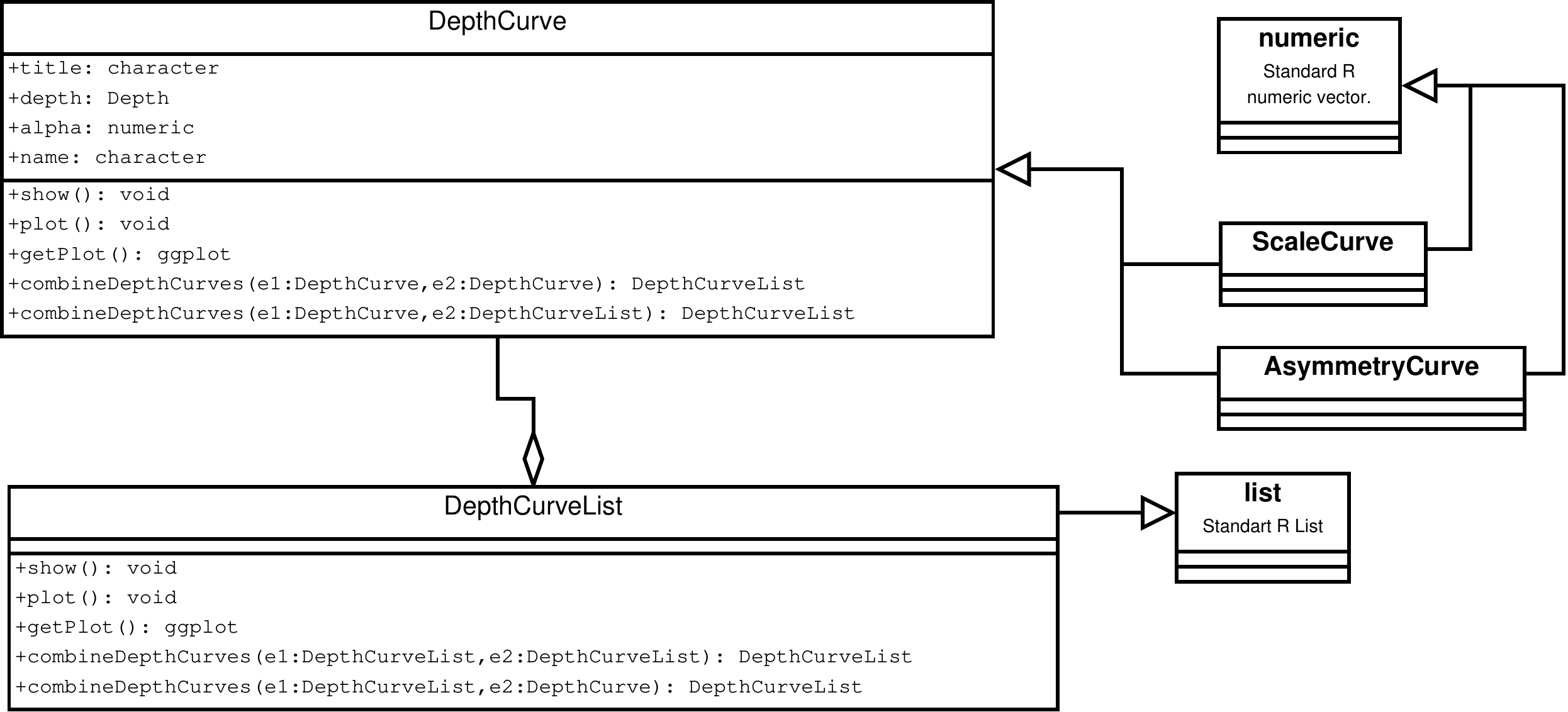}
\caption{Class structure for \code{DepthCurve}.}
\label{fig28}
\end{figure}
EXAMPLE 26
\begin{Sinput}
R> n <- 200
R> mat_list <- replicate(n, matrix(rnorm(200), ncol = 2), simplify = FALSE)
R> scurves <- lapply(mat_list, scaleCurve)
R> scurves <- Reduce(combineDepthCurves,scurves)
R> p <- getPlot(scurves)
R> p + theme(legend.position="none") +
+   scale_color_manual(values = rep("black",n))
\end{Sinput}

\section{Empirical research using the package}
For illustrating the usefulness of the \pkg{DepthProc} package in socioeconomic researches, let us consider an issue of nonparametric evaluation of the \emph{Fourth Millennium Development Goal} of The United Nations (4MG). The main aim of the goal was reducing the under-five-months child mortality by two-thirds, between 1990--2015. Using some selected multivariate techniques that are available within our \pkg{DepthProc} package, we answer \textbf{a question, "if during the period of 1990--2015, differences between the developed and the developing countries have really decreased?"}.

In the study, we jointly considered following variables:
\begin{itemize}
\item{\textbf{Infant mortality rate (under five months) per 1000 live births ($Y_1$)}}
\item{\textbf{Infant mortality rate (0--1 year) per 1000 live births ($Y_2$)}}
\item{\textbf{Children (one-year old) immunized against measles, percentage ($Y_3$)}}
\end{itemize}
Datasets were obtained from \url{http://mdgs.un.org/unsd/mdg/Data.aspx} and are made available within the package.
\begin{figure}
\centering
\begin{minipage}{.45\textwidth}
  \centering
  \includegraphics[width=.95\linewidth]{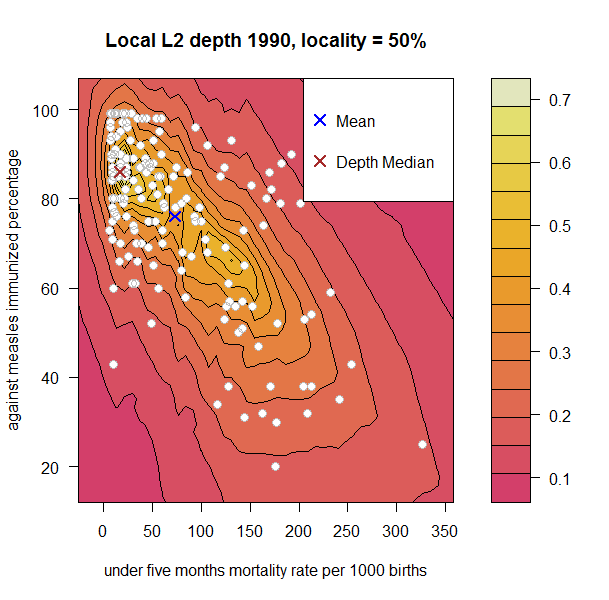}
  \caption{1990---$L^2$ sample depth contour plot $Y_1$ vs. $Y_3$}
    \label{fig41}
\end{minipage}%
\mbox{\hspace{0.1cm}} 
\begin{minipage}{.45\textwidth}
  \centering
  \includegraphics[width=.95\linewidth]{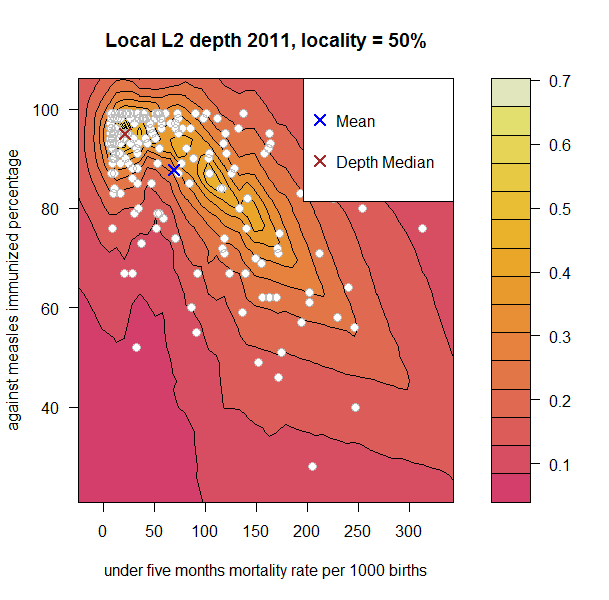}
        \caption{2011---$L^2$ sample depth contour plot $Y_1$ vs. $Y_3$}
        \label{fig42}
\end{minipage}
\begin{minipage}{.45\textwidth}
  \centering
  \includegraphics[width=.95\linewidth]{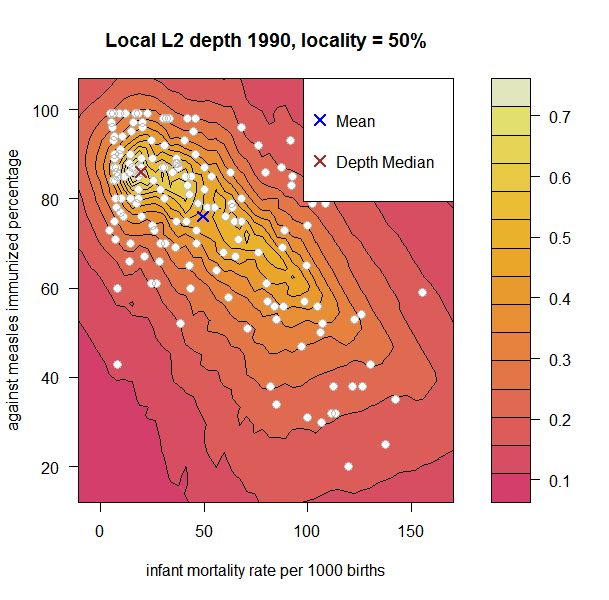}
  \caption{1990---$L^2$ sample depth contour plot $Y_2$ vs. $Y_3$}
    \label{fig43}
\end{minipage}%
\mbox{\hspace{0.1cm}} 
\begin{minipage}{.45\textwidth}
  \centering
  \includegraphics[width=.95\linewidth]{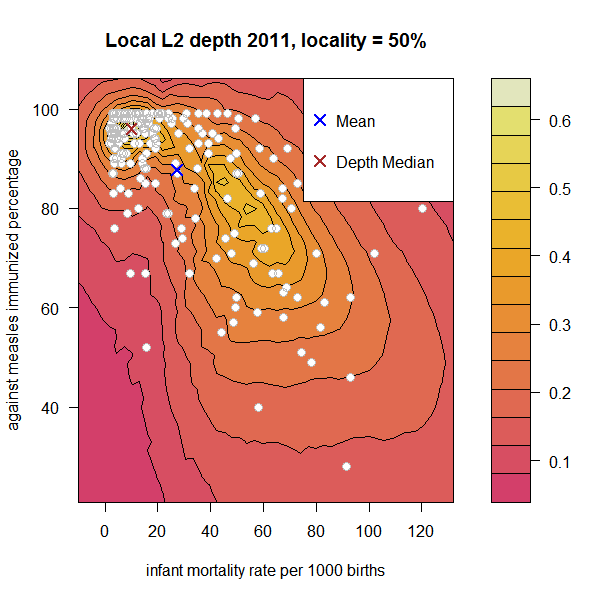}
      \caption{2011---$L^2$ sample depth contour plot $Y_2$ vs. $Y_3$}
      \label{fig44}
   \end{minipage}
   \end{figure}
Figure 41 shows weighted $L^2$ depth contour with locality parameter $\beta=0.5$ for countries in 1990 considered with regard to variables $Y_1$ and $Y_3$, whereas Figure 42 presents the same issue but in 2011. Figure 43 shows the weighted $L^2$ depth contour with locality parameter $\beta=0.5$ for countries in 1990 considered with regard to variables $Y_2$ and $Y_3,$ whereas Figure 44 presents the same issue but for 2011. Although we can notice a socio-economic development between 1990 and 2011, the clusters of developed and developing countries are still evident in 2011 as they were in 1990. For assessing changes in the location of the centers and scatters of the data between 1990 and 2011, we calculated $L^2$ \textbf{medians} and $L^2$ \textbf{weighted covariance matrices for $(Y_1,Y_2,Y_3)$} that are presented below:\\
\begin{minipage}{.45\textwidth}
  \centering
   \vskip0.5mm
    MED(1990): (73.7; 55.2; 78.0)
    \vskip0.5mm
    MED(1995): (59.7; 45.7; 76.0)
    \vskip 0.5mm
    MED(2000): (53.7; 42.0; 85.0)
    \vskip 0.5mm
    MED(2005): (40.2; 32.6; 86.0)
    \vskip 0.5mm
    MED(2010): (33.6; 27.8; 89.0)
\end{minipage}%
\mbox{\hspace{0.1cm}} 
\begin{minipage}{.45\textwidth}
  \centering
    $$COV_{L^2}(1990)=\left( \begin{matrix}
2420.8 & 1453.9 &-396.3 \\
1453.9 & 903.4  &-238.6 \\
-396.3 & -238.6 & 228.3 \\
\end{matrix} \right)$$
\vskip 0.5mm
$$COV_{L^2}(2010)=\left( \begin{matrix}
 738.5 & 493.9 & -158.5 \\
 493.9 & 337.7 & -104.9 \\
 -158.5 & -104.9 & 121.2 \\
\end{matrix} \right)$$
\end{minipage}
\vskip 0.5mm
Figure 45 presents a DD-plot for inspecting the location changes between 1990 and 2011 for countries considered with respect to variables $Y_1, Y_2, Y_3$ and Figure 46 presents the DD-plot for inspecting the scale changes for the same data.
\begin{figure}
\centering
\begin{minipage}{.45\textwidth}
  \centering
  \includegraphics[width=.95\linewidth]{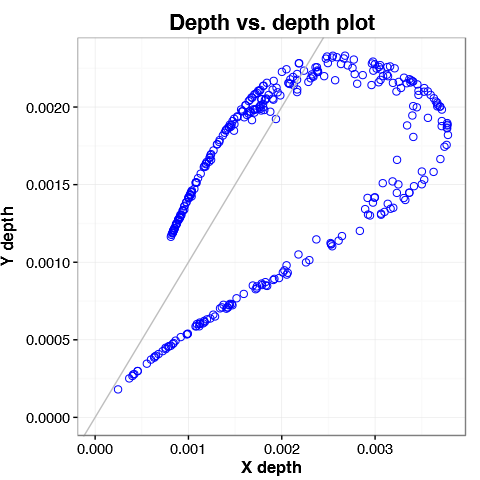}
  \caption{DD-plot for inspecting location differences.}
    \label{fig45}
\end{minipage}%
\mbox{\hspace{0.1cm}} 
\begin{minipage}{.45\textwidth}
  \centering
  \includegraphics[width=.95\linewidth]{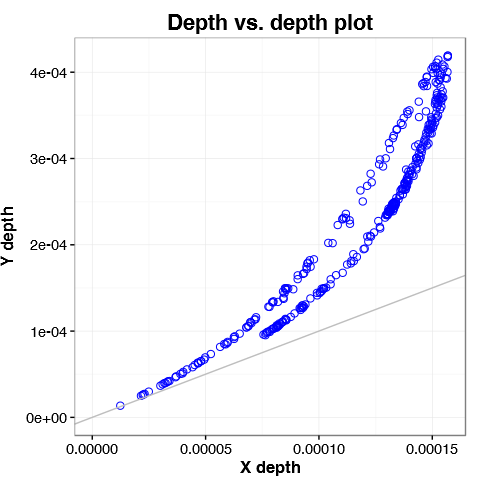}
  \caption{DD-plot for inspecting scale differences.}
    \label{fig46}
\end{minipage}
\end{figure}
We performed the multivariate Wilcoxon test (using $L^2$ depth) for scale change detection for the variables $Y_1,Y_2,Y_3$ in 1990 and 2011, induced by projection depth, and obtained: W=21150 and p-value=0.0046. We can therefore conclude that both the scale and the location have changed.

Figure 47 presents scale curves for the countries considered in the period of 1990--2011 jointly with respect to all variables whereas Figure 48 presents the Student depth contour plots for variable $Y_1$ in 1990--2011.
\begin{figure}
\centering
\begin{minipage}{.45\textwidth}
  \centering
  \includegraphics[width=.95\linewidth]{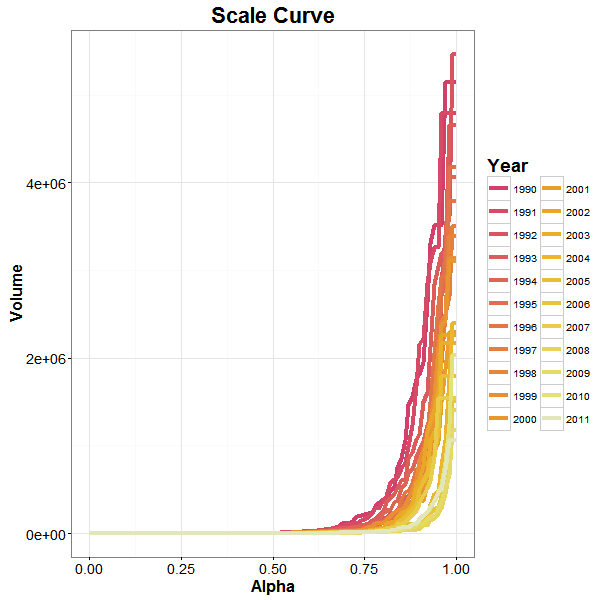}
        \caption{Scale curves for $(Y_1,Y_2,Y_3)$ 1990--2011.}
        \label{fig47}
\end{minipage}%
\mbox{\hspace{0.1cm}} 
\begin{minipage}{.45\textwidth}
  \centering
  \includegraphics[width=.95\linewidth]{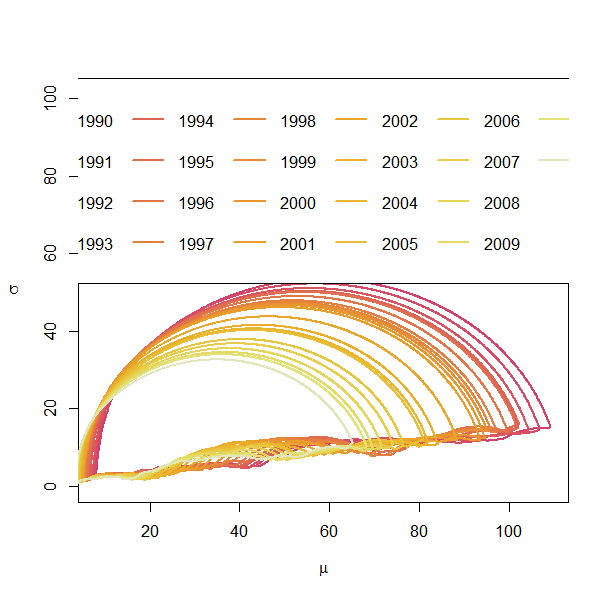}
       \caption{Student depth contour plots for $Y_1$ 1990--2011}
        \label{fig48}
\end{minipage}
\end{figure}
\begin{figure}
\centering
\begin{minipage}[t]{.45\textwidth}
  \centering
  \includegraphics[width=.95\linewidth]{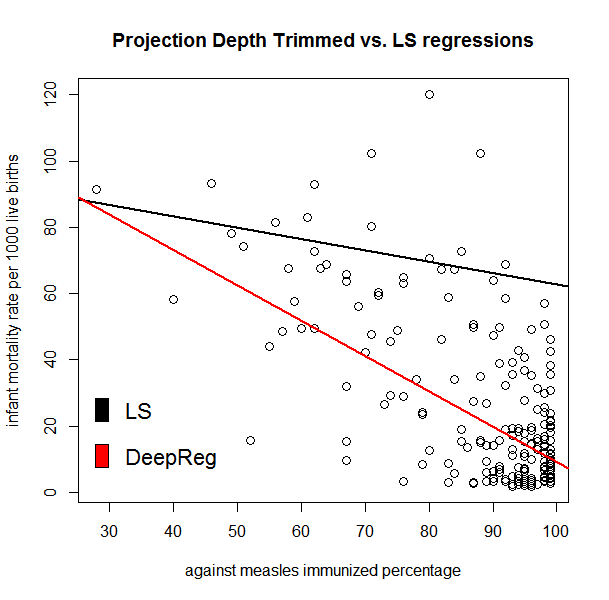}
  \caption{Deepest regression.}
  \label{fig49}
\end{minipage}
\mbox{\hspace{0.1cm}}
\begin{minipage}[t]{.45\textwidth}
  \centering
  \includegraphics[width=.95\linewidth]{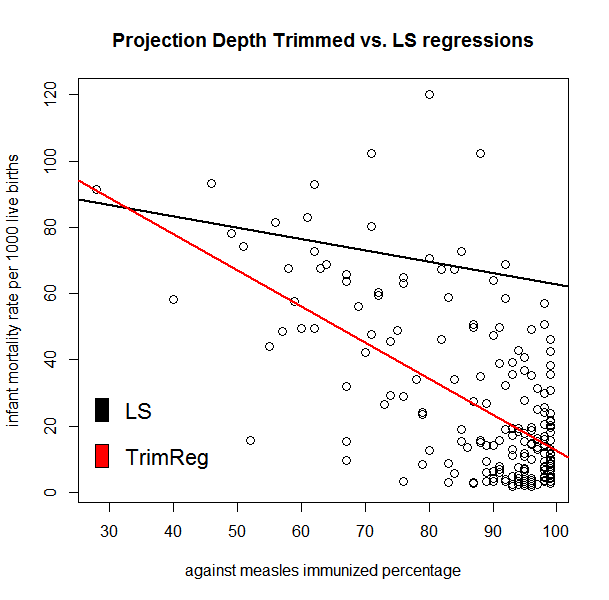}
  \caption{Least squares regression for the projection depth-trimmed data.}
  \label{fig50}
\end{minipage}
\end{figure}
Figures 49--50 present comparisons of least squares and two robust estimators of simple regression applied to express a relation between an infant mortality rate and children (one-year old) immunized against measles, in percentage. The robust estimators lead to stronger recommendations of vaccination than least squares estimator.
\vskip1mm
\textbf{The results of the analysis lead us to following conclusions:}
\begin{enumerate}
 \item{There are big chances for obtaining the 4MG. In the  year 2010, the decrease in the under-five-months child mortality was about 40\% with robust estimates being used.}
 \item{For the considered variables, both multivariate as well as univariate, scatters decreased in 1990--2011.}
 \item{The dispersion among countries, considered jointly with respect to variables $(Y_1,Y_2,Y_3),$ significantly decreased in 1990--2011, the clusters of \emph{the rich} and \emph{the poor} countries are still easily distinguishable.}
 \item{A comparison of Student depth medians of \emph{Children under-5-months mortality rate per 1000 live births} in 1990--2011 indicates the significant one-dimensional tendency for obtaining the 4MG.}
\item{The calculated simple deepest regressions for the variables and additional socio economic variables show clear relations between the 4MG indicators and other economic variables representing the economic development (for example, GDP per Capita).}
 \item{The data depth concept offers a complex family of powerful and user-friendly tools for nonparametric and robust analysis of socioeconomic multivariate data.}
 \end{enumerate}
Further considerations related to the issue can be found in \cite{Kosiorowska}.
\section{Summary}
This paper presents \pkg{R} package \pkg{DepthProc} that offers a selection of user-friendly robust, multivariate statistical methods originating from the DDC. Statistical procedures offered by the DDC very often significantly outperform the "classical" statistical methods. We hope that the existence of the \pkg{DepthProc} \pkg{R} package may increase the popularity of the DDC within economists and potential users in near future. \\
Theory and applications of the DDC are still being developed by many researchers. Recent findings presented in the literature concerning the DDC involve, among other, proposals of depths on infinite dimensional spaces, new algorithms for exact and approximate depth calculation, and new clustering and classification procedures for functional objects. The \pkg{DepthProc} package consists of a range of relatively simple but very powerful and user-friendly statistical tools, which are dedicated for conducting robust economic analysis. These tools may be successfully used for studying the new phenomena that appear in the current e-economy as well as for the analysis of classical economic issue, such as the evaluation of social inequalities.\\

Our plans for the future development of the package focus around the concept of local depth and depths for functional data (\cite{zakopanemediany}, \cite{zakopaneSVM}, \cite{Func_out}). In this context, we are working on the clustering and classification issues.  In a further perspective, we are planing to incorporate the DDC notions into the theory of economics and, in particular, into the theory of dynamic cooperative games (\cite{SGH}) where a notion of a center is of paramount importance for understanding nature of social choices. Decreasing the computational complexity of the procedures used within the package is our main aim, in the context of enlarging its popularity among analysts.
\section{General install info}
FOR A DEVELOPER VERSION (for Windows we need RTools)
\begin{Sinput}
R>require("devtools")
R>install_github("DepthProc", "zzawadz", subdir = "pkg")
\end{Sinput}
FOR INSTALLING The DepthProc for Windows from CRAN
\begin{Sinput}
R>install.packages("DepthProc")
\end{Sinput}
\section*{Acknowledgement}
Daniel Kosiorowski thanks the Polish NCS financial support DEC-011/03/B/HS4/01138 and the Faculty of Management of CUE grant 2017 and 2018 for preserving scientific resources.\\
The authors greatly appreciate the thoughtful and constructive remarks of the reviewers, which led to distinctive improvements in the paper and the package.
\bibliographystyle{jss}
\bibliography{literatura_JSS}
\end{document}